\documentclass[11pt,a4paper]{article}
\usepackage[english]{babel}
\usepackage{cite}

\usepackage{jheppub}
\usepackage{epsf}
\usepackage{amssymb}
\usepackage{amsmath}
\usepackage{amsfonts}
\usepackage{psfrag,epsfig,graphicx,graphics}
\usepackage{subcaption}
\usepackage{wasysym}
\usepackage{setspace}

\title{Interplay of the CGC and TMD frameworks to all orders in kinematic twist}
\author[a]{Tolga Altinoluk,}
\author[b,c]{Renaud Boussarie}
\author[c]{and Piotr Kotko}

\affiliation[a]{National Centre for Nuclear Research, 
00-681 Warsaw, Poland}
\affiliation[b]{Physics Department, Brookhaven National Laboratory, 
Upton, NY 11973, USA}
\affiliation[c]{Institute of Nuclear Physics Polish Academy of Sciences, 
 PL-31342 Krakow, Poland}

\emailAdd{tolga.altinoluk@ncbj.gov.pl}
\emailAdd{rboussarie@bnl.gov}
\emailAdd{piotr.kotko@ifj.edu.pl}

\abstract{A framework for an improved TMD (iTMD) factorization scheme at small $x$, involving off-shell perturbative subamplitudes, was recently developed as an interpolation between the TMD $k_t \ll Q$ regime and the BFKL $k_t \sim Q$ regime. In this article, we study the relation between CGC and iTMD amplitudes. We first show how the dipole-size expansion of CGC amplitudes resembles the twist expansion of a TMD amplitude. Then, by isolating kinematic twists, we prove that iTMD amplitudes are obtained with infinite kinematic twist accuracy by simply getting rid of all genuine twist contributions in a CGC amplitude. Finally we compare the amplitudes obtained via a proper kinematic twist expansion to those obtained via a more standard dilute expansion to show the relation between the iTMD framework and the dilute low $x$ framework.}

\begin{document}

\pagestyle{empty} \newpage{}

\mbox{%
}

\maketitle

\pagestyle{plain}

\setcounter{page}{1}

\section{Introduction}

Factorization is one of the most crucial features of QCD: all perturbative
QCD studies rely on this separation between a hard partonic subamplitude
and long distance matrix elements. This separation is justified in
the presence of a sufficiently large scale $Q$ in the observable,
for which $\alpha_{s}(Q)$ is small enough for perturbation theory
to apply. However large logarithms can arise from QCD
dynamics and compensate the smallness of $\alpha_{s}(Q),$ which makes
the resummation of such logarithms necessary.

For most observables, two different factorization schemes
can be employed, depending on the center-of-mass energy $s$
 of the
process. For processes with the center-of-mass energy comparable to the large scale 
of the process ($s\sim Q$), collinear factorization is applied and the large $\log(Q)$
terms are resummed via the Dokshitzer-Gribov-Lipatov-Altarelli-Parisi
(DGLAP) evolution equations \cite{Gribov:1972ri, Altarelli:1977zs, Dokshitzer:1977sg}. On the other hand, processes with the center-of-mass energy much larger than any other scale 
($s\gg Q$) are treated in the so-called low-$x$ regime. In this case, $k_{t}$-factorization
applies and large $\log(s)$ terms are resummed. 

Several descriptions of $k_{T}$-factorization for low-$x$ physics
have been developed over the last couple of decades, starting with the well known Balitsky-Fadin-Lipatov-Kuraev
(BFKL) framework \cite{Kuraev:1977fs, Balitsky:1978ic}.
The most
recent low-$x$ frameworks, namely the dipole model \cite{Mueller:1989st, Mueller:1993rr, Mueller:1994gb} and the shockwave framework \cite{Balitsky:1995ub, Balitsky:1998kc, Balitsky:1998ya} rely on a semi-classical approach, where low
$x$ gluon fields are treated as external fields. With such a treatment,
all interactions with the external field can be resummed into path-ordered
Wilson line operators which then constitute the building blocks of
these low-\textbf{$x$ }formalisms. Remarkably, due to this resummation
of all interactions, perturbative results from this framework were
found to be compatible with previous results for the semi-classical
treatment of scattering off dense targets \cite{McLerran:1993ni, McLerran:1993ka, McLerran:1994vd} which include gluon saturation effects from multiple
scatterings. All of these recent frameworks are equivalent, and logarithms
are resummed via the Balitsky/Jalilian-Marian-Iancu-McLerran-Weigert-Leonidov-Kovner (B-JIMWLK) hierarchy of evolution equations \cite{JalilianMarian:1997jx, JalilianMarian:1997gr, JalilianMarian:1997dw, Kovner:1999bj, Kovner:2000pt, Weigert:2000gi, CGC, Ferreiro:2001qy}, or in
the mean field approximation by the Balitsky-Kovchegov (BK) equation \cite{Balitsky:1995ub, Kovchegov}. Nowadays, the weak coupling non-perturbative realization of the saturation in QCD is referred to as  the Color Glass Condensate (CGC) \cite{CGC, Ferreiro:2001qy}. Throughout this paper, we refer to CGC as a unified picture (Balitsky formalism/ Mueller's dipole picture/CGC) of the small-$x$ QCD.

The fact that the CGC generalizes the BFKL framework was established
early on at Leading-Logarithmic (LL) accuracy \cite{Mueller_BFKL, Chen:1995pa} and made more explicit in \cite{Caron-Huot:2013fea}, then at Next-to-Leading-Logarithmic accuracy (NLL) in \cite{Balitsky:2008zza} and more explicitly in  \cite{Fadin:2007ee, Fadin:2007de, Fadin:2011jg}. This equivalence relies on the expansion
of the path-ordered Wilson lines in powers of the gluon field for
small values of $gA$, what is known as the dilute limit.

Although it is not a true all-order factorization scheme, as opposed
to collinear factorization for several simple processes \cite{Collins:2011zzd}, the CGC framework applies in principle to any low-$x$
or high-density process regardless of the number of observed scales.
In contrast, collinear factorization in its most common form is not
valid for processes involving not only a hard scale $Q$, but also
a second, smaller scale. 
In the present context the most interesting case 
is when that smaller scale is related to the transverse momentum 
of a parton inside a hadron.
The collinear distributions were generalized
for such processes, leading to the Transverse Momentum Dependent (TMD)
factorization scheme \cite{Collins:2011zzd,Collins:1983pk,Boer:1999si,Brodsky:2002cx,Collins:2002kn,Belitsky:2002sm,Bomhof:2004aw,TMD}.

For a process with center-of-mass energy $s$, a hard scale $Q$,
and a hard yet softer transverse momentum scale $\left|\boldsymbol{k}\right|\apprge\Lambda_{QCD},$
the respective application ranges of CGC and TMD schemes are $s\gg Q\apprge\left|\boldsymbol{k}\right|$
and $s\sim Q\gg\left|\boldsymbol{k}\right|$. A matching of these
schemes in the overlapping regime where $\left|\boldsymbol{k}\right|/Q$
and $Q/s$ are both small was proven in \cite{Dominguez:2010xd,Dominguez:2011wm}. Since then, gluon TMDs in the CGC have been
at stake in many recent studies (see for example \cite{Marquet:2016cgx, Marquet:2017xwy, Altinoluk:2018byz, Petreska:2018cbf} ). Indeed the
measurement of TMD parton distributions offers great insight in the
3D structure of hadrons, yet these distributions are not fully universal
and thus they require case-by-case studies. Studying them at low-$x$
allows one to use standard CGC tools like the McLerran-Venugopalan
(MV) model \cite{McLerran:1993ni, McLerran:1993ka, McLerran:1994vd}, Golec-Biernat-W\"usthoff 
(GBW) parametrization \cite{GolecBiernat:1998js} or numerical
solutions to the B-JIMWLK hierarchy of equations \cite{Marquet:2016cgx} for the description of these complicated
TMD distributions.

Notable attention has been drawn to polarized TMDs and to their role in angular distributions at low-$x$ \cite{Metz:2011wb, Akcakaya:2012si, Dumitru:2016jku, Boer:2017xpy, Marquet:2017xwy, Altinoluk:2018byz}, and the relation between process-independence
breaking in TMD factorization and the Wilson lines which are natural
built-in features of the CGC \cite{Petreska:2018cbf}.
 
On the other hand, the CGC framework in the so-called dilute limit
also matches BFKL results, which were built for processes with different kinematics,
where $s\gg Q\sim\left|\boldsymbol{k}\right|.$
A new scheme for TMD factorization at low-$x$, which is referred to as the improved TMD scheme (iTMD), was built in \cite{Kotko:2015ura, vanHameren:2016ftb} as an attempt to interpolate between both $\left|\boldsymbol{k}\right|\ll Q$ and $\left|\boldsymbol{k}\right|\sim Q$
limits. This framework aims at resumming some powers of $\left|\boldsymbol{k}\right|/Q$
by taking into account non-zero $\boldsymbol{k}$ in the hard subamplitude.
In practice, as we will show in this article, it resums all kinematic
twist corrections to the hard subamplitude which couples to the leading-twist TMD operator, leaving genuine
twist corrections aside.
For an alternative approach for twist studies in the saturation regime, see~\citep{Bartels:2009tu}.

The purpose of this paper is to study the relation between CGC and
iTMD amplitudes, with a comparison with dilute BFKL amplitudes as
well.
 It is organized as follows. In section~\ref{sec:PowExp}, we
consider the first corrections to the correlation limit in a CGC amplitude
and compare them to the first power corrections in the TMD factorization,
and show how both expansions are related to one another. Then in section~\ref{sec:CGCexpansion},
we start with the most generic form for $1\rightarrow2$ processes
in the CGC and expand it in powers of the dipole size. We extract
the pure kinematic twist corrections and resum them to infinite accuracy.
{\it This leads to the main result of this article: a completely generic infinite-twist
CGC amplitude in Eq.~(\ref{eq:genCrossSec}) in an all-body Wandzura-Wilczek
approximation (i.e. where all genuine twist corrections are neglected)}.
In section~\ref{sec:Dilute}, we start again from the generic CGC amplitude
and perform a more standard dilute expansion, leading
to a generic dilute CGC amplitude in Eq.~(\ref{eq:DilCSuPDF}). 
Section~\ref{sec:iTMD} is devoted to a short review of the iTMD framework and 
to recalculating the iTMD cross sections in a form that can be compared with the CGC
all-kinematic-twists result. In section~\ref{sec:Applications}, we apply the generic kinematic twist resummed CGC result for different processes and compare them to the iTMD predictions. We find a perfect match between the kinematic twist resummed cross sections for each process and the corresponding iTMD results. Moreover, we also compare the dilute limit of the generic CGC cross sections with the kinematic twist ressumed cross sections by simply setting all distributions to the same value and find a perfect matching as well. Finally, in section~\ref{sec:Discussions} we summarize and discuss our findings for this study.

%%%%%%%%%%%%%%%%%%%%%%%%%%%%%%%%%%%%%%%%%%%%%%%%%%%%%%%%%%%%%%%%%%%%%%%%
%%%%%%%%%%%%%%%%%%%%%%%%%%%%%%%%%%%%%%%%%%%%%%%%%%%%%%%%%%%%%%%%%%%%%%%%
\section*{Notations and conventions}

We define  two lightlike vectors $n_1$ and $n_2$ such that $n_1 \cdot n_2=1$, and light cone directions $+$ and $-$ such that $n_1 \cdot k=k^-$, $n_2\cdot k= k^+$. The projectile (resp. target) is assumed to have a large momentum $\sim \sqrt{s}$ along the $+$ (resp. $-$) direction. In the CGC calculations we use the lightcone gauge $A^+=0$. Transverse components are denoted with a $\perp$  subscript in Minkowski space and by bold characters in Euclidean space.
Therefore, for two vectors $k$ and $x$, we write
\begin{equation}
k\cdot x = k^+ x^- + k^- x^+ + k_\perp\cdot x_\perp = k^+ x^- +k^-x^+ -\textbf{k}\cdot \textbf{x}
\end{equation}
The CGC part of this paper relies on the separation of the gluon fields in the QCD Lagrangian depending on their $+$ momentum between \textit{fast} fields ($k^+>e^{-Y}p^+$) and \textit{slow} fields ($k^+<e^{-Y}p^+$). In the eikonal approximation, the slow fields have the shockwave form
\begin{equation}
A^\mu(x) = \delta(x^+)\mathbf{B}(x_\perp) n_2^\mu + O(s^{-1/2}),
\end{equation}
where $\mathbf{B}$ is a function of $x_\perp$  only. In the semi-classical approximation for the slow fields, treated as external fields for the projectile, interactions with the target are resummed into path-ordered Wilson lines
\begin{equation}
[a^+,b^+]_{\mathbf{x}} = \mathcal{P}\exp\left[ig\int_{a^+}^{b^+}dz^+ A^-(z^+,0,x_\perp)\right],\label{WilsonLine}
\end{equation}
and we write 
\begin{equation}
\label{Wilson2}
U_{\mathbf{x}}=[-\infty,+\infty]_{\mathbf{x}}.
\end{equation}
CGC Wilson line operators carry a color representation, in which case we define $U_{\mathbf{x}}^R$ as the Wilson line obtained from Eq.~(\ref{WilsonLine}) by replacing $A^-(x) \rightarrow T_R^a A_a^-(x)$.
Finally, we use the CGC brackets to describe the normalized forward actions of Wilson line operators on target states $\left|P\right\rangle$. For an operator $\mathcal{O}$  we define the brackets as:
\begin{equation}
\label{brackets}
\left\langle \mathcal{O} \right\rangle \equiv \frac{\left\langle P \left| \mathcal{O} \right| P \right\rangle}{\left\langle P | P \right\rangle}.
\end{equation}

%%%%%%%%%%%%%%%%%%%%%%%%%%%%%%%%%%%%%%%%%%%%%%%%%%%%%%%%%%%%%%%%%%%%%%%%
%%%%%%%%%%%%%%%%%%%%%%%%%%%%%%%%%%%%%%%%%%%%%%%%%%%%%%%%%%%%%%%%%%%%%%%%

\section{Correlation limit and TMD power expansion\label{sec:PowExp}}

In this work we study processes that describe the production of a pair of particles with a large invariant mass from a single particle in an external shockwave field built from the target gluons. We consider the case when both outgoing particles are tagged and their transverse momenta 
are fully reconstructed. The produced particles carry longitudinal momenta $p_1^+$ and $p_2^+$, and transverse momenta $\boldsymbol{p}_{1}$  and $\boldsymbol{p}_{2}$. The two important combinations of these momenta are the sum of the two transverse momenta $\boldsymbol{k}$ 
\begin{equation}
\label{k}
\boldsymbol{k}\equiv\boldsymbol{p}_{1}+\boldsymbol{p}_{2}
\end{equation}
and  the transverse-boost invariant momentum $\boldsymbol{q}$ which is defined as 
\begin{equation}
\label{q}
\boldsymbol{q}\equiv\frac{p_{2}^{+}\boldsymbol{p}_{1}-p_{1}^{+}\boldsymbol{p}_{2}}{p_{1}^{+}+p_{2}^{+}}\; . 
\end{equation}
The hard scale $Q$ of the process is given by the invariant mass of the outgoing pair which is directly related to the transverse boost invariant momentum: 
\begin{equation}
Q^{2}=\frac{\left(p_{1}^{+}+p_{2}^{+}\right)^{2}}{2p_{1}^{+}p_{2}^{+}}\boldsymbol{q}^{2}=\frac{\boldsymbol{q}^{2}}{2z\bar{z}},\label{eq:Hardscale}
\end{equation}
where
\begin{equation}
\label{frac}
z\equiv\frac{p_{1}^{+}}{p_{1}^{+}+p_{2}^{+}}\equiv1-\bar{z}\, .
\end{equation}

As discussed in detail in \cite{Dominguez:2011wm}, one can get the gluon TMDs through CGC calculations in certain limit which is usually referred to as "back-to-back correlation limit". In this limit, the two transverse scales $\left|\boldsymbol{k}\right|$ and $\left|\boldsymbol{q}\right|$ are well separated, i.e. $\left|\boldsymbol{q}\right|\gg\left|\boldsymbol{k}\right|$. In the CGC framework, the transverse boost invariant momentum $\boldsymbol{q}$ is Fourier conjugate to the transverse size of the produced pair (dipole size) $\boldsymbol{r}$ and the total transverse momentum is conjugate to the impact parameter $\boldsymbol{b}$. Therefore, the back-to-back correlation limit corresponds to the case $\left|\boldsymbol{r}\right|\ll\left|\boldsymbol{b}\right|$ in coordinate space allowing a Taylor expansion of the CGC observables in the dipole size $\boldsymbol{r}$. 

We start by clarifying the power expansion employed here and in the rest of this section we consider a simple process in the back-to-back correlation limit to utilize the small dipole size expansion in the CGC framework and compare it with the power expansion in the TMD factorization framework to clarify the relation between the two procedures.  

%%%%%%%%%%%%%%%%%%%%%%%%%%%%%%%%%%%%%%%%%%%%%%%%%%%%%%%%%%%%%%%%%%%%%%%%
\subsection{Power expansion at the amplitude level}
The TMD framework involves gauge invariant \textit{light ray}
operators~\citep{Balitsky:1987bk}, for which the distinction between kinematic twists and
genuine twists is convenient. For a set of gauge invariant twist $p$
operators\footnote{Note that in a light ray OPE, the gauge links in the operators are
not taken into account in the counting of twists.} $\mathcal{O}_{p}^{\left(i\right)}$ associated with the hard part
$\mathcal{H}_{p}^{\left(i\right)}$, the $n$-th power of $k_\perp$ in the cross
section is given by the sum over $p\in \{0 \cdots n\}$ of the $p$-th
power in $\mathcal{H}_{n-p}^{\left(i\right)}$ convoluted with
$\mathcal{O}_{n-p}^{\left(i\right)}$ and summed over all $i$. 

%  \\
For inclusive observables, power corrections are split between amplitudes
and complex conjugate amplitudes. However for the sake of this article,
which aims at comparing CGC and iTMD results, it is actually sufficient
to study power corrections at the amplitude level. Rather than using
full, gauge invariant, inclusive operators, it is also enough for
the comparison to use "half"-operators
at the amplitude level, knowing how they would get combined into gauge
invariant inclusive operators at the cross section level.\\
In the particular cases studied in this article, $\mathcal{O}_{p}^{\left(i\right)}$
will be a set of $p$-body gluon light ray half-operators
\begin{equation}
\mathcal{O}_{p}^{\left(i\right)}\left(x_{1},...,x_{p}\right)=\left[\pm\infty,x_{1}\right]F^{-j_{1}}\left(x_{1}\right)\left[x_{1},x_{2}\right]F^{-j_{2}}\left(x_{2}\right)...\left[x_{p-1},x_{p}\right]F^{-j_{p}}\left(x_{p}\right)\left[x_{p},\pm\infty\right].\label{eq:OpeP}
\end{equation}
We refer to $\mathcal{O}_{p}^{\left(i\right)}$ as a $p$-body
operator, with $\mathcal{O}_{1}^{\left(i\right)}$ being the set of
leading 1-body operators, which would combine into the leading twist
(2-body in the standard counting) TMDs at the cross section level.
Then the $n$-th power correction is given by the sum of $p$-th power in the $\left(n-p\right)$-body hard part, convoluted with the $\left(n-p\right)$-body
operator. Corrections from the hard parts are kinematic twists, while
higher-body operators lead to genuine twist corrections. In particular,
fully kinematic twists, that are the main focus of this study,
are given by successive $k_\perp$-derivatives of the 1-body hard part.

%%%%%%%%%%%%%%%%%%%%%%%%%%%%%%%%%%%%%%%%%%%%%%%%%%%%%%%%%%%%%%

\subsection{Dipole size expansion for $\gamma\rightarrow q\bar{q}$ in the CGC\label{subsec:GenExp}}

It is informative to start by computing the first few corrections
to the correlation limit in the CGC. As a simple example, let us consider the amplitude for the photoproduction of a quark-antiquark dijet which reads 
\begin{eqnarray}
\label{eq:GammatoQQbarini}
\hspace{-0.5cm}
\mathcal{A}_{\gamma\rightarrow q\bar{q}} & =&
\left(2\pi\right)\delta\left(p_{q}^{+}+p_{\bar{q}}^{+}-p_{\gamma}^{+}\right)\int d^{2}\boldsymbol{b}\, d^{2}\boldsymbol{r}\, e^{-i\left(\boldsymbol{q}\cdot\boldsymbol{r}\right)-i\left(\boldsymbol{k}\cdot\boldsymbol{b}\right)}
 \frac{r_{\perp}^{\mu}}{\boldsymbol{r}^{2}}\left[\left(U_{\boldsymbol{b}+\bar{z}\boldsymbol{r}}U_{\boldsymbol{b}-z\boldsymbol{r}}^{\dagger}\right)-\mathbf{1}\right]
\phi_{\mu}
\end{eqnarray} 
where the Wilson lines $U_{\boldsymbol{b}+\bar{z}\boldsymbol{r}}$ are defined in Eq. \eqref{Wilson2} with Eq. \eqref{WilsonLine}. Here, $\phi_\mu$ is the tensor part of the amplitude that encodes the Dirac structure for this process and it is defined as 
\begin{align}
\phi_{\mu}^{\left(\gamma\rightarrow q\bar{q}\right)} & =i\frac{e_{q}}{2\pi}\varepsilon_{p\perp}^{\sigma}\bar{u}_{p_{q}}\left[2zg_{\perp\mu\sigma}-\left(\gamma_{\perp\mu}\gamma_{\perp\sigma}\right)\right]\gamma^{+}v_{p_{\bar{q}}}.\label{eq:GammatoQQbartens}
\end{align}
In the correlation limit, it is straightforward to expand this amplitude in powers of the small dipole size $\boldsymbol{r}$ and keep the first two terms in the expansion. After performing a simple integration by parts, the result can be written as 
\begin{align}
\mathcal{A}_{\gamma\rightarrow q\bar{q}} & =i\frac{e_{q}}{2\pi}\boldsymbol{\varepsilon}_{p}^{l}\left(2\pi\right)\delta\left(p_{q}^{+}+p_{\bar{q}}^{+}-p_{g}^{+}\right)\int\!d^{2}\boldsymbol{b}\,d^{2}\boldsymbol{r}\,e^{-i\left(\boldsymbol{q}\cdot\boldsymbol{r}\right)-i\left(\boldsymbol{k}\cdot\boldsymbol{b}\right)}\nonumber \\
 & \times\frac{\boldsymbol{r}^{i}\boldsymbol{r}^{j}}{\boldsymbol{r}^{2}}\bar{u}_{p_{q}}\left(2z\delta^{il}+\boldsymbol{\gamma}^{i}\boldsymbol{\gamma}^{l}\right)\gamma^{+}v_{p_{\bar{q}}}\left[\frac{1}{2}\boldsymbol{r}^{k}\left(\partial^{j}U_{\boldsymbol{b}}\right)\left(\partial^{k}U_{\boldsymbol{b}}^{\dagger}\right)\right.\label{eq:Photo2Terms}\\
 & +\left.\bar{z}\left(\partial^{j}U_{\boldsymbol{b}}\right)U_{\boldsymbol{b}}^{\dagger}\left(1+\frac{1}{2}\left(i\bar{z}\boldsymbol{k}\cdot\boldsymbol{r}\right)\right)-zU_{\boldsymbol{b}}\left(\partial^{j}U_{\boldsymbol{b}}^{\dagger}\right)\left(1-\frac{1}{2}\left(iz\boldsymbol{k}\cdot\boldsymbol{r}\right)\right)\right] \,. \nonumber 
\end{align}
$O(1)$ terms in Eq.\eqref{eq:Photo2Terms} give the well-known back-to-back result which reads
\begin{eqnarray}
\label{eq:B2B}
\mathcal{A}_{\gamma\rightarrow q\bar{q}}^{\left(b2b\right)} & =i\frac{e_{q}}{2\pi}\boldsymbol{\varepsilon}_{p}^{l}\left(2\pi\right)\delta\left(p_{q}^{+}+p_{\bar{q}}^{+}-p_{g}^{+}\right)\int\!d^{2}\boldsymbol{b}\,d^{2}\boldsymbol{r}\,e^{-i\left(\boldsymbol{q}\cdot\boldsymbol{r}\right)-i\left(\boldsymbol{k}\cdot\boldsymbol{b}\right)}\nonumber \\
 & \times\frac{\boldsymbol{r}^{i}\boldsymbol{r}^{j}}{\boldsymbol{r}^{2}}\bar{u}_{p_{q}}\left(2z\delta^{il}+\boldsymbol{\gamma}^{i}\boldsymbol{\gamma}^{l}\right)\gamma^{+}v_{p_{\bar{q}}}
 \left[\bar{z}\left(\partial^{j}U_{\boldsymbol{b}}\right)U_{\boldsymbol{b}}^{\dagger}-zU_{\boldsymbol{b}}\left(\partial^{j}U_{\boldsymbol{b}}^{\dagger}\right)\right],
\end{eqnarray}
that has been proven to match the leading twist TMD amplitude. The rest of the terms are $O(\boldsymbol{r})$ in Eq.\eqref{eq:Photo2Terms} that are corrections to the back-to-back result:
\begin{align}
\mathcal{A}_{\gamma\rightarrow q\bar{q}}^{\left(nb2b\right)} & =i\frac{e_{q}}{2\pi}\boldsymbol{\varepsilon}_{p}^{l}\left(2\pi\right)\delta\left(p_{q}^{+}+p_{\bar{q}}^{+}-p_{g}^{+}\right)\int\!d^{2}\boldsymbol{b}\,d^{2}\boldsymbol{r}\,e^{-i\left(\boldsymbol{q}\cdot\boldsymbol{r}\right)-i\left(\boldsymbol{k}\cdot\boldsymbol{b}\right)}\nonumber \\
 & \times\frac{1}{2}\frac{\boldsymbol{r}^{i}\boldsymbol{r}^{j}}{\boldsymbol{r}^{2}}\bar{u}_{p_{q}}\left(2z\delta^{il}+\boldsymbol{\gamma}^{i}\boldsymbol{\gamma}^{l}\right)\gamma^{+}v_{p_{\bar{q}}}\label{eq:N2B2B}\\
 & \times\left[\boldsymbol{r}^{k}\left(\partial^{j}U_{\boldsymbol{b}}\right)\left(\partial^{k}U_{\boldsymbol{b}}^{\dagger}\right)+\bar{z}\left(\partial^{j}U_{\boldsymbol{b}}\right)U_{\boldsymbol{b}}^{\dagger}\left(i\bar{z}\boldsymbol{k}\cdot\boldsymbol{r}\right)-zU_{\boldsymbol{b}}\left(\partial^{j}U_{\boldsymbol{b}}^{\dagger}\right)\left(-\left(iz\boldsymbol{k}\cdot\boldsymbol{r}\right)\right)\right].\nonumber 
\end{align}
We would like to emphasize that the next-to-back-to-back term, Eq.~\eqref{eq:N2B2B}, has a very interesting form. Noting the fact that a derivative acting on a CGC Wilson line extracts a gluon field, one can immediately conclude that the first term in the brackets is a 2-body half-operator.
On the other hand, one can manipulate the last two terms using the fact that 
\begin{eqnarray}
i\bar{z}\boldsymbol{r}^{l}e^{-i\left(\boldsymbol{q}\cdot\boldsymbol{r}\right)}&=&-\frac{\partial}{\partial\boldsymbol{p}_{q}^{l}}e^{-i\left(\boldsymbol{q}\cdot\boldsymbol{r}\right)}\, ,\\
-iz\boldsymbol{r}^{l}e^{-i\left(\boldsymbol{q}\cdot\boldsymbol{r}\right)}&=&-\frac{\partial}{\partial\boldsymbol{p}_{\bar{q}}^{l}}e^{-i\left(\boldsymbol{q}\cdot\boldsymbol{r}\right)}\, ,
\end{eqnarray}
so that the next-to-back-to-back term can be written as 
\begin{align}
\mathcal{A}_{\gamma\rightarrow q\bar{q}}^{\left(nb2b\right)} & =i\frac{e_{q}}{2\pi}\boldsymbol{\varepsilon}_{p}^{l}\left(2\pi\right)\delta\left(p_{q}^{+}+p_{\bar{q}}^{+}-p_{g}^{+}\right)\int\!d^{2}\boldsymbol{b}\,d^{2}\boldsymbol{r}\,e^{-i\left(\boldsymbol{k}\cdot\boldsymbol{b}\right)}\nonumber \\
 & \times\frac{1}{2}\frac{\boldsymbol{r}^{i}\boldsymbol{r}^{j}}{\boldsymbol{r}^{2}}\bar{u}_{p_{q}}\left(2z\delta^{il}+\boldsymbol{\gamma}^{i}\boldsymbol{\gamma}^{l}\right)\gamma^{+}v_{p_{\bar{q}}}\label{eq:N2B2Bder}\\
 & \times\left[\boldsymbol{r}^{k}\left(\partial^{j}U_{\boldsymbol{b}}\right)\left(\partial^{k}U_{\boldsymbol{b}}^{\dagger}\right)-\bar{z}\left(\partial^{j}U_{\boldsymbol{b}}\right)U_{\boldsymbol{b}}^{\dagger}\left(\boldsymbol{k}\cdot\frac{\partial}{\partial\boldsymbol{p}_{q}}\right)+zU_{\boldsymbol{b}}\left(\partial^{j}U_{\boldsymbol{b}}^{\dagger}\right)\left(\boldsymbol{k}\cdot\frac{\partial}{\partial\boldsymbol{p}_{\bar{q}}}\right)\right]e^{-i\left(\boldsymbol{q}\cdot\boldsymbol{r}\right)}.\nonumber 
\end{align}
At this point we can make a diagram-by-diagram correspondence with TMD factorization. Naturally,  $\left(\partial^{j}U_{\boldsymbol{b}}\right)U_{\boldsymbol{b}}^{\dagger}$
terms correspond to the diagram where the TMD gluon hits the quark,
while $U_{\boldsymbol{b}}\left(\partial^{j}U_{\boldsymbol{b}}^{\dagger}\right)$
terms correspond to the diagram where it hits the antiquark. For such
diagrams, it is easy to see that the dependance on $\boldsymbol{k}$
and $\boldsymbol{p}_{q}$ (resp. $\boldsymbol{k}$ and $\boldsymbol{p}_{\bar{q}}$)
is only in the intermediate quark (resp. antiquark) propagator $G\left(\boldsymbol{k}+\boldsymbol{p}_{q}\right)$
(resp. $G\left(\boldsymbol{k}-\boldsymbol{p}_{\bar{q}}\right)$). Thus for those diagrams we have  
\begin{eqnarray}
\boldsymbol{k}\cdot\frac{\partial}{\partial\boldsymbol{p}_{q}}&=&\boldsymbol{k}\cdot\frac{\partial}{\partial\boldsymbol{k}}\, ,
\\
\boldsymbol{k}\cdot\frac{\partial}{\partial\boldsymbol{p}_{\bar{q}}}&=&-\boldsymbol{k}\cdot\frac{\partial}{\partial\boldsymbol{k}}\, .
\end{eqnarray}
Hence, the next-to-back-to-back contribution can be cast into the following form: 
\begin{align}
\mathcal{A}_{\gamma\rightarrow q\bar{q}}^{\left(nb2b\right)} & =\int\!d^{2}\boldsymbol{b}\,e^{-i\left(\boldsymbol{k}\cdot\boldsymbol{b}\right)}\left[\left(\partial^{j}U_{\boldsymbol{b}}\right)\left(\partial^{k}U_{\boldsymbol{b}}^{\dagger}\right)\mathcal{H}_{2}^{jk}+\left(\partial^{j}U_{\boldsymbol{b}}\right)U_{\boldsymbol{b}}^{\dagger}\left(\boldsymbol{k}\cdot\frac{\partial}{\partial\boldsymbol{k}}\right)\mathcal{H}_{1}^{j}\right]\,,\label{eq:N2B2Bform}
\end{align}
where $\mathcal{H}_{2}^{jk}$ is a 2-body hard subamplitude, and $\mathcal{H}_{1}^{j}$
is a 1-body hard subamplitude (given by the sum of the two diagrams
discussed above). 

\subsection{TMD power corrections to $\gamma\rightarrow q\bar{q}$\label{subsec:photoTMD}}

For the photoproduction of a quark-antiquark dijet, the 1-body amplitude\footnote{We write the amplitude in an operator form, similarly to what is done in the CGC. The true amplitude is given by the action of this operator on target states. } for TMD factorization has the following form:  
\begin{align}
\mathcal{A}_{1}\left(\boldsymbol{k}\right) & =ig\int\!\frac{d^{2}\boldsymbol{k}_{1}}{\left(2\pi\right)^{2}}\left(2\pi\right)^{2}\delta^{2}\left(\boldsymbol{k}_{1}-\boldsymbol{k}\right)\mathcal{H}_{1}^{i}\left(\boldsymbol{k}_{1}\right)\int\!db_{1}^{+}d^{2}\boldsymbol{b}_{1}e^{-i\left(\boldsymbol{k}_{1}\cdot\boldsymbol{b}_{1}\right)}\label{eq:N2B2BTMD}\\
 & \times\left[-\infty,b_{1}^{+}\right]_{\boldsymbol{b}_{1}}F^{-i}\left(b_{1}\right)\left[b_{1}^{+},-\infty\right]_{\boldsymbol{b}_{1}},\nonumber 
\end{align}
where $\mathcal{H}_{1}^{i}\left(\boldsymbol{k}_{1}\right)$ is a hard
subamplitude. Power corrections are obtained via the Taylor expansion
of this hard part. Up to the first correction, rewriting the TMD operator
as the derivative of a Wilson line, it reads: 
\begin{align}
\mathcal{A}_{1}\left(\boldsymbol{k}\right) & \simeq\int\!d^{2}\boldsymbol{b}\,e^{-i\left(\boldsymbol{k}\cdot\boldsymbol{b}\right)}\left[\mathcal{H}_{1}^{i}\left(\boldsymbol{0}\right)-\left(\boldsymbol{k}\cdot\frac{\partial}{\partial\boldsymbol{k}}\mathcal{H}_{1}^{i}\right)\left(\boldsymbol{0}\right)\right]\left(\partial^{i}U_{\boldsymbol{b}}\right)U_{\boldsymbol{b}}^{\dagger}\,.\label{eq:1BTMD}
\end{align}
The 2-body amplitude for the same process can be written as 
\begin{align}
\mathcal{A}_{2}\left(\boldsymbol{k}\right) & =g^{2}\int\frac{d^{2}\boldsymbol{k}_{1}}{\left(2\pi\right)^{2}}\frac{d^{2}\boldsymbol{k}_{2}}{\left(2\pi\right)^{2}}\left(2\pi\right)^{2}\delta^{2}\left(\boldsymbol{k}_{1}+\boldsymbol{k}_{2}-\boldsymbol{k}\right)\mathcal{H}_{2}^{ij}\left(\boldsymbol{k}_{1},\boldsymbol{k}_{2}\right) \int\!db_1^+db_2^+ d^{2}\boldsymbol{b}_{1}d^{2}\boldsymbol{b}_{2}\label{eq:2BTMD}\\
 & \times e^{-i\left(\boldsymbol{k}_{1}\cdot\boldsymbol{b}_{1}\right)-i\left(\boldsymbol{k}_{2}\cdot\boldsymbol{b}_{2}\right)}\left[-\infty,b_{1}^{+}\right]_{\boldsymbol{b}_{1}}\left\{F^{-i}\left( b_{1}\right)\left[b_{1},b_{2}\right]F^{-j}\left(b_{2}\right)\right\}_{b_1^-=b_2^-=0}\left[b_{2}^{+},-\infty\right]_{\boldsymbol{b}_{2}}.\nonumber 
\end{align}
Taking the leading term in the Taylor expansion of the hard part yields
\begin{align}
\mathcal{A}_{2}\left(\boldsymbol{k}\right) & =g^{2}\int\!db_{1}^{+}db_{2}^{+}\int\!d^{2}\boldsymbol{b}_{1}d^{2}\boldsymbol{b}_{2}\,\delta^{2}\left(\boldsymbol{b}_{1}-\boldsymbol{b}_{2}\right)e^{-i\left(\boldsymbol{k}\cdot\boldsymbol{b}_{2}\right)}\mathcal{H}_{2}^{ij}\left(\boldsymbol{0},\boldsymbol{0}\right)\nonumber \\
 & \times\left[-\infty,b_{1}^{+}\right]_{\boldsymbol{b}_{1}}\left\{F^{-i}\left(b_{1}\right)\left[b_{1},b_{2}\right]F^{-j}\left(b_{2}\right)\right\}_{b_1^-=b_2^-=0}\left[b_{2}^{+},-\infty\right]_{\boldsymbol{b}_{2}}.\label{eq:2BTMD2}
\end{align}
Using the $\delta$-function of the impact parameters $\boldsymbol{b_1}$ and $\boldsymbol{b_2}$ which sets these two transverse coordinates to the same value, one can rewrite the gauge link $[b_1,b_2]$ as $\left[b_{1}^{+},+\infty\right]_{\boldsymbol{b}_{1}}\left[+\infty,b_{2}^{+}\right]_{\boldsymbol{b}_{2}}$. 
This allows us to rewrite the operator as derivatives of Wilson lines and the leading term in the Taylor expansion of the 2-body amplitude for photoproduction of a quark-antiquark dijet reads
\begin{align}
\mathcal{A}_{2}\left(\boldsymbol{k}\right) & =\int\!d^{2}\boldsymbol{b}\,e^{-i\left(\boldsymbol{k}\cdot\boldsymbol{b}\right)}\mathcal{H}_{2}^{ij}\left(\boldsymbol{0},\boldsymbol{0}\right)\left(\partial^{i}U_{\boldsymbol{b}}\right)\left(\partial^{j}U_{\boldsymbol{b}}^{\dagger}\right).\label{eq:2BTMDp0}
\end{align}
The comparison between Eqs.~(\ref{eq:1BTMD}), (\ref{eq:2BTMDp0}) and
the CGC result given in Eq.~(\ref{eq:N2B2Bform}) shows a strong similarity
between the small-dipole expansion in the CGC and the power expansion
in the TMD framework. A more general matching could be conjectured.
In this paper, we only focus on kinematic twist corrections and 
compare the 1-body contributions from the CGC to those obtained in the
TMD framework with infinite power accuracy via the iTMD
scheme developed in \cite{Kotko:2015ura}. Comparisons for higher-body
terms are left for further studies.

\section{Kinematic twist resummation for a generic $1\to2$ process in the CGC\label{sec:CGCexpansion}}

In the previous section, we have calculated the next-to-back-to-back corrections for a specific process ($\gamma\to q\bar q$) in the CGC framework and showed how one can isolate the 1-body and 2-body terms in this contribution. Our main goal in this section is to generalize this procedure to all orders in the small dipole size expansion. We isolate the 1-body contribution from the higher-body contributions, and then resum the 1-body contributions that appear in higher orders in the small dipole size expansion.  

\begin{figure}[hbt]
\begin{centering}
\includegraphics[scale=0.5]{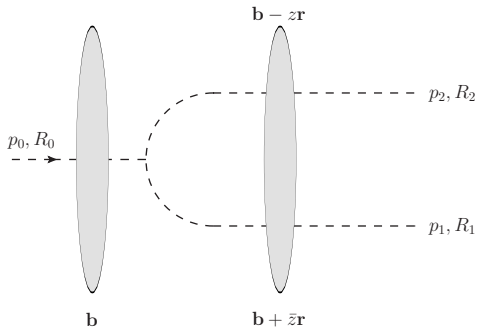}
\par\end{centering}
\caption{Generic ($0\rightarrow 12$) process in an external shockwave field. The gray blobs represent the dressing of each line crossing it by Wilson line operators, resumming any number of eikonal scatterings with the external field.}
\label{fig:Generic}  
\end{figure}

We would like to apply our results to several different $1\to2$ processes in the CGC framework. Therefore, we start from a generic CGC amplitude for a $1\to2$ process from which one can easily deduce all these different processes that are computed using effective Feynman rules in a shockwave background field \cite{Balitsky:1995ub, Balitsky:1998kc, Balitsky:1998ya, McLerran:1993ni, McLerran:1993ka, McLerran:1994vd} given in Appendix~\ref{sec:Feymanrules}. For this generic process, as before, we consider the case when the outgoing pair of particle has a large invariant mass, and the incoming particle is on the mass shell. For each $(p_{0}\rightarrow p_{1}p_{2})$ process, we use the same longitudinal momentum fractions ($z$ and $\bar z$) introduced in Eq.~\eqref{frac}, the total transverse momentum $\mathbf{k}$ of the produced particles defined in Eq.~\eqref{k} and the transverse boost invariant momentum $\mathbf{q}$ that is defined in Eq. \eqref{q}. The generic CGC amplitude (see Fig.~\ref{fig:Generic})  in this case reads 
\begin{align}
\mathcal{A}_{0\rightarrow12} 
&
 =\left(2\pi\right)\delta\left(p_{1}^{+}+p_{2}^{+}-p_{0}^{+}\right)\int\!d^{2}\boldsymbol{b}\,d^{2}\boldsymbol{r}\,e^{-i\left(\boldsymbol{q}\cdot\boldsymbol{r}\right)-i\left(\boldsymbol{k}\cdot\boldsymbol{b}\right)}
\nonumber \\
& \times
 \frac{r_{\perp}^{\mu}}{\boldsymbol{r}^{2}}\left[\left(U_{\boldsymbol{b}+\bar{z}\boldsymbol{r}}^{R_{1}}T^{R_{0}}U_{\boldsymbol{b}-z\boldsymbol{r}}^{R_{2}}\right)-\left(U_{\boldsymbol{b}}^{R_{1}}T^{R_{0}}U_{\boldsymbol{b}}^{R_{2}}\right)\right]\phi_{\mu},\label{eq:Genini}
\end{align}
where $\phi_{\mu}$ is a Dirac structure which does not depend on
coordinates, and $\left(R_{1},R_{0},R_{2}\right)$ are color representations.
This is a well known form in small-$x$ kinematics: the interaction
with the target can be factorized out in the eikonal limit, and it
contains all information on color flow. The spin structure factorizes
in the massless case due to transverse boost invariance: the mere
topology of a diagram is sufficient to predict its momentum structure,
or equivalently in coordinate space its dipole-size dependence. One
can easily check that the amplitudes listed in Appendix~\ref{sec:CGCamplitudes}
have the form of Eq.~(\ref{eq:Genini}).

The expression for the generic CGC amplitude for a $1\to2$ process expanded to the $n$-th power of $\boldsymbol{r}$ is obtained by performing a Taylor series expansion of the Wilson line operators in $\mathcal{A}_{0\rightarrow12}$ which can be simply written as 
\begin{align}
\mathcal{A}_{0\rightarrow12}^{\left(n\right)} & =\left(2\pi\right)\delta\left(p_{1}^{+}+p_{2}^{+}-p_{0}^{+}\right)\int\!d^{2}\boldsymbol{b}\,d^{2}\boldsymbol{r}\,e^{-i\left(\boldsymbol{q}\cdot\boldsymbol{r}\right)-i\left(\boldsymbol{k}\cdot\boldsymbol{b}\right)}\frac{r_{\perp}^{\mu}\phi_{\mu}}{\boldsymbol{r}^{2}}\label{eq:GenNth}\\
 & \times\frac{1}{n!}r_{\perp}^{\alpha_{1}}...r_{\perp}^{\alpha_{n}}\sum_{m=0}^{n}\left(\begin{array}{c}
n\\
m
\end{array}\right)\bar{z}^{m}\left(-z\right)^{n-m}\left(\partial_{\alpha_{1}}...\partial_{\alpha_{m}}U_{\boldsymbol{b}}^{R_{1}}\right)T^{R_{0}}\left(\partial_{\alpha_{m+1}}...\partial_{\alpha_{n}}U_{\boldsymbol{b}}^{R_{2}}\right).\nonumber 
\end{align}
The rest of our discussion relies on a symmetry hypothesis based on our experience of BFKL and CGC amplitudes. In the CGC, diagrams with scattering only on one line give $(U^{R_1}-\mathbf{1}^{R_1})\mathbf{1}^{R_2}$ and $\mathbf{1}^{R_1} (U^{R_2}-\mathbf{1}^{R_2})$ contributions, which once summed up with the symmetric contribution $(U^{R_1}-\mathbf{1}^{R_1})(U^{R_2}-\mathbf{1}^{R_2})$ lead to the gauge invariant dipole $U^{R_1}U^{R_2}-\mathbf{1}^{R_1}\mathbf{1}^{R_2}$. In BFKL computations, diagrams with one gluon on each line give the impact factor $\varphi(k_{1\perp},k_{2\perp})+\varphi(k_{2\perp},k_{1\perp})$ while diagrams with both gluons on one line give counterterms $-\varphi(k_{1\perp}+k_{2\perp},0_\perp)$ and $-\varphi(0_\perp,k_{1\perp}+k_{2\perp})$. The latter insure the cancellation of the full impact factor for $k_{1\perp}=0_\perp$ and for $k_{2\perp}=0_\perp$ and thus gauge invariance in the BFKL sense. \\
By analogy, keeping in mind that one derivative equals one gluon in the TMD, we assume that contributions with no derivative on one line must be a gauge-invariance restoring term for the 1-body contributions, i.e. a kinematic twist, which we extract with the following procedure. 

We assume that the $n$-body contribution to the amplitude, for
$n>1$, does not contain the least symmetric Wilson line operators,
in terms of derivatives. In other words, our statement is that no
$U(\partial_{i_{1}}...\partial_{i_{n}}U^{\dagger})$ or $(\partial_{i_{1}}...\partial_{i_{n}}U)U^{\dagger}$
term contributes to gauge invariant amplitudes. Operators with the least symmetric derivative structures need to be integrated by parts
using 
\begin{align}
 & \int\!d^{2}\boldsymbol{b}\,e^{-i\left(\boldsymbol{k}\cdot\boldsymbol{b}\right)}\left(\partial_{\alpha_{1}}...\partial_{\alpha_{m}}U_{\boldsymbol{b}}^{R_{1}}\right)T^{R_{0}}\left(\partial_{\alpha_{m+1}}...\partial_{\alpha_{n}}U_{\boldsymbol{b}}^{R_{2}}\right)\label{eq:IbP}\\
 & =\int\!d^{2}\boldsymbol{b}\,e^{-i\left(\boldsymbol{k}\cdot\boldsymbol{b}\right)}\left[-ik_{\perp\alpha_{n}}\left(\partial_{\alpha_{1}}...\partial_{\alpha_{m}}U_{\boldsymbol{b}}^{R_{1}}\right)T^{R_{0}}\left(\partial_{\alpha_{m+1}}...\partial_{\alpha_{n-1}}U_{\boldsymbol{b}}^{R_{2}}\right)\right.\nonumber \\
 & -\left.\left(\partial_{\alpha_{1}}...\partial_{\alpha_{m+1}}U_{\boldsymbol{b}}^{R_{1}}\right)T^{R_{0}}\left(\partial_{\alpha_{m+2}}...\partial_{\alpha_{n}}U_{\boldsymbol{b}}^{R_{2}}\right)\right]\nonumber 
\end{align}
or the other way around, depending on which Wilson line has more derivatives acting on it. By employing this procedure, we make sure that the non-symmetric operators are reduced to a more symmetric contribution and a contribution with less derivatives acting on the Wilson line operators. One can then proceed recursively in order to isolate all the 1-body contributions from the higher-body terms. However, we should emphasize that a stronger hypothesis is required in order to study genuine twist corrections, which are left for future studies. Nevertheless, as mentioned earlier in this study we focus on the kinematic twists. 

In order to clarify our discussion, let us consider the case for $n=4$. The generic CGC amplitude for a $1\to2$ process, when expanded to $O(\boldsymbol{r}^4)$, after employing the procedure described above, reads 
\begin{align}
\mathcal{A}_{0\rightarrow12} & =\left(2\pi\right)\delta\left(p_{1}^{+}+p_{2}^{+}-p_{0}^{+}\right)\int\!d^{2}\boldsymbol{b}\,d^{2}\boldsymbol{r}\,e^{-i\left(\boldsymbol{q}\cdot\boldsymbol{r}\right)-i\left(\boldsymbol{k}\cdot\boldsymbol{b}\right)}\frac{r_{\perp}^{\mu}\phi_{\mu}}{\boldsymbol{r}^{2}}\nonumber \\
 & \times\left\{ r_{\perp}^{\alpha_{1}}\left[\bar{z}\left(\partial_{\alpha_{1}}U_{\boldsymbol{b}}^{R_{1}}\right)T^{R_{0}}U_{\boldsymbol{b}}^{R_{2}}\left(1+\frac{i\bar{z}\left(\boldsymbol{k}\cdot\boldsymbol{r}\right)}{2!}+\frac{\left(i\bar{z}\left(\boldsymbol{k}\cdot\boldsymbol{r}\right)\right)^{2}}{3!}+\frac{\left(i\bar{z}\left(\boldsymbol{k}\cdot\boldsymbol{r}\right)\right)^{3}}{4!}\right)\right.\right.\nonumber \\
 & \left.-zU_{\boldsymbol{b}}^{R_{1}}T^{R_{0}}\left(\partial_{\alpha_{1}}U_{\boldsymbol{b}}^{R_{2}}\right)\left(1+\frac{-iz\left(\boldsymbol{k}\cdot\boldsymbol{r}\right)}{2!}+\frac{\left(-iz\left(\boldsymbol{k}\cdot\boldsymbol{r}\right)\right)^{2}}{3!}+\frac{\left(-iz\left(\boldsymbol{k}\cdot\boldsymbol{r}\right)\right)^{3}}{4!}\right)\right]\nonumber \\
 & -r_{\perp}^{\alpha_{1}}r_{\perp}^{\alpha_{2}}\left(\partial_{\alpha_{1}}U_{\boldsymbol{b}}^{R_{1}}\right)T^{R_{0}}\left(\partial_{\alpha_{2}}U_{\boldsymbol{b}}^{R_{2}}\right)\left(\frac{1}{2!}+\frac{-i\left(z-\bar{z}\right)\left(\boldsymbol{k}\cdot\boldsymbol{r}\right)}{3!}+\frac{\left(-i\left(z-\bar{z}\right)\left(\boldsymbol{k}\cdot\boldsymbol{r}\right)\right)^{2}}{4!}\right)\nonumber \\
 & +r_{\perp}^{\alpha_{1}}r_{\perp}^{\alpha_{2}}r_{\perp}^{\alpha_{3}}\left[z\left(\partial_{\alpha_{1}}U_{\boldsymbol{b}}^{R_{1}}\right)T^{R_{0}}\left(\partial_{\alpha_{2}}\partial_{\alpha_{3}}U_{\boldsymbol{b}}^{R_{2}}\right)\left(\frac{1}{3!}-\frac{2\left(iz\left(\boldsymbol{k}\cdot\boldsymbol{r}\right)\right)}{4!}\right)\right.\label{eq:4thPow}\\
 & \left.-\bar{z}\left(\partial_{\alpha_1}\partial_{\alpha_2}U_{\boldsymbol{b}}^{R_{1}}\right)T^{R_{0}}\left(\partial_{\alpha_3}U_{\boldsymbol{b}}^{R_{2}}\right)\left(\frac{1}{3!}+\frac{2\left(i\bar{z}\left(\boldsymbol{k}\cdot\boldsymbol{r}\right)\right)}{4!}\right)\right]\nonumber \\
 & \left.+r_{\perp}^{\alpha_{1}}r_{\perp}^{\alpha_{2}}r_{\perp}^{\alpha_{3}}r_{\perp}^{\alpha_{4}}\left(\partial_{\alpha_{1}}\partial_{\alpha_{2}}U_{\boldsymbol{b}}^{R_{1}}\right)T^{R_{0}}\left(\partial_{\alpha_{3}}\partial_{\alpha_{4}}U_{\boldsymbol{b}}^{R_{2}}\right)\frac{1}{4!}\right\} .\nonumber 
\end{align}
As emphasized multiple times earlier, our aim in thus work is to study the $\left(\partial_{\alpha_{1}}U_{\boldsymbol{b}}^{R_{1}}\right)T^{R_{0}}U_{\boldsymbol{b}}^{R_{2}}$ and $U_{\boldsymbol{b}}^{R_{1}}T^{R_{0}}\left(\partial_{\alpha_{1}}U_{\boldsymbol{b}}^{R_{2}}\right)$ terms and perform an all-order dipole size resummation for them. This amounts to the Wandzura-Wilczek approximation for all twists \cite{Wandzura:1997qf}. Here after, we denote all the amplitudes and the cross sections obtained from the CGC calculations by adopting the Wandzura-Wilczek approximation with the superscript $WW$. With our symmetry argument, it is easy to obtain a generic form for the $n$-th power in the amplitude, by performing ($n-1$) integrations by parts on the least symmetric terms. Summing up such contributions for all $n$ leads to  
\begin{align}
\mathcal{A}_{0\rightarrow12}^{WW} & =\left(2\pi\right)\delta\left(p_{1}^{+}+p_{2}^{+}-p_{0}^{+}\right)\int\!d^{2}\boldsymbol{b}\,d^{2}\boldsymbol{r}\,e^{-i\left(\boldsymbol{q}\cdot\boldsymbol{r}\right)-i\left(\boldsymbol{k}\cdot\boldsymbol{b}\right)}\frac{r_{\perp}^{\mu}\phi_{\mu}}{\boldsymbol{r}^{2}}\label{eq:1BodyNth}\\
 & \times r_{\perp}^{\alpha_{1}}\left[\bar{z}\left(\partial_{\alpha_{1}}U_{\boldsymbol{b}}^{R_{1}}\right)T^{R_{0}}U_{\boldsymbol{b}}^{R_{2}}\sum_{n}\frac{\left[i\bar{z}\left(\boldsymbol{k}\cdot\boldsymbol{r}\right)\right]^{n}}{\left(n+1\right)!}-zU_{\boldsymbol{b}}^{R_{1}}T^{R_{0}}\left(\partial_{\alpha_{1}}U_{\boldsymbol{b}}^{R_{2}}\right)\sum_{n}\frac{\left[-iz\left(\boldsymbol{k}\cdot\boldsymbol{r}\right)\right]^{n}}{\left(n+1\right)!}\right].\nonumber 
\end{align}
It is now straightforward to perform the resummation explicitly which results in the following form
\begin{align}
\mathcal{A}_{0\rightarrow12}^{WW} & =\left(2\pi\right)\delta\left(p_{1}^{+}+p_{2}^{+}-p_{0}^{+}\right)\int\!d^{2}\boldsymbol{b}\,d^{2}\boldsymbol{r}\,e^{-i\left(\boldsymbol{q}\cdot\boldsymbol{r}\right)-i\left(\boldsymbol{k}\cdot\boldsymbol{b}\right)}\frac{r_{\perp}^{\mu}\phi_{\mu}}{\boldsymbol{r}^{2}}\label{eq:1BodyResum}\\
 & \times r_{\perp}^{\alpha_{1}}\left[\bar{z}\left(\partial_{\alpha_{1}}U_{\boldsymbol{b}}^{R_{1}}\right)T^{R_{0}}U_{\boldsymbol{b}}^{R_{2}}\frac{e^{i\bar{z}\left(\boldsymbol{k}\cdot\boldsymbol{r}\right)}-1}{i\bar{z}\left(\boldsymbol{k}\cdot\boldsymbol{r}\right)}-zU_{\boldsymbol{b}}^{R_{1}}T^{R_{0}}\left(\partial_{\alpha_{1}}U_{\boldsymbol{b}}^{R_{2}}\right)\frac{e^{-iz\left(\boldsymbol{k}\cdot\boldsymbol{r}\right)}-1}{-iz\left(\boldsymbol{k}\cdot\boldsymbol{r}\right)}\right].\nonumber 
\end{align}
The integral over the dipole size $\boldsymbol{r}$ is factorized from the rest of the expression and can be performed explicitly by considering the following integral
\begin{align}
I^{ij}\left(\boldsymbol{p}\right) & \equiv\int\!d^{d}\boldsymbol{r}\frac{\boldsymbol{r}^{i}\boldsymbol{r}^{j}}{\boldsymbol{r}^{2}}\frac{e^{-i\left(\boldsymbol{p}\cdot\boldsymbol{r}\right)}-1}{\left(\boldsymbol{p}\cdot\boldsymbol{r}\right)}e^{-i\left(\boldsymbol{q}\cdot\boldsymbol{r}\right)},\label{eq:Integralini}
\end{align}
for $\mathbf{p}=\bar{z}\mathbf{k}$ or $\mathbf{p}=-z\mathbf{k}$. The details of the calculation can be found in Appendix~\ref{sec:Integral} and the result reads
\begin{equation}
I^{ij}\left(\boldsymbol{p}\right)=-2\frac{i\pi}{\boldsymbol{p}^{2}}\left(\boldsymbol{p}^{i}\delta^{jl}+\boldsymbol{p}^{j}\delta^{il}-\boldsymbol{p}^{l}\delta^{ij}\right)\left(\frac{\boldsymbol{q}^{l}+\boldsymbol{p}^{l}}{\left(\boldsymbol{q}+\boldsymbol{p}\right)^{2}}-\frac{\boldsymbol{q}^{l}}{\boldsymbol{q}^{2}}\right).\label{eq:Integralfin}
\end{equation}
Plugging this result  into Eq.~(\ref{eq:1BodyResum}) and reintroducing the transverse momenta of the produced particles ($\boldsymbol{p_1}, \boldsymbol{p_2}$)  leads to the final expression for the generic CGC amplitude for a $1\to2$ process in the Wandzura-Wilczek approximation:
\begin{align}
\mathcal{A}_{0\rightarrow12}^{WW} & =\left(2\pi\right)^{2}\delta\left(p_{1}^{+}+p_{2}^{+}-p_{0}^{+}\right)\int\!d^{2}\boldsymbol{b}\,e^{-i\left(\boldsymbol{k}\cdot\boldsymbol{b}\right)}\frac{\phi^{i}}{\boldsymbol{k}^{2}}\left(\boldsymbol{k}^{i}\delta^{jl}+\boldsymbol{k}^{j}\delta^{il}-\boldsymbol{k}^{l}\delta^{ij}\right)\nonumber \\
 & \times\left[\left(\frac{\boldsymbol{q}^{l}}{\boldsymbol{q}^{2}}+\frac{\boldsymbol{p}_{2}^{l}}{\boldsymbol{p}_{2}^{2}}\right)\left(\partial^{j}U_{\boldsymbol{b}}^{R_{1}}\right)T^{R_{0}}U_{\boldsymbol{b}}^{R_{2}}+\left(\frac{\boldsymbol{q}^{l}}{\boldsymbol{q}^{2}}-\frac{\boldsymbol{p}_{1}^{l}}{\boldsymbol{p}_{1}^{2}}\right)U_{\boldsymbol{b}}^{R_{1}}T^{R_{0}}\left(\partial^{j}U_{\boldsymbol{b}}^{R_{2}}\right)\right].\label{eq:GenResumWithInt}
\end{align}
Using the generic CGC amplitude given in Eq. \eqref{eq:GenResumWithInt}, the generic cross section can be calculated in a straightforward manner and the result reads
\begin{align}
 & \frac{d\sigma_{0\rightarrow12}^{WW}}{dy_{1}dy_{2}d^{2}\boldsymbol{p}_{1}d^{2}\boldsymbol{p}_{2}}\nonumber \\
 & =\frac{\left(2\pi\right)}{16C_{0}p_{0}^{+}}\delta\left(p_{1}^{+}+p_{2}^{+}-p_{0}^{+}\right)\left(\phi^{i}\phi^{i^{\prime}\ast}\right)\int\!\frac{d^{2}\boldsymbol{b}}{\left(2\pi\right)^{2}}\frac{d^{2}\boldsymbol{b}^{\prime}}{\left(2\pi\right)^{2}}e^{i\boldsymbol{k}\cdot\left(\boldsymbol{b}^{\prime}-\boldsymbol{b}\right)}\nonumber \\
 & \times\frac{1}{\boldsymbol{k}^{4}}\left(\boldsymbol{k}^{i}\delta^{jl}+\boldsymbol{k}^{j}\delta^{il}-\boldsymbol{k}^{l}\delta^{ij}\right)\left(\boldsymbol{k}^{i^{\prime}}\delta^{j^{\prime}l^{\prime}}+\boldsymbol{k}^{j^{\prime}}\delta^{i^{\prime}l^{\prime}}-\boldsymbol{k}^{l^{\prime}}\delta^{i^{\prime}j^{\prime}}\right)\nonumber \\
 & \times\left\{ \left(\frac{\boldsymbol{q}^{l}}{\boldsymbol{q}^{2}}+\frac{\boldsymbol{p}_{2}^{l}}{\boldsymbol{p}_{2}^{2}}\right)\left(\frac{\boldsymbol{q}^{l^{\prime}}}{\boldsymbol{q}^{2}}+\frac{\boldsymbol{p}_{2}^{l^{\prime}}}{\boldsymbol{p}_{2}^{2}}\right)
 \left\langle  
 \mathrm{Tr}\left[\left(\partial^{j}U_{\boldsymbol{b}}^{R_{1}}\right)T^{R_{0}}U_{\boldsymbol{b}}^{R_{2}}U_{\boldsymbol{b}^{\prime}}^{R_{2}\dagger}T^{R_{0}\dagger}\left(\partial^{j^{\prime}}U_{\boldsymbol{b}^{\prime}}^{R_{1}\dagger}\right)\right]
\right\rangle
 \right.\nonumber \\
 & +\left(\frac{\boldsymbol{q}^{l}}{\boldsymbol{q}^{2}}+\frac{\boldsymbol{p}_{2}^{l}}{\boldsymbol{p}_{2}^{2}}\right)\left(\frac{\boldsymbol{q}^{l^{\prime}}}{\boldsymbol{q}^{2}}-\frac{\boldsymbol{p}_{1}^{l^{\prime}}}{\boldsymbol{p}_{1}^{2}}\right)
  \left\langle 
 \mathrm{Tr}\left[\left(\partial^{j}U_{\boldsymbol{b}}^{R_{1}}\right)T^{R_{0}}U_{\boldsymbol{b}}^{R_{2}}\left(\partial^{j^{\prime}}U_{\boldsymbol{b}^{\prime}}^{R_{2}\dagger}\right)T^{R_{0}\dagger}U_{\boldsymbol{b}^{\prime}}^{R_{1}\dagger}\right]
 \right\rangle
 \label{eq:genCrossSec}\\
 & +\left(\frac{\boldsymbol{q}^{l}}{\boldsymbol{q}^{2}}-\frac{\boldsymbol{p}_{1}^{l}}{\boldsymbol{p}_{1}^{2}}\right)\left(\frac{\boldsymbol{q}^{l^{\prime}}}{\boldsymbol{q}^{2}}+\frac{\boldsymbol{p}_{2}^{l^{\prime}}}{\boldsymbol{p}_{2}^{2}}\right)
 \left\langle
 \mathrm{Tr}\left[U_{\boldsymbol{b}}^{R_{1}}T^{R_{0}}\left(\partial^{j}U_{\boldsymbol{b}}^{R_{2}}\right)U_{\boldsymbol{b}^{\prime}}^{R_{2}\dagger}T^{R_{0}\dagger}\left(\partial^{j^{\prime}}U_{\boldsymbol{b}^{\prime}}^{R_{1}\dagger}\right)\right]
\right\rangle
 \nonumber \\
 & \left.+\left(\frac{\boldsymbol{q}^{l}}{\boldsymbol{q}^{2}}-\frac{\boldsymbol{p}_{1}^{l}}{\boldsymbol{p}_{1}^{2}}\right)\left(\frac{\boldsymbol{q}^{l^{\prime}}}{\boldsymbol{q}^{2}}-\frac{\boldsymbol{p}_{1}^{l^{\prime}}}{\boldsymbol{p}_{1}^{2}}\right)
\left\langle 
\mathrm{Tr}\left[U_{\boldsymbol{b}}^{R_{1}}T^{R_{0}}\left(\partial^{j}U_{\boldsymbol{b}}^{R_{2}}\right)\left(\partial^{j^{\prime}}U_{\boldsymbol{b}^{\prime}}^{R_{2}\dagger}\right)T^{R_{0}\dagger}U_{\boldsymbol{b}^{\prime}}^{R_{1}\dagger}\right]
\right\rangle
 \right\} ,\nonumber 
\end{align}
where the factor $\frac{1}{2C_{0}}$ originates from the spin and color averaging over the incoming state and $\left\langle \cdots\right\rangle$ is defined in Eq. \eqref{brackets}. The color Fierz factor $C_{0}$ is $N_{c}$ for a quark, $\left(N_{c}^{2}-1\right)$ for a gluon and $1$ for a photon. 

Eq.~(\ref{eq:genCrossSec}) is the main result of this paper. It is the generic CGC cross section for a $1\to 2$ process that resums all kinematic twists.  By introducing the proper color structure and the proper Dirac structure for a specific $1\to 2$ process, one can get the kinematic twist resummed CGC cross section for that specific process. In the following sections, we study several of such specific processes and show that the results match exactly the ones obtained through the iTMD calculations. 

\section{Dilute limit of a generic $1\to2$ process in the CGC\label{sec:Dilute}}

For very high values of the center-of-mass energy $s$ or for dense targets, multiple scatterings are expected to occur. In practice, for values of $\left|\boldsymbol{k}\right|$ of the order of the target saturation scale $Q_{s}^2\sim\left(A/x\right)^{1/3}$, it is expected for the target fields $A^{-}$ to scale like $1/g$ due to a high gluon occupation number, so that $gA^{-}$ must be resummed into the path-ordered Wilson line operators $U_{\boldsymbol{b}}^{R}$ which are the natural building blocks of the CGC or shockwave formalisms.

The regime where $\left|\boldsymbol{k}\right|\gg Q_{s}$, is referred to as the dilute limit. In this limit $gA^{-}$ is expected to be small and therefore one is allowed to expand Wilson line operators in gluon fields (or in Reggeon fields for more involved analysis, as \cite{Caron-Huot:2013fea, Altinoluk:2013rua}) or equivalently to use a dilute formalism like BFKL. 

In this section, we consider the dilute limit of the CGC by expanding the Wilson line operators in the generic CGC amplitude for a $1\to2$ process whose expression is given in Eq. \eqref{eq:Genini}. The generic Wilson line operator, when expanded in powers of the strong coupling constant $g$, in arbitrary representation $R$ reads
\begin{equation}
U_{\boldsymbol{x}}^{R}=1+igT_{R}^{a}\int\!dx^{+}A_{a}^{-}\left(x^{+}\!,0,\boldsymbol{x}\right)+O\left(g^{2}\right).\label{eq:WilExp}
\end{equation}
with $T_R^a$ being the $SU(N_c)$ generator in the representation $R$. Then, in the dilute limit, the generic CGC amplitude given in Eq. \eqref{eq:Genini} can be written as 
\begin{align}
\mathcal{A}_{0\rightarrow12}^{gA\sim0} & =ig\left(2\pi\right)\delta\left(p_{1}^{+}+p_{2}^{+}-p_{0}^{+}\right)\int\!d^{2}\boldsymbol{b}\,d^{2}\boldsymbol{r}\,e^{-i\left(\boldsymbol{q}\cdot\boldsymbol{r}\right)-i\left(\boldsymbol{k}\cdot\boldsymbol{b}\right)}\frac{r_{\perp}^{\mu}}{\boldsymbol{r}^{2}}\phi_{\mu}\nonumber \\
 & \times\int\!dz^{+}\left\{ T_{R_{0}}T_{R_{2}}^{a}\left[A_{a}^{-}\left(z^{+}\!,0,\boldsymbol{b}-z\boldsymbol{r}\right)-A_{a}^{-}\left(z^{+}\!,0,\boldsymbol{b}\right)\right]\right.\label{eq:DilAmpIni}\\
 & \left.+T_{R_{1}}^{a}T_{R_{0}}\left[A_{a}^{-}\left(z^{+}\!,0,\boldsymbol{b}+\bar{z}\boldsymbol{r}\right)-A_{a}^{-}\left(z^{+}\!,0,\boldsymbol{b}\right)\right]\right\} \,.\nonumber 
\end{align}
After introducing the incoming target state $P$ and the target remnant states $X$, and using the translation invariance of the  $\left\langle X\left|(...)\right|P\right\rangle $ matrix elements, one can easily integrate over the impact parameter which yields to the following form of the matrix element:
\begin{align}
\left\langle X\left|\mathcal{A}_{0\rightarrow12}^{gA\sim0}\right|P\right\rangle  & =ig\left(2\pi\right)^{4}\delta\left(k+P_{X}-P-p_{0}\right)\int\!d^{2}\boldsymbol{r}\,e^{-i\left(\boldsymbol{q}\cdot\boldsymbol{r}\right)}\frac{r_{\perp}^{\mu}}{\boldsymbol{r}^{2}}\phi_{\mu}\left\langle X\left|A_{a}^{-}\left(0\right)\right|P\right\rangle \nonumber \\
 & \times\left[T_{R_{0}}T_{R_{2}}^{a}\left(e^{-iz\left(\boldsymbol{k}\cdot\boldsymbol{r}\right)}-1\right)+T_{R_{1}}^{a}T_{R_{0}}\left(e^{i\bar{z}\left(\boldsymbol{k}\cdot\boldsymbol{r}\right)}-1\right)\right].\label{eq:DilAmpStates}
\end{align}
In Eq. \eqref{eq:DilAmpStates}, the integral over the dipole size $\boldsymbol{r}$ can be performed in a straightforward manner by using the well known integral 
\begin{align}
\int\!d^{2}\boldsymbol{r}\frac{r_{\perp}^{\mu}}{\boldsymbol{r}^{2}}e^{-i\left(\boldsymbol{\ell}\cdot\boldsymbol{r}\right)} & =-2i\pi\frac{\ell_{\perp}^{\mu}}{\boldsymbol{\ell}^{2}},\label{eq:DilInt}
\end{align}
which finally leads to the following form of the dilute amplitude
\begin{align}
\left\langle X\left|\mathcal{A}_{0\rightarrow12}^{gA\sim0}\right|P\right\rangle  & =2\pi g\left(2\pi\right)^{4}\delta\left(k+P_{X}-P-p_{0}\right)\left\langle X\left|A_{a}^{-}\left(0\right)\right|P\right\rangle \label{eq:DilAmpIntegrated}\\
 & \times\left[T_{R_{0}}T_{R_{2}}^{a}\left(\frac{p_{1\perp}^{\mu}}{\boldsymbol{p}_{1}^{2}}-\frac{q_{\perp}^{\mu}}{\boldsymbol{q}^{2}}\right)-T_{R_{1}}^{a}T_{R_{0}}\left(\frac{p_{2\perp}^{\mu}}{\boldsymbol{p}_{2}^{2}}+\frac{q_{\perp}^{\mu}}{\boldsymbol{q}^{2}}\right)\right]\phi_{\mu}.\nonumber 
\end{align}
The cross section in the dilute limit can be easily obtained from Eq. \eqref{eq:DilAmpIntegrated}, and the result reads
\begin{align}
\frac{d\sigma_{0\rightarrow12}^{gA\sim0}}{dy_{1}dy_{2}d^{2}\boldsymbol{p}_{1}d^{2}\boldsymbol{p}_{2}} & =\frac{\alpha_{s}}{4s}\delta\left(p_{1}^{+}+p_{2}^{+}-p_{0}^{+}\right)\int\!\frac{db^{+}d^{2}\boldsymbol{b}}{\left(2\pi\right)^{2}}e^{-i\left(\boldsymbol{k}\cdot\boldsymbol{b}\right)}\left\langle P\left|A_{c}^{-}\left(b\right)A_{a}^{-}\left(0\right)\right|P\right\rangle _{b^{-}=0}\left(\phi^{i}\phi^{j\ast}\right)\nonumber \\
 & \times\mathrm{Tr}\left\{ \left[T_{R_{0}}T_{R_{2}}^{a}\left(\frac{\boldsymbol{p}_{1}^{i}}{\boldsymbol{p}_{1}^{2}}-\frac{\boldsymbol{q}^{i}}{\boldsymbol{q}^{2}}\right)-T_{R_{1}}^{a}T_{R_{0}}\left(\frac{\boldsymbol{p}_{2}^{i}}{\boldsymbol{p}_{2}^{2}}+\frac{\boldsymbol{q}^{i}}{\boldsymbol{q}^{2}}\right)\right]\right.\label{eq:DilCSbrackets}\\
 & \left.\times\left[T_{R_{2}}^{c\dagger}T_{R_{0}}^{\dagger}\left(\frac{\boldsymbol{p}_{1}^{j}}{\boldsymbol{p}_{1}^{2}}-\frac{\boldsymbol{q}^{j}}{\boldsymbol{q}^{2}}\right)-T_{R_{0}}^{\dagger}T_{R_{1}}^{c\dagger}\left(\frac{\boldsymbol{p}_{2}^{j}}{\boldsymbol{p}_{2}^{2}}+\frac{\boldsymbol{q}^{j}}{\boldsymbol{q}^{2}}\right)\right]\right\} .\nonumber 
\end{align}
Finally, it is customary to introduce the unintegrated parton distribution function (uPDF) $\mathcal{G}\left(\boldsymbol{k}\right)$ that is defined as 
\begin{equation}
\int\!db^{+}\int\!\frac{d^{2}\boldsymbol{b}}{\left(2\pi\right)^{2}}e^{-i\left(\boldsymbol{k}\cdot\boldsymbol{b}\right)}\left\langle P\left|A_{a}^{-}\left(b\right)A_{c}^{-}\left(0\right)\right|P\right\rangle   =\left(2\pi\right)P^{-}\frac{\mathcal{G}_{ac}\left(\boldsymbol{k}\right)}{\boldsymbol{k}^{2}}\, 
\label{eq:uPDF}
\end{equation}
with
\begin{equation}
\delta^{ac}\mathcal{G}_{ac}\left(\boldsymbol{k}\right)  =\mathcal{G}\left(\boldsymbol{k}\right).\nonumber 
\end{equation}
Averaging over the spin and color states of the incoming parton or photon, we arrive to the generic form of the cross section in the dilute limit:
\begin{align}
\frac{d\sigma_{0\rightarrow12}^{gA\sim0}}{dy_{1}dy_{2}d^{2}\boldsymbol{p}_{1}d^{2}\boldsymbol{p}_{2}} & =\frac{\left(2\pi\right)}{16C_{0}p_{0}^{+}}\alpha_{s}\delta\left(p_{1}^{+}+p_{2}^{+}-p_{0}^{+}\right)\left(\phi^{i}\phi^{j\ast}\right)\frac{\mathcal{G}_{ac}\left(\boldsymbol{k}\right)}{\boldsymbol{k}^{2}}\nonumber \\
 & \times\left\{ \left(\frac{\boldsymbol{q}^{i}}{\boldsymbol{q}^{2}}+\frac{\boldsymbol{p}_{2}^{i}}{\boldsymbol{p}_{2}^{2}}\right)\left(\frac{\boldsymbol{q}^{j}}{\boldsymbol{q}^{2}}+\frac{\boldsymbol{p}_{2}^{j}}{\boldsymbol{p}_{2}^{2}}\right)\mathrm{Tr}\left(T_{R_{1}}^{a}T_{R_{0}}T_{R_{0}}^{\dagger}T_{R_{1}}^{c\dagger}\right)\right.\nonumber \\
 & +\left(\frac{\boldsymbol{q}^{i}}{\boldsymbol{q}^{2}}+\frac{\boldsymbol{p}_{2}^{i}}{\boldsymbol{p}_{2}^{2}}\right)\left(\frac{\boldsymbol{q}^{j}}{\boldsymbol{q}^{2}}-\frac{\boldsymbol{p}_{1}^{j}}{\boldsymbol{p}_{1}^{2}}\right)\mathrm{Tr}\left(T_{R_{1}}^{a}T_{R_{0}}T_{R_{2}}^{c\dagger}T_{R_{0}}^{\dagger}\right)\label{eq:DilCSuPDF}\\
 & +\left(\frac{\boldsymbol{q}^{i}}{\boldsymbol{q}^{2}}-\frac{\boldsymbol{p}_{1}^{i}}{\boldsymbol{p}_{1}^{2}}\right)\left(\frac{\boldsymbol{q}^{j}}{\boldsymbol{q}^{2}}+\frac{\boldsymbol{p}_{2}^{j}}{\boldsymbol{p}_{2}^{2}}\right)\mathrm{Tr}\left(T_{R_{0}}T_{R_{2}}^{a}T_{R_{0}}^{\dagger}T_{R_{1}}^{c\dagger}\right)\nonumber \\
 & \left.+\left(\frac{\boldsymbol{q}^{i}}{\boldsymbol{q}^{2}}-\frac{\boldsymbol{p}_{1}^{i}}{\boldsymbol{p}_{1}^{2}}\right)\left(\frac{\boldsymbol{q}^{j}}{\boldsymbol{q}^{2}}-\frac{\boldsymbol{p}_{1}^{j}}{\boldsymbol{p}_{1}^{2}}\right)\mathrm{Tr}\left(T_{R_{0}}T_{R_{2}}^{a}T_{R_{2}}^{c\dagger}T_{R_{0}}^{\dagger}\right)\right\} ,\nonumber 
\end{align}
with $C_0$ being the factor that one obtains via color averaging, as introduced previously in section~\ref{sec:CGCexpansion}. We would like to draw attention to the similarity between dilute limit of the generic cross section given in Eq. \eqref{eq:DilCSuPDF} and the kinematic-twist-resummed cross section given in Eq. \eqref{eq:genCrossSec}. We discuss the implications of this similarity in section~\ref{sec:Discussions}.

\section{Small-$x$ Improved TMD factorization (iTMD)\label{sec:iTMD}}

In the following section we briefly recall the small-$x$ improved
TMD factorization constructed in \citep{Kotko:2015ura}. Although
the framework is more general, here we focus on dijets in $pA$
and $\gamma A$ collisions. This section is organized as follows.
We first list and explain the general form of the formulas for
dijets in $pA$ collisions. Next, we shall put the iTMD formulation
into the context of the TMD factorization theorems to better clarify
the terminology. In the end of this section, we shall give the formulas for the cross section for all channels in a form that can be compared
with the CGC framework.

\subsection{Framework}

The iTMD factorization formula for $pA$ collisions  has the form of a hybrid generalized
$k_{T}$-factorization. That is: (i) the incoming dilute projectile
is described by the collinear PDF as it is probed at large $x$ -- so called hybrid approach \citep{Dumitru:2005gt},
(ii) the target is probed at small $x$  and is described by a
set of process-dependent TMD gluon distributions,
(iii) the hard factors are constructed from off-shell gauge invariant
matrix elements. Thanks to (i), the formula for the cross section
can be written as
\begin{equation}
d\sigma_{pA\rightarrow 2j+X}=\sum_{q}f_{q/H}\otimes d\sigma_{qA\rightarrow qg}+f_{g/H}\otimes\left[d\sigma_{gA\rightarrow gg}+n_{f}d\sigma_{gA\rightarrow q\overline{q}}\right]\,,
\end{equation}
where $f_{a/H}$ is the collinear PDF for parton $a=q,g$ (we can
safely neglect antiquarks in this approximation), $\otimes$ denotes
the convolution in the longitudinal fraction $x_{p}$ of the proton
momentum carried by  parton $a$, $n_{f}$ is the number of
flavors. The remaining objects are cross sections for scattering a parton $a$ off the target to produce the given final states.
They can be generically written as follows:
\begin{equation}
\frac{d\sigma_{aA\rightarrow bc}}{d^{2}\boldsymbol{p}_{1}d^{2}\boldsymbol{p}_{2}dy_{1}dy_{2}}=
p_{0}^{+}\delta\left(p_{1}^{+}+p_{2}^{+}-p_{0}^{+}\right)
\frac{1}{\bar{s}^{2}}\sum_{i=1,2}\tilde{\mathcal{H}}_{ag^{*}\rightarrow bc}^{\left(i\right)}\left(\boldsymbol{p}_{1},\boldsymbol{p}_{2},z\right)\Phi_{ag\rightarrow bc}^{\left(i\right)}\left(x_{A},\boldsymbol{k}\right)\,,\label{eq:iTMD2}
\end{equation}
where $\bar{s}=x_{p}x_{A}s$, $\tilde{\mathcal{H}}_{ag^{*}\rightarrow bc}^{\left(i\right)}$
are off-shell gauge invariant hard factors and $\Phi_{ag\rightarrow bc}^{\left(i\right)}$
are unpolarized TMD gluon distributions in the target. The sum over $i$
corresponds to two inequivalent color flows that exist for each channel.

The TMD gluon distributions $\Phi_{ag\rightarrow bc}^{\left(i\right)}$
are linear combinations \citep{Kotko:2015ura} (Table~\ref{tab:Phi})
of the basic distributions with the following operator definitions
\citep{TMD}:
\begin{equation}
\begin{aligned}\mathcal{F}_{qg}^{(1)}\left(x,\left|\boldsymbol{k}\right|\right) & =2\int\frac{d\xi^{+}d^{2}\boldsymbol{\xi}}{\left(2\pi\right)^{3}P^{-}}\,e^{ixP^{-}\xi^{+}-i\boldsymbol{k}\cdot\boldsymbol{\xi}}\,\left\langle P\right|\mathrm{Tr}\left[\hat{F}^{i-}\left(\xi\right)\mathcal{U}^{[-]\dagger}\hat{F}^{i-}\left(0\right)\mathcal{U}^{[+]}\right]\left|P\right\rangle \,\end{aligned}
,
\end{equation}
\begin{equation}
\begin{aligned}\mathcal{F}_{qg}^{(2)}\left(x,\left|\boldsymbol{k}\right|\right) & =2\int\frac{d\xi^{+}d^{2}\boldsymbol{\xi}}{\left(2\pi\right)^{3}P^{-}}\,e^{ixP^{-}\xi^{+}-i\boldsymbol{k}\cdot\boldsymbol{\xi}}\,\left\langle P\right|\frac{\mathrm{Tr}\left[\mathcal{U}^{[\square]}\right]}{N_{c}}\mathrm{Tr}\left[\hat{F}^{i-}\left(\xi\right)\mathcal{U}^{[+]\dagger}\hat{F}^{i-}\left(0\right)\mathcal{U}^{[+]}\right]\left|P\right\rangle \,,\end{aligned}
\end{equation}
\begin{equation}
\begin{aligned}\mathcal{F}_{gg}^{(1)}\left(x,\left|\boldsymbol{k}\right|\right) & =2\int\frac{d\xi^{+}d^{2}\boldsymbol{\xi}}{\left(2\pi\right)^{3}P^{-}}\,e^{ixP^{-}\xi^{+}-i\boldsymbol{k}\cdot\boldsymbol{\xi}}\,\left\langle P\right|\frac{\mathrm{Tr}\left[\mathcal{U}^{[\square]\dagger}\right]}{N_{c}}\mathrm{Tr}\left[\hat{F}^{i-}\left(\xi\right)\mathcal{U}^{[-]\dagger}\hat{F}^{i-}\left(0\right)\mathcal{U}^{[+]}\right]\left|P\right\rangle \,,\end{aligned}
\end{equation}
\begin{equation}
\begin{aligned}\mathcal{F}_{gg}^{(2)}\left(x,\left|\boldsymbol{k}\right|\right) & =2\int\frac{d\xi^{+}d^{2}\boldsymbol{\xi}}{\left(2\pi\right)^{3}P^{-}}\,e^{ixP^{-}\xi^{+}-i\boldsymbol{k}\cdot\boldsymbol{\xi}}\,\frac{1}{N_{c}}\left\langle P\right|\mathrm{Tr}\left[\hat{F}^{i-}\left(\xi\right)\mathcal{U}^{[\square]\dagger}\right]\mathrm{Tr}\left[\hat{F}^{i-}\left(0\right)\mathcal{U}^{[\square]}\right]\left|P\right\rangle \,,\end{aligned}
\end{equation}
\begin{equation}
\mathcal{F}_{gg}^{(3)}\left(x,\left|\boldsymbol{k}\right|\right)=2\int\frac{d\xi^{+}d^{2}\boldsymbol{\xi}}{\left(2\pi\right)^{3}P^{-}}\,e^{ixP^{-}\xi^{+}-i\boldsymbol{k}\cdot\boldsymbol{\xi}}\,\left\langle P\right|\mathrm{Tr}\left[\hat{F}^{i-}\left(\xi\right)\mathcal{U}^{[+]\dagger}\hat{F}^{i-}\left(0\right)\mathcal{U}^{[+]}\right]\left|P\right\rangle \,,
\end{equation}
\begin{equation}
\mathcal{F}_{gg}^{(4)}\left(x,\left|\boldsymbol{k}\right|\right)=2\int\frac{d\xi^{+}d^{2}\boldsymbol{\xi}}{\left(2\pi\right)^{3}P^{-}}\,e^{ixP^{-}\xi^{+}-i\boldsymbol{k}\cdot\boldsymbol{\xi}}\,\left\langle P\right|\mathrm{Tr}\left[\hat{F}^{i-}\left(\xi\right)\mathcal{U}^{[-]\dagger}\hat{F}^{i-}\left(0\right)\mathcal{U}^{[-]}\right]\left|P\right\rangle \,,
\end{equation}
\begin{equation}
\mathcal{F}_{gg}^{(5)}\left(x,\left|\boldsymbol{k}\right|\right)=2\int\frac{d\xi^{+}d^{2}\boldsymbol{\xi}}{\left(2\pi\right)^{3}P^{-}}\,e^{ixP^{-}\xi^{+}-i\boldsymbol{k}\cdot\boldsymbol{\xi}}\,\left\langle P\right|\mathrm{Tr}\left[\hat{F}^{i-}\left(\xi\right)\mathcal{U}^{[\square]\dagger}\mathcal{U}^{[+]\dagger}\hat{F}^{i-}\left(0\right)\mathcal{U}^{[\square]}\mathcal{U}^{[+]}\right]\left|P\right\rangle \,,
\end{equation}
\begin{multline}
\mathcal{F}_{gg}^{(6)}\left(x,\left|\boldsymbol{k}\right|\right)=2\int\frac{d\xi^{+}d^{2}\boldsymbol{\xi}}{\left(2\pi\right)^{3}P^{-}}\,e^{ixP^{-}\xi^{+}-i\boldsymbol{k}\cdot\boldsymbol{\xi}}\\
\,\left\langle P\right|\frac{\mathrm{Tr}\left[\mathcal{U}^{[\square]}\right]}{N_{c}}\frac{\mathrm{Tr}\left[\mathcal{U}^{[\square]\dagger}\right]}{N_{c}}\mathrm{Tr}\left[\hat{F}^{i-}\left(\xi\right)\mathcal{U}^{[+]\dagger}\hat{F}^{i-}\left(0\right)\mathcal{U}^{[+]}\right]\left|P\right\rangle \,,
\end{multline}
with $\hat{F}\left(\xi\right)=t^{a}F^{a}\left(\xi^{+},\xi^{-}=0,\boldsymbol{\xi}\right)$.
The staple-like Wilson lines appearing above are defined as
\begin{multline}
\mathcal{U}^{\left[\pm\right]}=\left[\left(0^{+},0^{-},\boldsymbol{0}\right),\left(\pm\infty,0^{-},\boldsymbol{0}\right)\right]\\
\left[\left(\pm\infty,0^{-},\boldsymbol{0}\right),\left(\pm\infty,0^{-},\boldsymbol{\xi}\right)\right]\left[\left(\pm\infty,0^{-},\boldsymbol{\xi}\right),\left(\xi^{+},0^{-},\boldsymbol{\xi}\right)\right]\,.\label{eq:Staples}
\end{multline}
 The Wilson loop is made from two staples glued together:
\begin{equation}
\mathcal{U}^{\left[\square\right]}=\mathcal{U}^{\left[-\right]\dagger}\mathcal{U}^{\left[+\right]}\,.\label{eq:WilsonLoopDef}
\end{equation}

\begin{table}
\begin{centering}
\begin{tabular}{c|c|c}
\hline 
$i$  & 1 & 2 \tabularnewline
\hline 
${\displaystyle \Phi_{gg^{*}\to gg}^{(i)}}$  & %
\begin{minipage}[c][3.cm]{6cm}%
\vspace{-0.5cm}
\begin{multline*}
\frac{1}{2N_{c}^{2}}\big(N_{c}^{2}\mathcal{F}_{gg}^{\left(1\right)}-2\mathcal{F}_{gg}^{\left(3\right)}\\
+\mathcal{F}_{gg}^{\left(4\right)}+\mathcal{F}_{gg}^{\left(5\right)}+N_{c}^{2}\mathcal{F}_{gg}^{\left(6\right)}\big)
\end{multline*}
\end{minipage}  & %
\begin{minipage}[c][3.cm]{6cm}%
\vspace{-0.5cm}
\begin{multline*}
\frac{1}{N_{c}^{2}}\big(N_{c}^{2}\mathcal{F}_{gg}^{\left(2\right)}-2\mathcal{F}_{gg}^{\left(3\right)}\\
+\mathcal{F}_{gg}^{\left(4\right)}+\mathcal{F}_{gg}^{\left(5\right)}+N_{c}^{2}\mathcal{F}_{gg}^{\left(6\right)}\big)
\end{multline*}
\end{minipage} \tabularnewline
\hline 
${\displaystyle \Phi_{gg^{*}\to q\overline{q}}^{(i)}}$  & %
\begin{minipage}[c][2.cm]{6cm}%
\[
\frac{1}{N_{c}^{2}-1}\left(N_{c}^{2}\mathcal{F}_{gg}^{\left(1\right)}-\mathcal{F}_{gg}^{\left(3\right)}\right)
\]
\end{minipage}  & %
\begin{minipage}[c][2.cm]{6cm}%
\[
-N_{c}^{2}\mathcal{F}_{gg}^{\left(2\right)}+\mathcal{F}_{gg}^{\left(3\right)}
\]
\end{minipage} \tabularnewline
\hline 
${\displaystyle \Phi_{qg^{*}\to qg}^{(i)}}$  & %
\begin{minipage}[c][2.cm]{6cm}%
\[
\mathcal{F}_{qg}^{\left(1\right)}
\]
\end{minipage}  & %
\begin{minipage}[c][2.cm]{6cm}%
\[
\frac{1}{N_{c}^{2}-1}\left(-\mathcal{F}_{qg}^{\left(1\right)}+N_{c}^{2}\mathcal{F}_{qg}^{\left(2\right)}\right)
\]
\end{minipage} \tabularnewline
\hline 
${\displaystyle \Phi_{\gamma g^{*}\to q\overline{q}}^{(i)}}$ & %
\begin{minipage}[c][2cm]{6cm}%
\[
\mathcal{F}_{gg}^{\left(3\right)}
\]
\end{minipage}  & ---\tabularnewline
\hline
${\displaystyle \Phi_{q g^{*}\to \gamma q}^{(i)}}$ & %
\begin{minipage}[c][2cm]{6cm}%
\[
\mathcal{F}_{qg}^{\left(1\right)}
\]
\end{minipage}  & ---\tabularnewline
\hline 
\end{tabular}
\par\end{centering}
\caption{The TMD gluon distributions corresponding to the hard factors $\tilde{\mathcal{H}}^{\left(i\right)}$.
\label{tab:Phi}}
\end{table}

\begin{figure}
\begin{centering}
\includegraphics[width=10cm]{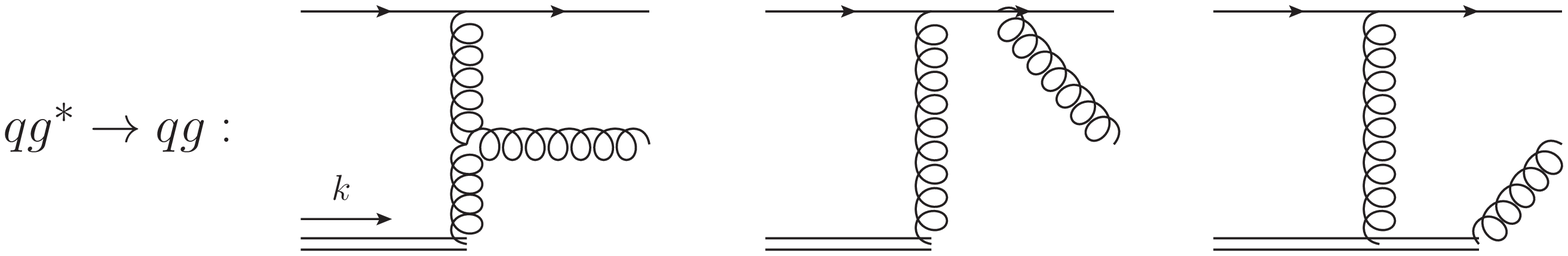}
\par\end{centering}
\begin{centering}
\vspace{0.6cm}
\par\end{centering}
\begin{centering}
\includegraphics[width=10cm]{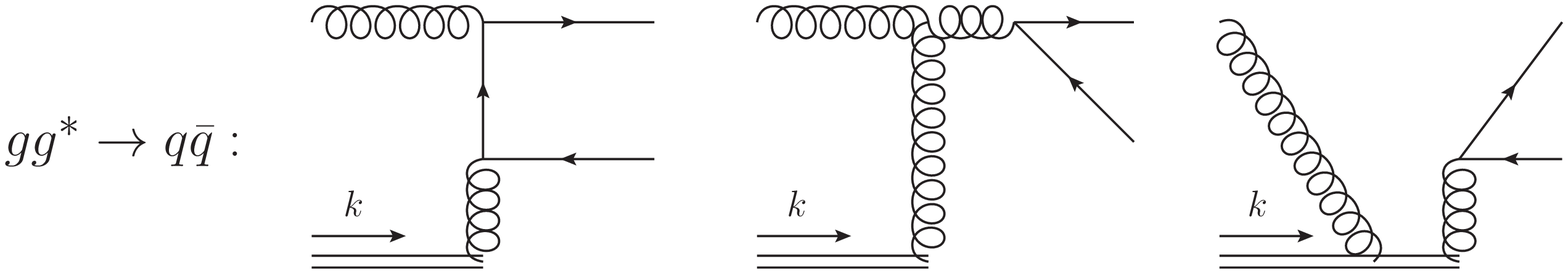}
\par\end{centering}
\begin{centering}
\vspace{0.6cm}
\par\end{centering}
\begin{centering}
\includegraphics[width=10cm]{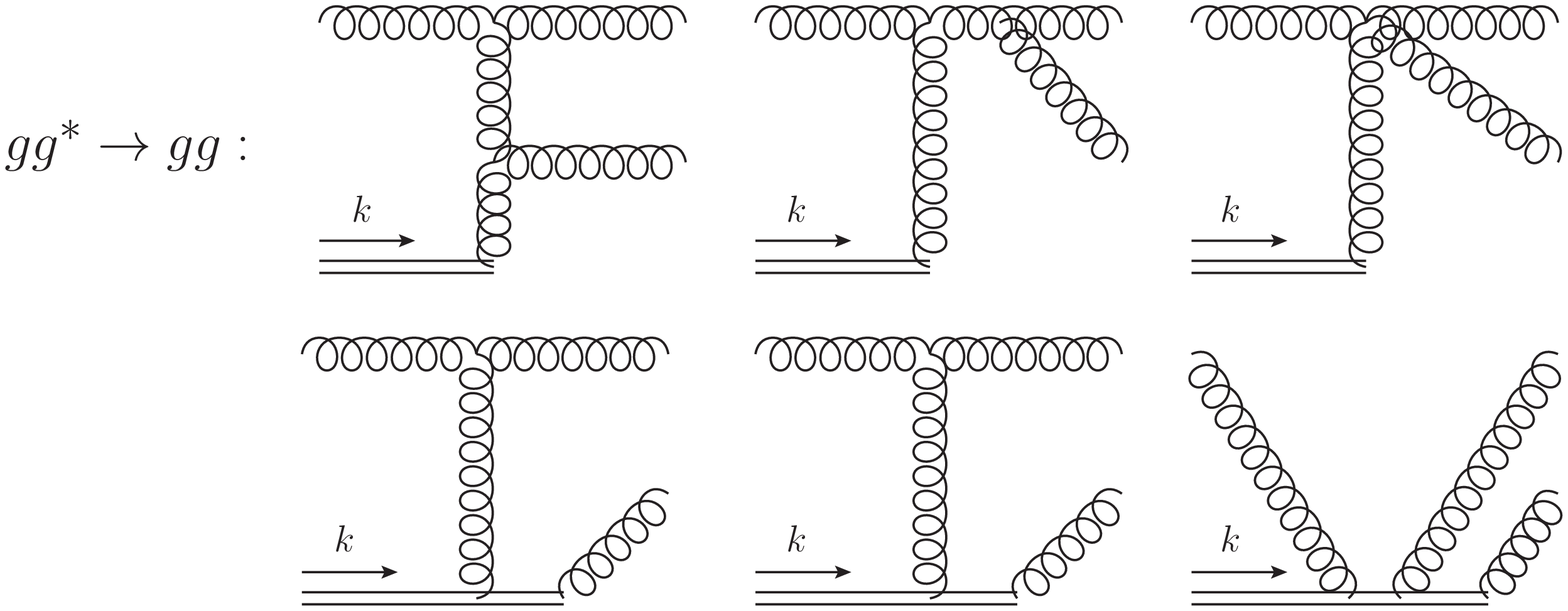}
\par\end{centering}
\caption{Diagrams contributing to the gauge invariant hard factors $\tilde{\mathcal{H}}_{ag^{*}\rightarrow bc}^{\left(i\right)}$
for various channels.
 We show only planar color-ordered diagrams, i.e. the planar diagrams with fixed ordering of external legs,
as they are enough to reconstruct the hard factors contributing to the
in-equivalent color flows (see Section 6 of \citep{Kotko:2015ura} on how to reconstruct the hard factors from color-ordered amplitudes and  \citep{Mangano:1990by} for a general review of color decompositions). 
The off-shell gluon has momentum $k$.
The double line corresponds to the Wilson line propagator in momentum
space, which couples to gluons via the $igt^{a}P^{\mu}$ vertex. The
double line propagator with a momentum $p$ is $-i/\left(p\cdot P+i\epsilon\right)$.
These diagrams have to be multiplied by $\boldsymbol{k}^{2}/g$ --
for all the details see \cite{Kotko2014a}. We do not display the
diagrams for processes with a photon since they do not require the use
of a Wilson line, despite the off-shellnes of the gluon. \label{fig:WilsonLine_diags}
}
\end{figure}

The off-shell gauge invariant hard factors $\tilde{\mathcal{H}}^{\left(i\right)}$
involve incoming off-shell gluons with momentum $k=x_A P+k_{\perp}$, $k^{2}=-\boldsymbol{k}^{2}$,
coupled eikonally to the target via a TMD correlator. In general, such Feynman diagrams are
not gauge invariant when calculated using the standard QCD Feynman
rules. There are several ways, to deal with this. First, one could
use the Lipatov effective action and resulting vertices in the quasi-multi-Regge
kinematics \citep{Antonov:2004hh}. In \citep{VanHameren2012,VanHameren2013a,vanHameren:2014iua,vanHameren:2015bba}
other methods have been developed, based on the spinor helicity formalism,
especially convenient to deal with multiparticle processes and to
guarantee fast computer implementation. The method \citep{VanHameren2013a}
has been recently extended to loop level \citep{VanHameren2017}.
The easiest way to understand the diagrammatic content of the hard
factors is probably provided by the method \citep{Kotko2014a} which
defines the gauge invariant off-shell amplitudes as partonic matrix
elements of straight infinite Wilson line operators. In case of the
hard factors involving one off-shell gluon needed here the Wilson
line has a direction along $P^{-}$. The diagrams contributing to
each channel for pA collisions are given in Fig.~\ref{fig:WilsonLine_diags}.

The form of the generalized factorization (\ref{eq:iTMD2}) appears
as follows. First the color structure is separated from the kinematic
part of the amplitude by means of the color decomposition \citep{Mangano:1990by}.
The amplitudes with the color structure separated contain only planar
diagrams with fixed ordering of the external legs. The TMD gluon distributions
$\Phi_{ag\rightarrow bc}^{\left(i\right)}$ are derived for the color
structures (squared) following the general procedure of resummation
of collinear gluons constructed in \citep{TMD}. The color
decomposition of amplitudes guarantees that each $\Phi_{ag\rightarrow bc}^{\left(i\right)}$
corresponds to a gauge invariant subset of diagrams. For more details
and application to multiparticle processes see \citep{Bury2018}.

The iTMD formula was constructed to agree with the $k_{T}$-factorization
for dijet production \citep{Deak:2009xt} in the limit of
 $\boldsymbol{k}^{2}\sim Q^{2}\gg Q_{s}^{2}$
and also with the leading power limit of the CGC expressions \citep{Dominguez:2011wm}
for $Q^{2}\gg\boldsymbol{k}^{2}\sim Q_{s}^{2}$. In the present
paper we further compare all the power corrections contained in the
framework. To this end, we need the small $x$ limit of the TMD gluon
distributions compliant with the CGC theory. They are obtained by
neglecting the $x$ dependence in the Fourier transforms and trading
the hadronic matrix elements to the averages over the color distributions
in the nucleus. In addition, lightcone gauge is used,
in which for the shockwave approximation the transverse components
of the gauge fields do not contribute due to EOM. This allows to neglect
the transverse parts of the staple gauge links (\ref{eq:Staples}).
Within the above approximation we have \citep{Dominguez:2011wm,Marquet:2016cgx}:
\begin{equation}
\label{F1_qg}
\mathcal{F}_{qg}^{\left(1\right)}=\frac{4}{g^{2}}\,\int\frac{d^{2}\boldsymbol{x}d^{2}\boldsymbol{y}}{\left(2\pi\right)^{3}}\,e^{-i\boldsymbol{k}\cdot\left(\boldsymbol{x}-\boldsymbol{y}\right)}\left\langle \mathrm{Tr}\left\{ \left(\partial_{i}U_{\boldsymbol{y}}\right)\left(\partial_{i}U_{\boldsymbol{x}}^{\dagger}\right)\right\} \right\rangle \,,
\end{equation}

\begin{equation}
\label{F2_qg}
\mathcal{F}_{qg}^{\left(2\right)}=-\frac{4}{g^{2}}\,\int\frac{d^{2}\boldsymbol{x}d^{2}\boldsymbol{y}}{\left(2\pi\right)^{3}}\,e^{-i\boldsymbol{k}\cdot\left(\boldsymbol{x}-\boldsymbol{y}\right)}\frac{1}{N_{c}}\left\langle \mathrm{Tr}\left\{ \left(\partial_{i}U_{\boldsymbol{x}}\right)U_{\boldsymbol{y}}^{\dagger}\left(\partial_{i}U_{\boldsymbol{y}}^{\dagger}\right)U_{\boldsymbol{x}}^{\dagger}\right\} \mathrm{Tr}\left\{ U_{\boldsymbol{y}}U_{\boldsymbol{x}}^{\dagger}\right\} \right\rangle \,.
\end{equation}
\begin{equation}
\label{F1_gg}
\mathcal{F}_{gg}^{\left(1\right)}=\frac{4}{g^{2}}\,\int\frac{d^{2}\boldsymbol{x}d^{2}\boldsymbol{y}}{\left(2\pi\right)^{3}}\,e^{-i\boldsymbol{k}\cdot\left(\boldsymbol{x}-\boldsymbol{y}\right)}\frac{1}{N_{c}}\left\langle \mathrm{Tr}\left\{ \left(\partial_{i}U_{\boldsymbol{y}}\right)\left(\partial_{i}U_{\boldsymbol{x}}^{\dagger}\right)\right\} \mathrm{Tr}\left\{ U_{\boldsymbol{x}}U_{\boldsymbol{y}}^{\dagger}\right\} \right\rangle \,,
\end{equation}

\begin{equation}
\label{F2_gg}
\mathcal{F}_{gg}^{\left(2\right)}=-\frac{4}{g^{2}}\,\int\frac{d^{2}\boldsymbol{x}d^{2}\boldsymbol{y}}{\left(2\pi\right)^{3}}\,e^{-i\boldsymbol{k}\cdot\left(\boldsymbol{x}-\boldsymbol{y}\right)}\frac{1}{N_{c}}\left\langle \mathrm{Tr}\left\{ \left(\partial_{i}U_{\boldsymbol{x}}\right)U_{\boldsymbol{y}}^{\dagger}\right\} \mathrm{Tr}\left\{ \left(\partial_{i}U_{\boldsymbol{y}}\right)U_{\boldsymbol{x}}^{\dagger}\right\} \right\rangle \,,
\end{equation}
\begin{equation}
\label{F3_gg}
\mathcal{F}_{gg}^{\left(3\right)}=-\frac{4}{g^{2}}\,\int\frac{d^{2}\boldsymbol{x}d^{2}\boldsymbol{y}}{\left(2\pi\right)^{3}}\,e^{-i\boldsymbol{k}\cdot\left(\boldsymbol{x}-\boldsymbol{y}\right)}\left\langle \mathrm{Tr}\left\{ \left(\partial_{i}U_{\boldsymbol{x}}\right)U_{\boldsymbol{y}}^{\dagger}\left(\partial_{i}U_{\boldsymbol{y}}\right)U_{\boldsymbol{x}}^{\dagger}\right\} \right\rangle \,,
\end{equation}

\begin{equation}
\label{F4_gg}
\mathcal{F}_{gg}^{\left(4\right)}=-\frac{4}{g^{2}}\,\int\frac{d^{2}\boldsymbol{x}d^{2}\boldsymbol{y}}{\left(2\pi\right)^{3}}\,e^{-i\boldsymbol{k}\cdot\left(\boldsymbol{x}-\boldsymbol{y}\right)}\left\langle \mathrm{Tr}\left\{ \left(\partial_{i}U_{\boldsymbol{x}}\right)U_{\boldsymbol{x}}^{\dagger}\left(\partial_{i}U_{\boldsymbol{y}}\right)U_{\boldsymbol{y}}^{\dagger}\right\} \right\rangle \,,
\end{equation}
\begin{equation}
\label{F5_gg}
\mathcal{F}_{gg}^{\left(5\right)}=-\frac{4}{g^{2}}\,\int\frac{d^{2}\boldsymbol{x}d^{2}\boldsymbol{y}}{\left(2\pi\right)^{3}}\,e^{-i\boldsymbol{k}\cdot\left(\boldsymbol{x}-\boldsymbol{y}\right)}\left\langle \mathrm{Tr}\left\{ \left(\partial_{i}U_{\boldsymbol{x}}\right)U_{\boldsymbol{y}}^{\dagger}U_{\boldsymbol{x}}U_{\boldsymbol{y}}^{\dagger}\left(\partial_{i}U_{\boldsymbol{y}}\right)U_{\boldsymbol{x}}^{\dagger}U_{\boldsymbol{y}}U_{\boldsymbol{x}}^{\dagger}\right\} \right\rangle \,,
\end{equation}
\begin{multline}
\label{F6_gg}
\mathcal{F}_{gg}^{\left(6\right)}=-\frac{4}{g^{2}}\,\int\frac{d^{2}\boldsymbol{x}d^{2}\boldsymbol{y}}{\left(2\pi\right)^{3}}\,e^{-i\boldsymbol{k}\cdot\left(\boldsymbol{x}-\boldsymbol{y}\right)} \\
\frac{1}{N_{c}^{2}}\left\langle \mathrm{Tr}\left\{ \left(\partial_{i}U_{\boldsymbol{x}}\right)U_{\boldsymbol{y}}^{\dagger}\left(\partial_{i}U_{\boldsymbol{y}}\right)U_{\boldsymbol{x}}^{\dagger}\right\} \mathrm{Tr}\left\{ U_{\boldsymbol{x}}U_{\boldsymbol{y}}^{\dagger}\right\} \mathrm{Tr}\left\{ U_{\boldsymbol{y}}U_{\boldsymbol{x}}^{\dagger}\right\} \right\rangle \,.
\end{multline}

For completeness, let us now put the iTMD formulation into the context
of the formal TMD factorization theorems \citep{Collins:2011zzd}.
First, one should understand that it 
does not involve an all-order factorization
theorem like the ones existing for the Drell-Yan process and semi-inclusive
DIS. These theorems are proved to leading power in the hard scale to any
logarithmic accuracy, while the iTMD framework resums the power corrections,
but its validity is limited to leading logarithms of energy. 
Next, the mentioned TMD factorization theorems involve
processes with at most two colored partons in the hard process (plus
soft/collinear contributions of course) and two TMD correlators (parton
distribution or fragmentation function). Because of the simplicity of the color structure, all Wilson lines
appearing due to the resummation of collinear gluons can be disentangled
and put into the gauge invariant definitions of the TMD objects. For
jet production processes in hadron-hadron collision, where formally
one has at least two TMD correlators and more than two colored partons, it is not possible. Thus, formally, even the
generalized factorization breaks down \citep{Rogers:2010dm}. However,
in the iTMD approach, which targets the collisions of a moderate-$x$ projectile and a low-$x$ target, there
is only one TMD correlator, thus, at least formally, this problem
does not occur. On the formal ground there is no all-order proof of
the hybrid approach so far.

Finally, let us comment on the evolution equations for the TMD gluon
distributions. The most adequate treatment would be using the renormalization
group equation at small and moderate $x$ developed in \citep{Balitsky2015a,Balitsky2016}.
It however still requires work to derive the complete set of equations,
not to mention solving them. An important feature of such procedure
would be that some Sudakov logarithms $\ln\boldsymbol{k}^2$ can be consistently resummed. 
For existing phenomenological applications using iTMD \citep{vanHameren:2016ftb,Kotko2017b}
the evolution was based on BK or B-JIMWLK and some Sudakov resummation
effects were estimated by means of a phenomenological model.

Below, we explicitly give formulas for the cross sections (\ref{eq:iTMD2})
in a form that can be directly compared with the CGC expressions.

\subsection{$qg^{*}\rightarrow qg$ channel}

We get
\begin{equation}  
\label{q2qg-ITMD}
\frac{d\sigma_{qA\rightarrow qg}}{d^{2}\boldsymbol{p}_{1}d^{2}\boldsymbol{p}_{2}dy_{1}dy_{2}}=
 p_{0}^{+}\delta\left(p_{1}^{+}+p_{2}^{+}-p_{0}^{+}\right)
\left[\mathcal{H}_{qg\rightarrow qg}^{\left(1\right)}\Phi_{qg\rightarrow qg}^{\left(1\right)}+\mathcal{H}_{qg\rightarrow qg}^{\left(2\right)}\Phi_{qg\rightarrow qg}^{\left(2\right)}\right]\,,
\end{equation}
with
\begin{equation}
\label{q2qg_1}
\mathcal{H}_{qg\rightarrow qg}^{\left(1\right)}=\alpha_{s}^{2}\,
\frac{z^2\left(1+z^{2}\right)}{2\boldsymbol{q}^{2}}
\Bigg\{\frac{z}{\boldsymbol{p}_{1}^{2}}+\frac{1}{N_{c}^{2}}\frac{\boldsymbol{q}^{2}-\overline{z}^{2}\boldsymbol{p}_{1}^{2}}{{z}\boldsymbol{p}_{1}^{2}\boldsymbol{p}_{2}^{2}}\Bigg\}\,,
\end{equation}
\begin{equation}
\label{q2qg_2}
\mathcal{H}_{qg\rightarrow qg}^{\left(2\right)}=\alpha_{s}^{2}\, \frac{N_{A}}{2N_{c}^{2}}\, \frac{{z}\left(1+{z}^{2}\right)}{\boldsymbol{p}_{1}^{2}\boldsymbol{p}_{2}^{2}}\,.
\end{equation}
Note, that the above hard factors $\mathcal{H}_{qg\rightarrow qg}^{\left(i\right)}$
are not exactly the ones in (\ref{eq:iTMD2}). The expressions are
however more compact in the above notation.

\subsection{$gg^{*}\rightarrow q\overline{q}$ channel}

\begin{equation}
\label{g2qbarq-ITMD}
\frac{d\sigma_{gA\rightarrow q\overline{q}}}{d^{2}\boldsymbol{p}_{1}d^{2}\boldsymbol{p}_{2}dy_{1}dy_{2}}
= p_{0}^{+}\delta\left(p_{1}^{+}+p_{2}^{+}-p_{0}^{+}\right) \left[\mathcal{H}_{gg\rightarrow q\overline{q}}^{\left(1\right)}\Phi_{gg\rightarrow q\overline{q}}^{\left(1\right)}+\mathcal{H}_{gg\rightarrow q\overline{q}}^{\left(2\right)}\Phi_{gg\rightarrow q\overline{q}}^{\left(2\right)}\right]\,,
\end{equation}
where $\mathcal{H}_{gg\rightarrow q\overline{q}}^{\left(1\right)}$,
$\mathcal{H}_{gg\rightarrow q\overline{q}}^{\left(2\right)}$ are
the reduced off-shell hard factors. They read
\begin{equation}
\label{g2qbarq_1}
\mathcal{H}_{gg\rightarrow q\overline{q}}^{\left(1\right)}=\frac{\alpha_{s}^{2}}{2N_{c}}
\, \, z\overline{z}\left(1-2z\overline{z}\right) \, \frac{\boldsymbol{p}_{1}^{2}\left(1-z\right)^{2}+\boldsymbol{p}_{2}^{2}z^{2}}{\boldsymbol{q}^2 \boldsymbol{p}_{1}^{2}\boldsymbol{p}_{2}^{2}}\,,
\end{equation}
\begin{equation}
\label{g2qbarq_2}
\mathcal{H}_{gg\rightarrow q\overline{q}}^{\left(2\right)}=\frac{\alpha_{s}^{2}}{2N_{c}^{2}C_{F}}
\, \left(z\overline{z}\right)^2\left(1-2z\overline{z}\right) \,  \frac{(\boldsymbol{p_1}\cdot\boldsymbol{p_2})}{\boldsymbol{q}^{2}\boldsymbol{p}_{1}^2\boldsymbol{p}_{2}^2}  \, .
\end{equation}

\subsection{$gg^{*}\rightarrow gg$ channel}

\begin{equation}
\label{g2gg-ITMD}
\frac{d\sigma_{gA\rightarrow gg}}{d^{2}\boldsymbol{p}_{1}d^{2}\boldsymbol{p}_{2}dy_{1}dy_{2}}
=  p_{0}^{+}\delta\left(p_{1}^{+}+p_{2}^{+}-p_{0}^{+}\right) \left[\mathcal{H}_{gg\rightarrow gg}^{\left(1\right)}\Phi_{gg\rightarrow gg}^{\left(1\right)}+\mathcal{H}_{gg\rightarrow gg}^{\left(2\right)}\Phi_{gg\rightarrow gg}^{\left(2\right)}\right]\,,
\end{equation}
with
\begin{equation}
\label{g2gg_1}
\mathcal{H}_{gg\rightarrow gg}^{\left(1\right)}=\alpha_{s}^{2} \frac{2N_{c}^{2}}{N_{A}}\, \left(1-z\overline{z}\right)^{2}\, \frac{\boldsymbol{p}_{1}^{2}\overline{z}^{2}+\boldsymbol{p}_{2}^{2}z^{2}}{\boldsymbol{q}^{2} \boldsymbol{p}_{1}^{2}\boldsymbol{p}_{2}^{2}}\,,
\end{equation}
\begin{equation}
\label{g2gg_2}
\mathcal{H}_{gg\rightarrow gg}^{\left(2\right)}=\alpha_{s}^{2}\frac{N_{c}^{2}}{N_{A}}\, \left(1-z\overline{z}\right)^{2} \, \frac{\boldsymbol{q}^{2}-\boldsymbol{p}_{1}^{2}\overline{z}^{2}-\boldsymbol{p}_{2}^{2}z^{2}}{\boldsymbol{q}^{2} \boldsymbol{p}_{1}^{2}\boldsymbol{p}_{2}^{2}}\,.
\end{equation}
Above, an additional symmetry factor of $1/2$ was included to account
for identical final states.

\subsection{$\gamma g^{*}\rightarrow q\overline{q}$ channel}

\begin{equation}
\label{gamma2qbarq-ITMD}
\frac{d\sigma_{\gamma A\rightarrow q\overline{q}}}{d^{2}\boldsymbol{p}_{1}d^{2}\boldsymbol{p}_{2}dy_{1}dy_{2}}=
 p_{0}^{+}\delta\left(p_{1}^{+}+p_{2}^{+}-p_{0}^{+}\right)
\mathcal{H}_{\gamma g^{*}\rightarrow gg}\mathcal{F}_{gg}^{\left(3\right)}\,,
\end{equation}
with
\begin{equation}
\label{gamma2qbarq_1}
\mathcal{H}_{\gamma g^{*}\rightarrow gg}=\alpha_{\mathrm{em}}\alpha_{s}\,\frac{z\overline{z}\left(1-2z\overline{z}\right)}{\boldsymbol{p}_{1}^{2}\boldsymbol{p}_{2}^{2}}\,.
\end{equation}

\subsection{$qg^{*}\rightarrow q\gamma$ channel}

\begin{equation}
\label{q2qgamma-ITMD}
\frac{d\sigma_{qA\rightarrow q\gamma}}{d^{2}\boldsymbol{p}_{1}d^{2}\boldsymbol{p}_{2}dy_{1}dy_{2}}=
 p_{0}^{+}\delta\left(p_{1}^{+}+p_{2}^{+}-p_{0}^{+}\right)
\mathcal{H}_{qg^{*}\rightarrow q\gamma}\mathcal{F}_{qg}^{\left(1\right)}\,,
\end{equation}
with
\begin{equation}
\label{q2qgamma_1}
\mathcal{H}_{qg^{*}\rightarrow q\gamma}=\alpha_{\mathrm{em}}\alpha_{s}\,\frac{1}{C_{A}}\,\frac{z\overline{z}^{2}\left(1+z^{2}\right)}{\boldsymbol{q}^{2}\boldsymbol{p}_{2}^{2}}\,.
\end{equation}

%%%%%%%%%%%%%%%%%%%%%%%%%%%%%%%%%%%%%%%%%%%%%%%%%%%%%%%%%%%%%%%%%%%%%%%
%%%%%%%%%%%%%%%%%%%%%%%%%%%%%%%%%%%%%%%%%%%%%%%%%%%%%%%%%%%%%%%%%%%%%%%
\section{From the generic CGC process to the specific cases \label{sec:Applications}}

In sections \ref{sec:CGCexpansion} and \ref{sec:Dilute}, we have computed both the kinematic-twist-resummed cross section and the dilute limit of the CGC cross section for a generic $1\to2$ process respectively. Our aim in this section is to get both of these cross sections for specific processes and compare these results with the ones that are obtained through iTMD framework in section~\ref{sec:iTMD}. 

To be more accurate, we consider the photoproduction of a dijet, as well as all possible channels for two particle production (dijet or photon-jet) in forward pp and pA collisions. Within the CGC framework, hybrid formalism \cite{Dumitru:2005gt} is the state of the art approach for these processes. It has been very successfully used to calculate the next-to-leading order single inclusive particle production \cite{Altinoluk:2011qy}-\cite{Ducloue:2017mpb}, heavy quark production \cite{Altinoluk:2015vax}, dijet production \cite{Marquet:2016cgx} and recently dijet+photon \cite{Altinoluk:2018byz,Altinoluk:2018uax} and trijet production \cite{Iancu:2018hwa} in forward pA collisions.

In the hybrid formalism, the final state particles are produced in the forward rapidity region so they can be treated in the collinear framework, i.e. the incoming partons are on-shell collinear partons and the partonic cross section calculated in this set up should be convoluted with the collinear parton distribution functions in order to get the hadronic cross sections. On the other hand, the target is assumed to be dense and the center-of-mass energy is large so it can be treated in the CGC framework. At the parton level, the set up that we have used for the calculation of the kinematic-twist-resummed cross section Eq. \eqref{eq:genCrossSec} and the dilute cross section Eq. \eqref{eq:DilCSuPDF} for a generic process is compatible with the hybrid formalism. Thus, we use those results to study the different channels and compare them with the ones obtained from iTMD framework in the rest of this section.

%%%%%%%%%%%%%%%%%%%%%%%%%%%%%%%%%%%%%%%%%%%%%%%%%%%%%%%%%%%%%%%%%%%%%%%%%%%%

\subsection{$q\rightarrow qg$ channel\label{subsec:qqgres}}

\begin{figure}[hbt]
\begin{centering}
\includegraphics[scale=0.5]{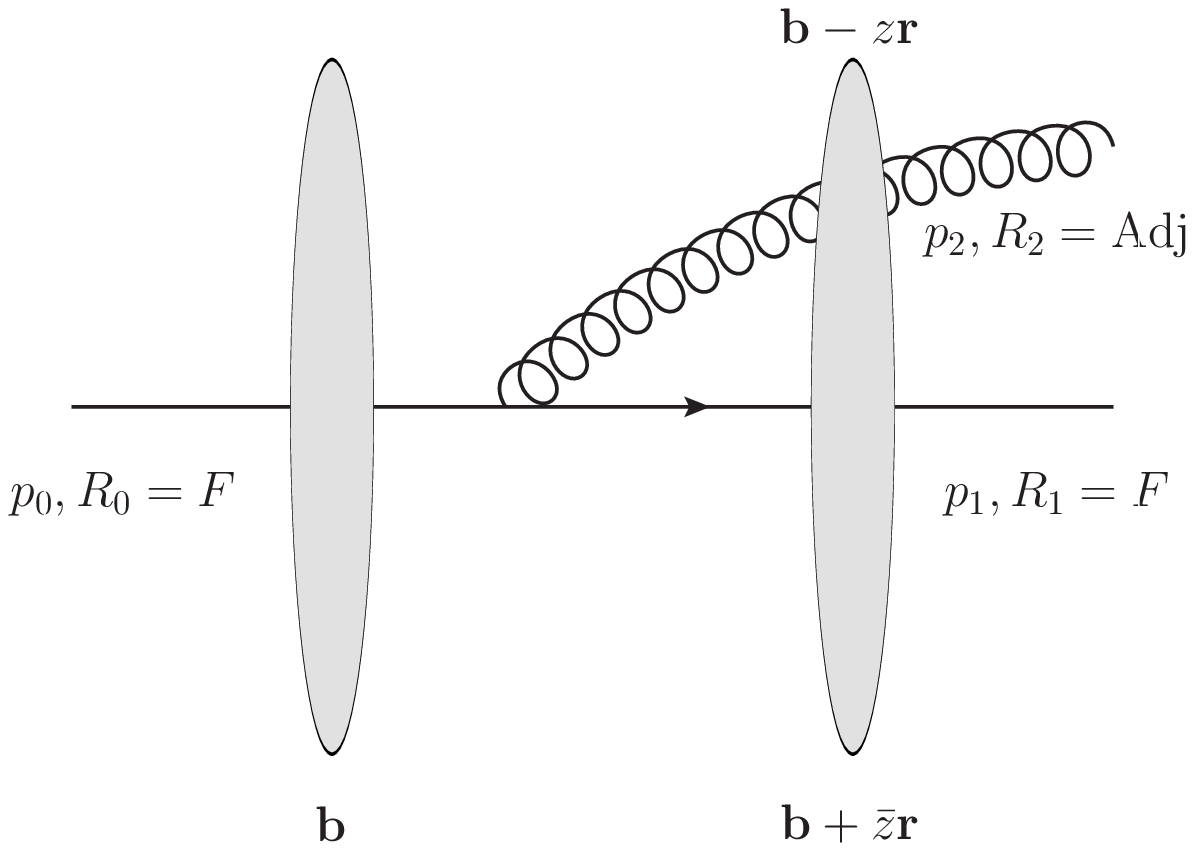}
\par\end{centering}
\caption{$q\to qg$ amplitude in an external shockwave background with the appropriate color representations. }
\label{fig:QtoQG}  
\end{figure}

Let us start our analysis by considering the $q\rightarrow qg$ channel (see Fig.~\ref {fig:QtoQG}). In this channel, the incoming quark splits into a quark-gluon pair at order $g_s$ which then scatters off the target via eikonal interaction. The CGC amplitude for this channel is given in Eq.~\eqref{eq:QtoQGini}. In order to be able to use the kinematic twist resummed  generic cross section Eq.~\eqref{eq:genCrossSec}, the first thing we need is the tensor part of the amplitude that encodes the Dirac structure of this channel and it is given by 
\begin{align}
\phi_{\mu}^{\left(q\rightarrow qg\right)} & =\frac{ig_{s}}{2\pi}\varepsilon_{p_{g}\perp}^{\sigma\ast}\bar{u}_{p_{q}}\left[2zg_{\perp\mu\sigma}+\bar{z}\left(\gamma_{\perp\mu}\gamma_{\perp\sigma}\right)\right]\gamma^{+}u_{p},\label{eq:QtoQGtens}
\end{align}
whose square for an unpolarized observable can be calculated in a straightforward manner and the result reads 
\begin{align}
\phi^{i\left(q\rightarrow qg\right)}\phi^{i^{\prime}\ast\left(q\rightarrow qg\right)} & =\delta^{ii^{\prime}}\left(\frac{g_{s}}{2\pi}\right)^{2}\left(p_{0}^{+}\right)^{2}16z\left(1+z^{2}\right).\label{eq:QtoQGtens2}
\end{align}
One can read off the color structure in this channel from Fig.~\ref{fig:QtoQG} and it is given by setting  $U_{\boldsymbol{b}}^{R_{1}}=U_{\boldsymbol{b}},$ $U_{\boldsymbol{b}}^{R_{2}}=U_{\boldsymbol{b}}^{ab}$ and $T^{R_{0}}=T^{b}$. This color structure leads to the following TMD operators   
\begin{align}
\mathcal{O}_{1}^{\left(q\rightarrow qg\right)} & =\left(\partial^{j}U_{\boldsymbol{b}}\right)T^{b}U_{\boldsymbol{b}}^{ab}U_{\boldsymbol{b}^{\prime}}^{ac\dagger}T^{c}\left(\partial^{j^{\prime}}U_{\boldsymbol{b}^{\prime}}^{\dagger}\right)\nonumber \\
\mathcal{O}_{2}^{\left(q\rightarrow qg\right)} & =\left(\partial^{j}U_{\boldsymbol{b}}\right)T^{b}U_{\boldsymbol{b}}^{ab}\left(\partial^{j^{\prime}}U_{\boldsymbol{b}^{\prime}}^{ac}\right)T^{c}U_{\boldsymbol{b}^{\prime}}^{\dagger}\label{eq:QtoQGope}\\
\mathcal{O}_{3}^{\left(q\rightarrow qg\right)} & =U_{\boldsymbol{b}}T^{b}\left(\partial^{j}U_{\boldsymbol{b}}^{ab}\right)U_{\boldsymbol{b}^{\prime}}^{ac\dagger}T^{c}\left(\partial^{j^{\prime}}U_{\boldsymbol{b}^{\prime}}^{\dagger}\right)\nonumber \\
\mathcal{O}_{4}^{\left(q\rightarrow qg\right)} & =U_{\boldsymbol{b}}T^{b}\left(\partial^{j}U_{\boldsymbol{b}}^{ab}\right)\left(\partial^{j^{\prime}}U_{\boldsymbol{b}^{\prime}}^{ac}\right)T^{c}U_{\boldsymbol{b}^{\prime}}^{\dagger}.\nonumber 
\end{align}
By using the identity that relates the adjoint and fundamental representations of a unitary matrix
\begin{equation}
U^{ab}(\boldsymbol{b})=2\, \mathrm{Tr}\big[ t^aU(\boldsymbol{b})t^bU^\dagger(\boldsymbol{b})\big]
\end{equation}
and the Fierz identity 
\begin{equation}
t^a_{\alpha\beta}t^a_{\sigma\lambda}=\frac{1}{2}\bigg[\delta_{\alpha\lambda}\delta_{\beta\sigma}-\frac{1}{N_c}\delta_{\alpha\beta}\delta_{\sigma\lambda}\bigg]
\end{equation}
one can easily get the following identities 
\begin{align}
T^{b}\left(\partial^{j}U_{\boldsymbol{b}}^{ab}\right) & =\left(\partial^{j}U_{\boldsymbol{b}}^{\dagger}\right)T^{a}U_{\boldsymbol{b}}+U_{\boldsymbol{b}}^{\dagger}T^{a}\left(\partial^{j}U_{\boldsymbol{b}}\right)\label{eq:DerGlu}\\
\left(\partial^{j^{\prime}}U_{\boldsymbol{b}^{\prime}}^{ac}\right)T^{c\dagger} & =\left(\partial^{j^{\prime}}U_{\boldsymbol{b}^{\prime}}^{\dagger}\right)T^{a}U_{\boldsymbol{b}^{\prime}}+U_{\boldsymbol{b}^{\prime}}^{\dagger}T^{a}\left(\partial^{j^{\prime}}U_{\boldsymbol{b}^{\prime}}\right),\nonumber 
\end{align}
The next step is to compute the color trace of the TMD operators that are listed in Eq. \eqref{eq:QtoQGope}. By using the identities given in Eq. \eqref{eq:DerGlu}, these traces can easily be computed and the result reads
\begin{align}
\mathrm{Tr}\,\Big[ \mathcal{O}_{1}^{\left(q\rightarrow qg\right)} \Big]& =-\frac{1}{2}\mathrm{Tr}\left[\left(\partial^{j}U_{\boldsymbol{b}}\right)U_{\boldsymbol{b}}^{\dagger}\left(\partial^{j^{\prime}}U_{\boldsymbol{b}^{\prime}}\right)U_{\boldsymbol{b}^{\prime}}^{\dagger}\right]\mathrm{Tr}\left(U_{\boldsymbol{b}}U_{\boldsymbol{b}^{\prime}}^{\dagger}\right)-\frac{1}{2N_{c}}\mathrm{Tr}\left[\left(\partial^{j}U_{\boldsymbol{b}}\right)\left(\partial^{j^{\prime}}U_{\boldsymbol{b}^{\prime}}^{\dagger}\right)\right]\nonumber \\
\mathrm{Tr}\, \Big[ \mathcal{O}_{2}^{\left(q\rightarrow qg\right)} \Big]& =\frac{1}{2}\mathrm{Tr}\left[\left(\partial^{j}U_{\boldsymbol{b}}\right)U_{\boldsymbol{b}}^{\dagger}\left(\partial^{j^{\prime}}U_{\boldsymbol{b}^{\prime}}\right)U_{\boldsymbol{b}^{\prime}}^{\dagger}\right]\mathrm{Tr}\left(U_{\boldsymbol{b}}U_{\boldsymbol{b}^{\prime}}^{\dagger}\right)\label{eq:QtoQGopeTr}\\
\mathrm{Tr}\, \Big[ \mathcal{O}_{3}^{\left(q\rightarrow qg\right)} \Big] & =\frac{1}{2}\mathrm{Tr}\left[\left(\partial^{j}U_{\boldsymbol{b}}\right)U_{\boldsymbol{b}}^{\dagger}\left(\partial^{j^{\prime}}U_{\boldsymbol{b}^{\prime}}\right)U_{\boldsymbol{b}^{\prime}}^{\dagger}\right]\mathrm{Tr}\left(U_{\boldsymbol{b}}U_{\boldsymbol{b}^{\prime}}^{\dagger}\right)\nonumber \\
\mathrm{Tr}\, \Big[ \mathcal{O}_{4}^{\left(q\rightarrow qg\right)} \Big]& =-\frac{1}{2}\mathrm{Tr}\left[\left(\partial^{j}U_{\boldsymbol{b}}\right)U_{\boldsymbol{b}}^{\dagger}\left(\partial^{j^{\prime}}U_{\boldsymbol{b}^{\prime}}\right)U_{\boldsymbol{b}^{\prime}}^{\dagger}\right]\mathrm{Tr}\left(U_{\boldsymbol{b}}U_{\boldsymbol{b}^{\prime}}^{\dagger}\right)+\frac{N_{c}}{2}\mathrm{Tr}\left[\left(\partial^{j}U_{\boldsymbol{b}}\right)\left(\partial^{j^{\prime}}U_{\boldsymbol{b}^{\prime}}^{\dagger}\right)\right].\nonumber 
\end{align}
Comparing the structure of the trace of the Wilson lines in Eq. \eqref{eq:QtoQGopeTr} and the definitions of the first two gluon TMDs in the quark channel $\mathcal{F}_{qg}^{\left(1\right)}$ and $\mathcal{F}_{qg}^{\left(2\right)}$ given in Eqs. \eqref{F1_qg} and \eqref{F2_qg} respectively, one can conclude that these are the two gluon TMDs which appear in this channel. Moreover, for convenience, we can define the following combinations of the gluon TMDs $\mathcal{F}_{qg}^{\left(1\right)}$ and $\mathcal{F}_{qg}^{\left(2\right)}$ :
\begin{align}
\Phi_{q\rightarrow qg}^{\left(1\right)}\left(\boldsymbol{k}\right) & \equiv\mathcal{F}_{qg}^{\left(1\right)}\left(\boldsymbol{k}\right)\label{gqqphi}\\
\nonumber \\
\Phi_{q\rightarrow qg}^{\left(2\right)}\left(\boldsymbol{k}\right) & \equiv\frac{N_{c}^{2}\mathcal{F}_{qg}^{\left(2\right)}\left(\boldsymbol{k}\right)-\mathcal{F}_{qg}^{\left(1\right)}\left(\boldsymbol{k}\right)}{N_{c}^{2}-1}\nonumber 
\end{align}
which are exactly the same combinations that one gets from the iTMD calculations given in the Table~\ref{tab:Phi}.

Finally, we can plug the square of the tensor part of the amplitude given in Eq. \eqref{eq:QtoQGtens2} and the Wilson line structure given in Eq. \eqref{eq:QtoQGopeTr} together with the definitions and the combinations of the gluon TMDs Eq. \eqref{gqqphi} in the generic kinematic twist resummed cross section Eq. \eqref{eq:genCrossSec} to get the cross section for $q\to qg$ channel as 
\begin{align}
\frac{d\sigma_{q\rightarrow qg}^{WW}}{dy_{1}dy_{2}d^{2}\boldsymbol{p}_{1}d^{2}\boldsymbol{p}_{2}} & =\alpha_{s}^{2}\frac{z\left(1+z^{2}\right)}{2\boldsymbol{p}_{1}^{2}\boldsymbol{p}_{2}^{2}}p_{0}^{+}\delta\left(p_{1}^{+}+p_{2}^{+}-p_{0}^{+}\right)\label{eq:QtoQGCrossSec}\\
 & \times\left[\left(z^{2}\frac{\boldsymbol{p}_{2}^{2}}{\boldsymbol{q}^{2}}+\frac{1}{N_{c}^{2}}\left(1-\bar{z}^{2}\frac{\boldsymbol{p}_{1}^{2}}{\boldsymbol{q}^{2}}\right)\right)\Phi_{q\rightarrow qg}^{\left(1\right)}\left(\boldsymbol{k}\right)+\left(\frac{N_{c}^{2}-1}{N_{c}^{2}}\right)\Phi_{q\rightarrow qg}^{\left(2\right)}\left(\boldsymbol{k}\right)\right]\nonumber 
\end{align}
which coincides exactly with Eq. \eqref{q2qg-ITMD} by using Eqs. \eqref{q2qg_1} and \eqref{q2qg_2}.

Our next order of business is to consider the dilute limit in the $q\to qg$ channel. Inserting the proper color representations in the dilute limit of the generic cross section given in Eq. \eqref{eq:DilCSuPDF}, we get
\begin{align}
 & \frac{d\sigma_{q\rightarrow qg}^{gA\sim0}}{dy_{1}dy_{2}d^{2}\boldsymbol{p}_{1}d^{2}\boldsymbol{p}_{2}}=\frac{\alpha_{s}}{16N_{c}\left(p_{0}^{+}\right)^{2}}\left(2\pi\right)\frac{\mathcal{G}_{ac}\left(\boldsymbol{k}\right)}{\boldsymbol{k}^{2}}\left(\phi^{i}\phi^{j\ast}\right)p_{0}^{+}\delta\left(p_{1}^{+}+p_{2}^{+}-p_{0}^{+}\right)\label{eq:qqgdilini}\\
 & \times\mathrm{Tr}\left[if^{bad}t^{d}\left(\frac{\boldsymbol{p}_{1}^{i}}{\boldsymbol{p}_{1}^{2}}-\frac{\boldsymbol{q}^{i}}{\boldsymbol{q}^{2}}\right)-t^{a}t^{b}\left(\frac{\boldsymbol{p}_{2}^{i}}{\boldsymbol{p}_{2}^{2}}+\frac{\boldsymbol{q}^{i}}{\boldsymbol{q}^{2}}\right)\right]\left[-if^{bce}t^{e}\left(\frac{\boldsymbol{p}_{1}^{j}}{\boldsymbol{p}_{1}^{2}}-\frac{\boldsymbol{q}^{j}}{\boldsymbol{q}^{2}}\right)-t^{b}t^{c}\left(\frac{\boldsymbol{p}_{2}^{j}}{\boldsymbol{p}_{2}^{2}}+\frac{\boldsymbol{q}^{j}}{\boldsymbol{q}^{2}}\right)\right],\nonumber 
\end{align}
with $\mathcal{G}_{ac}\left(\boldsymbol{k}\right)$ being the unintegrated parton distribution function defined in Eq.~\eqref{eq:uPDF}. Using the definition of the tensor part of the amplitude that encodes the Dirac structure in the $q\to qg$ channel given in Eq. \eqref{eq:QtoQGtens} and performing some color algebra, one simply gets the dilute limit of the cross section in this channel: 
\begin{align}
\frac{d\sigma_{q\rightarrow qg}^{gA\sim0}}{dy_{1}dy_{2}d^{2}\boldsymbol{p}_{1}d^{2}\boldsymbol{p}_{2}} & =\alpha_{s}^{2}\frac{z\left(1+z^{2}\right)}{2\boldsymbol{p}_{1}^{2}\boldsymbol{p}_{2}^{2}}p_{0}^{+}\delta\left(p_{1}^{+}+p_{2}^{+}-p_{0}^{+}\right)\mathcal{G}\left(\boldsymbol{k}\right)\left(1+z^{2}\frac{\boldsymbol{p}_{2}^{2}}{\boldsymbol{q}^{2}}-\frac{1}{N_{c}^{2}}\frac{\bar{z}^{2}\boldsymbol{p}_{1}^{2}}{\boldsymbol{q}^{2}}\right).\label{eq:qqgdilfin}
\end{align}
From Eq.~(\ref{eq:QtoQGCrossSec}) and (\ref{eq:qqgdilfin}), we
also get a straightforward matching between the improved TMD scheme
and the dilute scheme:
\begin{equation}
\sigma_{q\rightarrow qg}^{gA\sim0}=\left.\sigma_{q\rightarrow qg}^{WW}\right|_{\Phi_{q\rightarrow qg}^{\left(1\right)}=\Phi_{q\rightarrow qg}^{\left(2\right)}=\mathcal{G}}.\label{eq:qqgDilvsTMD}
\end{equation}
The substitution $\Phi_{q\rightarrow qg}^{\left(1\right)}=\Phi_{q\rightarrow qg}^{\left(2\right)}=\mathcal{G}$ in the iTMD scheme in the dilute limit can be simply justified as follows. For $|\boldsymbol{k}|\gg Q_s$ and large, the Fourier transforms in the operator definitions force the transverse separation between the fields to be small. In that limit the gauge links become identical, while the Wilson loops become trivial. This universal behaviour was tested numerically in \cite{vanHameren:2016ftb} and \cite{Marquet:2016cgx}.

%%%%%%%%%%%%%%%%%%%%%%%%%%%%%%%%%%%%%%%%%%%%%%%%%%%%%%%%%%%%%%%%%%%%%%%%%%%%

\subsection{$g\rightarrow q\bar{q}$ channel\label{subsec:gqqres}}
\begin{figure}[hbt]
\begin{centering}
\includegraphics[scale=0.5]{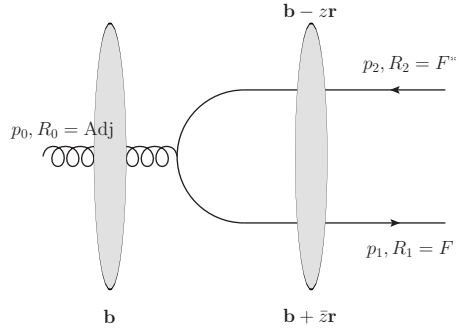}
\par\end{centering}
\caption{$g\to q\bar q$ amplitude in an external shockwave background with the appropriate color representations.}
\label{fig:GtoQQbar}  
\end{figure}

The next channel we consider is $g\to q\bar q$. In this channel, the incoming gluon splits into a quark-antiquark pair at order $g_s$, then it scatters through the target (see Fig.~\ref{fig:GtoQQbar}). The CGC amplitude for this channel is given in Eq. \eqref{eq:GtoQQbarini} and the tensor part of it reads
\begin{align}
\phi_{\mu}^{\left(g\rightarrow q\bar{q}\right)} & =-i\frac{g_{s}}{2\pi}\varepsilon_{p\perp}^{\sigma}\bar{u}_{p_{q}}\left[2zg_{\perp\mu\sigma}-\left(\gamma_{\perp\mu}\gamma_{\perp\sigma}\right)\right]\gamma^{+}v_{p_{\bar{q}}},\label{eq:GtoQQbartens}
\end{align}
whose square can be computed easily for an unpolarized observable: 
\begin{align}
\phi^{i\left(g\rightarrow q\bar{q}\right)}\phi^{i^{\prime}\ast\left(g\rightarrow q\bar{q}\right)} & =\delta^{ii^{\prime}}\left(\frac{g_{s}}{2\pi}\right)^{2}\left(p_{0}^{+}\right)^{2}16z\bar{z}\left(z^{2}+\bar{z}^{2}\right)\label{eq:GtoQQbartens2}
\end{align}
The color structure of this channel can be read off from Fig.~\ref{fig:GtoQQbar} and it is given by setting $U_{\boldsymbol{b}}^{R_{1}}=U_{\boldsymbol{b}},$ $U_{\boldsymbol{b}}^{R_{2}}=U_{\boldsymbol{b}}^{\dagger}$ and $T^{R_{0}}=T^{b}$. This color structure leads to the following gluon TMD operators that appears in the generic kinematic twist resummed cross section given in Eq.~\eqref{eq:genCrossSec}:
\begin{eqnarray}
\mathcal{O}_{1}^{\left(g\rightarrow q\bar{q}\right)} & =&\left(\partial^{j}U_{\boldsymbol{b}}\right)T^{b}U_{\boldsymbol{b}}^{\dagger}U_{\boldsymbol{b}^{\prime}}T^{b}\left(\partial^{j^{\prime}}U_{\boldsymbol{b}^{\prime}}^{\dagger}\right)\nonumber \\
\mathcal{O}_{2}^{\left(g\rightarrow q\bar{q}\right)} & =&\left(\partial^{j}U_{\boldsymbol{b}}\right)T^{R}U_{\boldsymbol{b}}^{\dagger}\left(\partial^{j^{\prime}}U_{\boldsymbol{b}^{\prime}}\right)T^{b}U_{\boldsymbol{b}^{\prime}}\label{eq:GtoQQbarope}\\
\mathcal{O}_{3}^{\left(g\rightarrow q\bar{q}\right)} & = & U_{\boldsymbol{b}}T^{b}\left(\partial^{j}U_{\boldsymbol{b}}^{\dagger}\right)U_{\boldsymbol{b}^{\prime}}T^{b}\left(\partial^{j^{\prime}}U_{\boldsymbol{b}^{\prime}}^{\dagger}\right)\nonumber \\
\mathcal{O}_{4}^{\left(g\rightarrow q\bar{q}\right)} & = & U_{\boldsymbol{b}}T^{b}\left(\partial^{j}U_{\boldsymbol{b}}^{\dagger}\right)\left(\partial^{j^{\prime}}U_{\boldsymbol{b}^{\prime}}\right)T^{b}U_{\boldsymbol{b}^{\prime}}^{\dagger}\nonumber 
\end{eqnarray}
One can easily compute the trace over the color indexes of the operators listed in Eq. \eqref{eq:GtoQQbarope} and the result reads 
\begin{align}
\mathrm{Tr}\, \Big[ \mathcal{O}_{1}^{\left(g\rightarrow q\bar{q}\right)} \Big]& =\frac{1}{2}\mathrm{Tr}\left[\left(\partial^{j}U_{\boldsymbol{b}}\right)\left(\partial^{j^{\prime}}U_{\boldsymbol{b}^{\prime}}^{\dagger}\right)\right]\mathrm{Tr}\left(U_{\boldsymbol{b}^{\prime}}U_{\boldsymbol{b}}^{\dagger}\right)+\frac{1}{2N_{c}}\mathrm{Tr}\left[\left(\partial^{j}U_{\boldsymbol{b}}\right)U_{\boldsymbol{b}}^{\dagger}\left(\partial^{j^{\prime}}U_{\boldsymbol{b}^{\prime}}\right)U_{\boldsymbol{b}^{\prime}}^{\dagger}\right]\nonumber \\
\mathrm{Tr} \, \Big[\mathcal{O}_{2}^{\left(g\rightarrow q\bar{q}\right)}\Big] & =\frac{1}{2}\mathrm{Tr}\left[\left(\partial^{j}U_{\boldsymbol{b}}\right)U_{\boldsymbol{b}^{\prime}}^{\dagger}\right]\mathrm{Tr}\left[U_{\boldsymbol{b}}^{\dagger}\left(\partial^{j^{\prime}}U_{\boldsymbol{b}^{\prime}}\right)\right]-\frac{1}{2N_{c}}\mathrm{Tr}\left[\left(\partial^{j}U_{\boldsymbol{b}}\right)U_{\boldsymbol{b}}^{\dagger}\left(\partial^{j^{\prime}}U_{\boldsymbol{b}^{\prime}}\right)U_{\boldsymbol{b}^{\prime}}^{\dagger}\right]\label{eq:GtoQQbaropeTr}\\
\mathrm{Tr} \, \Big[ \mathcal{O}_{3}^{\left(g\rightarrow q\bar{q}\right)} \Big]& =\frac{1}{2}\mathrm{Tr}\left[\left(\partial^{j}U_{\boldsymbol{b}}^{\dagger}\right)U_{\boldsymbol{b}^{\prime}}\right]\mathrm{Tr}\left[U_{\boldsymbol{b}}\left(\partial^{j^{\prime}}U_{\boldsymbol{b}^{\prime}}^{\dagger}\right)\right]-\frac{1}{2N_{c}}\mathrm{Tr}\left[\left(\partial^{j}U_{\boldsymbol{b}}\right)U_{\boldsymbol{b}}^{\dagger}\left(\partial^{j^{\prime}}U_{\boldsymbol{b}^{\prime}}\right)U_{\boldsymbol{b}^{\prime}}^{\dagger}\right]\nonumber \\
\mathrm{Tr} \, \Big[\mathcal{O}_{4}^{\left(g\rightarrow q\bar{q}\right)} \Big]& =\frac{1}{2}\mathrm{Tr}\left[\left(\partial^{j^{\prime}}U_{\boldsymbol{b}^{\prime}}\right)\left(\partial^{j}U_{\boldsymbol{b}}^{\dagger}\right)\right]\mathrm{Tr}\left(U_{\boldsymbol{b}}U_{\boldsymbol{b}^{\prime}}^{\dagger}\right)+\frac{1}{2N_{c}}\mathrm{Tr}\left[\left(\partial^{j}U_{\boldsymbol{b}}\right)U_{\boldsymbol{b}}^{\dagger}\left(\partial^{j^{\prime}}U_{\boldsymbol{b}^{\prime}}\right)U_{\boldsymbol{b}^{\prime}}^{\dagger}\right].\nonumber 
\end{align}
A comparison between the Wilson line structure in this channel given in Eq. \eqref{eq:GtoQQbaropeTr} and the definitions of the first three gluon TMDs in the gluon channel $\mathcal{F}_{gg}^{\left(1\right)}$, $\mathcal{F}_{gg}^{\left(2\right)}$ and $\mathcal{F}_{gg}^{\left(3\right)}$ given in Eqs. \eqref{F1_gg}, \eqref{F2_gg} and \eqref{F3_gg} suggests that these are the three gluon TMDs that appear in the $g\to q\bar q$ channel.  We define the following combinations of the TMDs which are the same combinations defined in Table~\ref{tab:Phi}:
\begin{align}
\Phi_{g\rightarrow q\bar{q}}^{\left(1\right)}\left(\boldsymbol{k}\right) & \equiv\frac{N_{c}^{2}\mathcal{F}_{gg}^{\left(1\right)}\left(\boldsymbol{k}\right)-\mathcal{F}_{gg}^{\left(3\right)}\left(\boldsymbol{k}\right)}{N_{c}^{2}-1}\label{eq:Phigqq}\\
\Phi_{g\rightarrow q\bar{q}}^{\left(2\right)}\left(\boldsymbol{k}\right) & \equiv-N_{c}^{2}\mathcal{F}_{gg}^{\left(2\right)}\left(\boldsymbol{k}\right)+\mathcal{F}_{gg}^{\left(3\right)}\left(\boldsymbol{k}\right)\nonumber 
\end{align}
Finally, the square of the tensor structure, Eq. \eqref{eq:GtoQQbartens2}, the Wilson line structure, Eq. \eqref{eq:GtoQQbaropeTr}, and the TMD definitions with the combinations given in Eq. \eqref{eq:Phigqq} are plugged in the generic kinematic twist resummed cross section given in Eq. \eqref{eq:genCrossSec}. The result can simply be written as  
\begin{align}
\frac{d\sigma_{g\rightarrow q\bar{q}}^{WW}}{dy_{1}dy_{2}d^{2}\boldsymbol{p}_{1}d^{2}\boldsymbol{p}_{2}}= & \frac{\alpha_{s}^{2}}{2N_{c}}p_{0}^{+}\delta\left(p_{1}^{+}+p_{2}^{+}-p_{0}^{+}\right)\frac{z\bar{z}\left(z^{2}+\bar{z}^{2}\right)}{\boldsymbol{q}^{2}}\label{eq:GtoQQbarCrosSec}\\
 & \times\left[\frac{\bar{z}^{2}}{\boldsymbol{p}_{2}^{2}}\Phi_{g\rightarrow q\bar{q}}^{\left(1\right)}\left(\boldsymbol{k}\right)+\frac{z^{2}}{\boldsymbol{p}_{1}^{2}}\Phi_{g\rightarrow q\bar{q}}^{\left(1\right)}\left(-\boldsymbol{k}\right)+z\bar{z}\frac{\left(\boldsymbol{p}_{1}\cdot\boldsymbol{p}_{2}\right)}{\boldsymbol{p}_{1}^{2}\boldsymbol{p}_{2}^{2}}\frac{\Phi_{g\rightarrow q\bar{q}}^{\left(2\right)}\left(\boldsymbol{k}\right)+\Phi_{g\rightarrow q\bar{q}}^{\left(2\right)}\left(-\boldsymbol{k}\right)}{\left(N_{c}^{2}-1\right)}\right]\nonumber 
\end{align}
which coincides exactly with Eq. \eqref{g2qbarq-ITMD} by using Eqs. \eqref{g2qbarq_1} and \eqref{g2qbarq_2}. 

The next step is to consider the dilute limit in the $g\to q\bar q$ channel. Introducing the proper color structure in the generic dilute cross section in Eq. \eqref{eq:DilCSuPDF}, we get
\begin{align}
\frac{d\sigma_{g\rightarrow q\bar{q}}^{gA\sim0}}{dy_{1}dy_{2}d^{2}\boldsymbol{p}_{1}d^{2}\boldsymbol{p}_{2}} & =\frac{\alpha_{s}}{16\left(N_{c}^{2}-1\right)\left(p_{0}^{+}\right)^{2}}\left(2\pi\right)p_{0}^{+}\delta\left(p_{1}^{+}+p_{2}^{+}-p_{0}^{+}\right)\frac{\mathcal{G}_{ac}\left(\boldsymbol{k}\right)}{\boldsymbol{k}^{2}}\left(\phi^{i}\phi^{j\ast}\right)\label{eq:gqqdilini}\\
 & \times\mathrm{Tr}\left[t^{b}t^{a}\left(\frac{\boldsymbol{p}_{1}^{i}}{\boldsymbol{p}_{1}^{2}}-\frac{\boldsymbol{q}^{i}}{\boldsymbol{q}^{2}}\right)+t^{a}t^{b}\left(\frac{\boldsymbol{p}_{2}^{i}}{\boldsymbol{p}_{2}^{2}}+\frac{\boldsymbol{q}^{i}}{\boldsymbol{q}^{2}}\right)\right]\left[t^{c}t^{b}\left(\frac{\boldsymbol{p}_{1}^{j}}{\boldsymbol{p}_{1}^{2}}-\frac{\boldsymbol{q}^{j}}{\boldsymbol{q}^{2}}\right)+t^{b}t^{c}\left(\frac{\boldsymbol{p}_{2}^{j}}{\boldsymbol{p}_{2}^{2}}+\frac{\boldsymbol{q}^{j}}{\boldsymbol{q}^{2}}\right)\right],\nonumber 
\end{align}
which after some color algebra leads to 
\begin{align}
\frac{d\sigma_{g\rightarrow q\bar{q}}^{gA\sim0}}{dy_{1}dy_{2}d^{2}\boldsymbol{p}_{1}d^{2}\boldsymbol{p}_{2}} & =\frac{\alpha_{s}^{2}}{N_{c}\left(N_{c}^{2}-1\right)}\frac{z\bar{z}\left(z^{2}+\bar{z}^{2}\right)}{2\boldsymbol{p}_{1}^{2}\boldsymbol{p}_{2}^{2}}p_{0}^{+}\delta\left(p_{1}^{+}+p_{2}^{+}-p_{0}^{+}\right)\mathcal{G}\left(\boldsymbol{k}\right)
\nonumber\\
& \times\, 
\left[N_{c}^{2}\left(z^{2}\frac{\boldsymbol{p}_{2}^{2}}{\boldsymbol{q}^{2}}+\bar{z}^{2}\frac{\boldsymbol{p}_{1}^{2}}{\boldsymbol{q}^{2}}\right)-1\right].\label{eq:gqqdilfin}
\end{align}
Finally, a comparison between the kinematic twist resummed cross section Eq. \eqref{eq:GtoQQbarCrosSec} and the dilute limit of the cross section given in Eq. \eqref{eq:gqqdilfin},  again leads to a straightforward matching between the iTMD scheme and the dilute scheme for $g\to q\bar q$ channel: 
\begin{equation}
\sigma_{g\rightarrow q\bar{q}}^{gA\sim0}=\left.\sigma_{g\rightarrow q\bar{q}}^{WW}\right|_{\Phi_{g\rightarrow q\bar{q}}^{\left(1\right)}=\Phi_{g\rightarrow q\bar{q}}^{\left(2\right)}=\mathcal{G}}.\label{eq:gqqDilvsTMD}
\end{equation}
%
%%%%%%%%%%%%%%%%%%%%%%%%%%%%%%%%%%%%%%%%%%%%%%%%%%%%%%%%%%%%%%%%%%%%%%%%%%%%

\subsection{$g\rightarrow gg$ channel\label{subsec:gggres}}

The next channel we consider is $g\to gg$. The CGC amplitude for this channel is given in Eq. \eqref{eq:GtoGGini}. The tensor part for this channel can simply be read off from Eq. \eqref{eq:GtoGGini} and it is given as 
\begin{align}
\phi_{\mu}^{\left(g\rightarrow gg\right)} & =\frac{2g_{s}p_{0}^{+}}{\pi}\varepsilon_{p\perp}^{\sigma_{0}}\varepsilon_{p_{g}\perp}^{\sigma_{1}\ast}\varepsilon_{q_{g}\perp}^{\sigma_{2}\ast}\left[zg_{\perp\sigma_{0}\sigma_{1}}g_{\perp\mu\sigma_{2}}-z\bar{z}g_{\perp\sigma_{1}\sigma_{2}}g_{\perp\mu\sigma_{0}}+\bar{z}g_{\perp\sigma_{0}\sigma_{2}}g_{\perp\mu\sigma_{1}}\right]\label{eq:GtoGGtens}
\end{align}
Its square can be computed in a straightforward manner with the result being 
\begin{align}
\phi^{i\left(g\rightarrow gg\right)}\phi^{i^{\prime}\ast\left(g\rightarrow gg\right)} & =\delta^{ii^{\prime}}\left(\frac{g_{s}}{2\pi}\right)^{2}\left(p_{0}^{+}\right)^{2}32\left(1-z\bar{z}\right)^{2}\label{eq:GtoGGtens2}
\end{align}
\begin{figure}[hbt]
\begin{centering}
\includegraphics[scale=0.5]{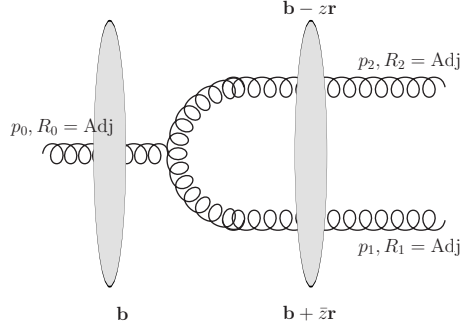}
\par\end{centering}
\caption{$g\to gg$ amplitude in an external shockwave background with the appropriate color representations.}
\label{fig:GtoGG}  
\end{figure}
The color structure of this channel is demonstrated in Fig. \ref{fig:GtoGG} and it is given by $U_{\boldsymbol{b}}^{R_{1}}=U_{\boldsymbol{b}}^{b_{1}a_{1}},$ $U_{\boldsymbol{b}}^{R_{2}}=U_{\boldsymbol{b}}^{b_{2}a_{2}}$ and $T^{R_{0}}=f^{a_{0}b_{1}b_{2}}$. This leads to the following TMD operators once it is inserted to the Wilson line structure of the generic kinematic twist resummed cross section in Eq. \eqref{eq:genCrossSec}:
\begin{align}
\mathcal{O}_{1}^{\left(g\rightarrow gg\right)} & =\left(\partial^{j}U_{\boldsymbol{b}}^{b_{1}a_{1}}\right)f^{a_{0}b_{1}b_{2}}U_{\boldsymbol{b}}^{b_{2}a_{2}}U_{\boldsymbol{b}^{\prime}}^{a_{2}c_{2}}f^{a_{0}c_{1}c_{2}}\left(\partial^{j^{\prime}}U_{\boldsymbol{b}^{\prime}}^{a_{1}c_{1}}\right)\nonumber \\
\mathcal{O}_{2}^{\left(g\rightarrow gg\right)} & =\left(\partial^{j}U_{\boldsymbol{b}}^{b_{1}a_{1}}\right)f^{a_{0}b_{1}b_{2}}U_{\boldsymbol{b}}^{b_{2}a_{2}}\left(\partial^{j^{\prime}}U_{\boldsymbol{b}^{\prime}}^{a_{2}c_{2}}\right)f^{a_{0}c_{1}c_{2}}U_{\boldsymbol{b}^{\prime}}^{a_{1}c_{1}}\label{eq:GtoGGope}\\
\mathcal{O}_{3}^{\left(g\rightarrow gg\right)} & =U_{\boldsymbol{b}}^{b_{1}a_{1}}f^{a_{0}b_{1}b_{2}}\left(\partial^{j}U_{\boldsymbol{b}}^{b_{2}a_{2}}\right)U_{\boldsymbol{b}^{\prime}}^{a_{2}c_{2}}f^{a_{0}c_{1}c_{2}}\left(\partial^{j^{\prime}}U_{\boldsymbol{b}^{\prime}}^{a_{1}c_{1}}\right)\nonumber \\
\mathcal{O}_{4}^{\left(g\rightarrow gg\right)} & =U_{\boldsymbol{b}}^{b_{1}a_{1}}f^{a_{0}b_{1}b_{2}}\left(\partial^{j}U_{\boldsymbol{b}}^{b_{2}a_{2}}\right)\left(\partial^{j^{\prime}}U_{\boldsymbol{b}^{\prime}}^{a_{2}c_{2}}\right)f^{a_{0}c_{1}c_{2}}U_{\boldsymbol{b}^{\prime}}^{a_{1}c_{1}}.\nonumber 
\end{align}
After a standard but cumbersome color algebra, the trace over the color indexes of the above TMD operators can be written in terms of the fundamental Wilson line operators as 
\begin{align}
\label{eq:O1gggTrace}
\mathrm{Tr}\, \Big[\mathcal{O}_{1}^{\left(g\rightarrow gg\right)} \Big]& =-\mathrm{Tr}\left[\left(\partial^{j}U_{\boldsymbol{b}}\right)U_{\boldsymbol{b}}^{\dagger}\left(\partial^{j^{\prime}}U_{\boldsymbol{b}^{\prime}}\right)U_{\boldsymbol{b}^{\prime}}^{\dagger}\right]\mathrm{Tr}\left(U_{\boldsymbol{b}}U_{\boldsymbol{b}^{\prime}}^{\dagger}\right)\mathrm{Tr}\left(U_{\boldsymbol{b}^{\prime}}U_{\boldsymbol{b}}^{\dagger}\right)
\nonumber \\
 &
  -\mathrm{Tr}\left[\left(\partial^{j}U_{\boldsymbol{b}}\right)U_{\boldsymbol{b}}^{\dagger}U_{\boldsymbol{b}^{\prime}}U_{\boldsymbol{b}}^{\dagger}\left(\partial^{j^{\prime}}U_{\boldsymbol{b}^{\prime}}\right)U_{\boldsymbol{b}^{\prime}}^{\dagger}U_{\boldsymbol{b}}U_{\boldsymbol{b}^{\prime}}^{\dagger}\right]\nonumber \\
 & +2\mathrm{Tr}\left[\left(\partial^{j}U_{\boldsymbol{b}}\right)U_{\boldsymbol{b}}^{\dagger}\left(\partial^{j^{\prime}}U_{\boldsymbol{b}^{\prime}}\right)U_{\boldsymbol{b}^{\prime}}^{\dagger}\right]
  -\mathrm{Tr}\left[\left(\partial^{j}U_{\boldsymbol{b}}^{\dagger}\right)U_{\boldsymbol{b}}\left(\partial^{j^{\prime}}U_{\boldsymbol{b}^{\prime}}^{\dagger}\right)U_{\boldsymbol{b}^{\prime}}\right] \\
 & +\frac{N_{c}}{2}
 \bigg\{
 \mathrm{Tr}\left[\left(\partial^{j}U_{\boldsymbol{b}}^{\dagger}\right)\left(\partial^{j^{\prime}}U_{\boldsymbol{b}^{\prime}}\right)\right]\mathrm{Tr}\left(U_{\boldsymbol{b}}U_{\boldsymbol{b}^{\prime}}^{\dagger}\right)
+ \mathrm{Tr}\left[\left(\partial^{j}U_{\boldsymbol{b}}\right)\left(\partial^{j^{\prime}}U_{\boldsymbol{b}^{\prime}}^{\dagger}\right)\right]\mathrm{Tr}\left(U_{\boldsymbol{b}^{\prime}}U_{\boldsymbol{b}}^{\dagger}\right)\bigg\},\nonumber 
\end{align}
\begin{align}
\label{eq:O2gggTrace}
\mathrm{Tr}\, \Big[ \mathcal{O}_{2}^{\left(g\rightarrow gg\right)} \Big]& =\mathrm{Tr}\left[\left(\partial^{j}U_{\boldsymbol{b}}\right)U_{\boldsymbol{b}}^{\dagger}\left(\partial^{j^{\prime}}U_{\boldsymbol{b}^{\prime}}\right)U_{\boldsymbol{b}^{\prime}}^{\dagger}\right]\mathrm{Tr}\left(U_{\boldsymbol{b}}U_{\boldsymbol{b}^{\prime}}^{\dagger}\right)\mathrm{Tr}\left(U_{\boldsymbol{b}^{\prime}}U_{\boldsymbol{b}}^{\dagger}\right)\nonumber \\
 & +\mathrm{Tr}\left[\left(\partial^{j}U_{\boldsymbol{b}}\right)U_{\boldsymbol{b}}^{\dagger}U_{\boldsymbol{b}^{\prime}}U_{\boldsymbol{b}}^{\dagger}\left(\partial^{j^{\prime}}U_{\boldsymbol{b}^{\prime}}\right)U_{\boldsymbol{b}^{\prime}}^{\dagger}U_{\boldsymbol{b}}U_{\boldsymbol{b}^{\prime}}^{\dagger}\right]\nonumber \\
 & +\mathrm{Tr}\left[\left(\partial^{j}U_{\boldsymbol{b}}^{\dagger}\right)U_{\boldsymbol{b}}\left(\partial^{j^{\prime}}U_{\boldsymbol{b}^{\prime}}^{\dagger}\right)U_{\boldsymbol{b}^{\prime}}\right]
 -2\mathrm{Tr}\left[\left(\partial^{j}U_{\boldsymbol{b}}\right)U_{\boldsymbol{b}}^{\dagger}\left(\partial^{j^{\prime}}U_{\boldsymbol{b}^{\prime}}\right)U_{\boldsymbol{b}^{\prime}}^{\dagger}\right] \\
 & +\frac{N_{c}}{2}
 \bigg\{
 \mathrm{Tr}\left[U_{\boldsymbol{b}^{\prime}}^{\dagger}\left(\partial^{j}U_{\boldsymbol{b}}\right)\right]\mathrm{Tr}\left[\left(\partial^{j^{\prime}}U_{\boldsymbol{b}^{\prime}}\right)U_{\boldsymbol{b}}^{\dagger}\right]
  +\mathrm{Tr}\left[U_{\boldsymbol{b}}\left(\partial^{j^{\prime}}U_{\boldsymbol{b}^{\prime}}^{\dagger}\right)\right]\mathrm{Tr}\left[\left(\partial^{j}U_{\boldsymbol{b}}^{\dagger}\right)U_{\boldsymbol{b}^{\prime}}\right]\bigg\},\nonumber 
\end{align}
\begin{align}
\label{eq:O3gggTrace}
\mathrm{Tr}\, \Big[ \mathcal{O}_{3}^{\left(g\rightarrow gg\right)} \Big]& =\mathrm{Tr}\left[\left(\partial^{j}U_{\boldsymbol{b}}\right)U_{\boldsymbol{b}}^{\dagger}\left(\partial^{j^{\prime}}U_{\boldsymbol{b}^{\prime}}\right)U_{\boldsymbol{b}^{\prime}}^{\dagger}\right]\mathrm{Tr}\left(U_{\boldsymbol{b}}U_{\boldsymbol{b}^{\prime}}^{\dagger}\right)\mathrm{Tr}\left(U_{\boldsymbol{b}^{\prime}}U_{\boldsymbol{b}}^{\dagger}\right)\nonumber \\
 & +\mathrm{Tr}\left[\left(\partial^{j}U_{\boldsymbol{b}}\right)U_{\boldsymbol{b}}^{\dagger}U_{\boldsymbol{b}^{\prime}}U_{\boldsymbol{b}}^{\dagger}\left(\partial^{j^{\prime}}U_{\boldsymbol{b}^{\prime}}\right)U_{\boldsymbol{b}^{\prime}}^{\dagger}U_{\boldsymbol{b}}U_{\boldsymbol{b}^{\prime}}^{\dagger}\right]\nonumber \\
 & +\mathrm{Tr}\left[\left(\partial^{j}U_{\boldsymbol{b}}^{\dagger}\right)U_{\boldsymbol{b}}\left(\partial^{j^{\prime}}U_{\boldsymbol{b}^{\prime}}^{\dagger}\right)U_{\boldsymbol{b}^{\prime}}\right]
 -2\mathrm{Tr}\left[\left(\partial^{j}U_{\boldsymbol{b}}\right)U_{\boldsymbol{b}}^{\dagger}\left(\partial^{j^{\prime}}U_{\boldsymbol{b}^{\prime}}\right)U_{\boldsymbol{b}^{\prime}}^{\dagger}\right] \\
 & +\frac{N_{c}}{2}
 \bigg\{
 \mathrm{Tr}\left[U_{\boldsymbol{b}}\left(\partial^{j^{\prime}}U_{\boldsymbol{b}^{\prime}}^{\dagger}\right)\right]\mathrm{Tr}\left[\left(\partial^{j}U_{\boldsymbol{b}}^{\dagger}\right)U_{\boldsymbol{b}^{\prime}}\right]
 +\mathrm{Tr}\left[\left(\partial^{j}U_{\boldsymbol{b}}\right)U_{\boldsymbol{b}^{\prime}}^{\dagger}\right]\mathrm{Tr}\left[\left(\partial^{j^{\prime}}U_{\boldsymbol{b}^{\prime}}\right)U_{\boldsymbol{b}}^{\dagger}\right] \bigg\},\nonumber 
\end{align}
and
\begin{align}
\label{eq:O4gggTrace}
\mathrm{Tr}\, \Big[ \mathcal{O}_{4}^{\left(g\rightarrow gg\right)} \Big] & =-\mathrm{Tr}\left[\left(\partial^{j}U_{\boldsymbol{b}}\right)U_{\boldsymbol{b}}^{\dagger}\left(\partial^{j^{\prime}}U_{\boldsymbol{b}^{\prime}}\right)U_{\boldsymbol{b}^{\prime}}^{\dagger}\right]\mathrm{Tr}\left(U_{\boldsymbol{b}}U_{\boldsymbol{b}^{\prime}}^{\dagger}\right)\mathrm{Tr}\left(U_{\boldsymbol{b}^{\prime}}U_{\boldsymbol{b}}^{\dagger}\right)\nonumber \\
 & -\mathrm{Tr}\left[\left(\partial^{j}U_{\boldsymbol{b}}\right)U_{\boldsymbol{b}}^{\dagger}U_{\boldsymbol{b}^{\prime}}U_{\boldsymbol{b}}^{\dagger}\left(\partial^{j^{\prime}}U_{\boldsymbol{b}^{\prime}}\right)U_{\boldsymbol{b}^{\prime}}^{\dagger}U_{\boldsymbol{b}}U_{\boldsymbol{b}^{\prime}}^{\dagger}\right]\nonumber \\
 & +2\mathrm{Tr}\left[\left(\partial^{j}U_{\boldsymbol{b}}\right)U_{\boldsymbol{b}}^{\dagger}\left(\partial^{j^{\prime}}U_{\boldsymbol{b}^{\prime}}\right)U_{\boldsymbol{b}^{\prime}}^{\dagger}\right]
 -\mathrm{Tr}\left[\left(\partial^{j}U_{\boldsymbol{b}}^{\dagger}\right)U_{\boldsymbol{b}}\left(\partial^{j^{\prime}}U_{\boldsymbol{b}^{\prime}}^{\dagger}\right)U_{\boldsymbol{b}^{\prime}}\right] \\
 & +\frac{N_{c}}{2}
 \bigg\{
 \mathrm{Tr}\left[\left(\partial^{j^{\prime}}U_{\boldsymbol{b}^{\prime}}\right)\left(\partial^{j}U_{\boldsymbol{b}}^{\dagger}\right)\right]\mathrm{Tr}\left(U_{\boldsymbol{b}}U_{\boldsymbol{b}^{\prime}}^{\dagger}\right)
 +\mathrm{Tr}\left[\left(\partial^{j}U_{\boldsymbol{b}}\right)\left(\partial^{j^{\prime}}U_{\boldsymbol{b}^{\prime}}^{\dagger}\right)\right]\mathrm{Tr}\left(U_{\boldsymbol{b}}^{\dagger}U_{\boldsymbol{b}^{\prime}}\right)\bigg\} \nonumber 
\end{align}
Comparing the Wilson line structures appearing in Eqs. \eqref{eq:O1gggTrace}, \eqref{eq:O2gggTrace}, \eqref{eq:O3gggTrace} and \eqref{eq:O4gggTrace} with the TMD definitions given in Eqs. \eqref{F4_gg}, \eqref{F5_gg}  and \eqref{F6_gg}, one can conclude that on top of the gluon TMDs $\mathcal{F}_{gg}^{\left(1\right)}$, $\mathcal{F}_{gg}^{\left(2\right)}$ and $\mathcal{F}_{gg}^{\left(3\right)}$ that have already appeared in the $g\to q\bar q$ channel, one also gets new gluon TMDs $\mathcal{F}_{gg}^{\left(4\right)}$, $\mathcal{F}_{gg}^{\left(5\right)}$ and $\mathcal{F}_{gg}^{\left(6\right)}$ in the $g\to gg$ channel.   
Again, for convenience, we define the following combinations of the TMDs
\begin{align}
\Phi_{gg\rightarrow gg}^{\left(1\right)}\left(\boldsymbol{k}\right) & =\frac{1}{2N_{c}^{2}}\left[N_{c}^{2}\mathcal{F}_{gg}^{\left(6\right)}\left(\boldsymbol{k}\right)+\mathcal{F}_{gg}^{\left(5\right)}\left(\boldsymbol{k}\right)+\mathcal{F}_{gg}^{\left(4\right)}\left(\boldsymbol{k}\right)-2\mathcal{F}_{gg}^{\left(3\right)}\left(\boldsymbol{k}\right)+N_{c}^{2}\left(\frac{\mathcal{F}_{gg}^{\left(1\right)}\left(\boldsymbol{k}\right)+\mathcal{F}_{gg}^{\left(1\right)}\left(-\boldsymbol{k}\right)}{2}\right)\right]\nonumber \\
\label{eq:Phigg}\\
\Phi_{gg\rightarrow gg}^{\left(2\right)}\left(\boldsymbol{k}\right) & =\frac{1}{N_{c}^{2}}\left[N_{c}^{2}\mathcal{F}_{gg}^{\left(6\right)}\left(\boldsymbol{k}\right)+\mathcal{F}_{gg}^{\left(5\right)}\left(\boldsymbol{k}\right)+\mathcal{F}_{gg}^{\left(4\right)}\left(\boldsymbol{k}\right)-2\mathcal{F}_{gg}^{\left(3\right)}\left(\boldsymbol{k}\right)+N_{c}^{2}\left(\frac{\mathcal{F}_{gg}^{\left(2\right)}\left(\boldsymbol{k}\right)+\mathcal{F}_{gg}^{\left(2\right)}\left(-\boldsymbol{k}\right)}{2}\right)\right],\nonumber 
\end{align}
which match exactly the combinations one get from iTMD calculations given in Table~\ref{tab:Phi}. 
After plugging these results into the generic kinematic  twist resummed cross section given in Eq. \eqref{eq:genCrossSec}, we get the result for the $g\to gg$ channel as
\begin{align}
\frac{d\sigma_{g\rightarrow gg}^{WW}}{dy_{1}dy_{2}d^{2}\boldsymbol{p}_{1}d^{2}\boldsymbol{p}_{2}} & =2\alpha_{s}^{2}\frac{N_{c}^{2}}{N_{c}^{2}-1}p_{0}^{+}\delta\left(p_{1}^{+}+p_{2}^{+}-p_{0}^{+}\right)\frac{\left(1-z\bar{z}\right)^{2}}{\boldsymbol{p}_{1}^{2}\boldsymbol{p}_{2}^{2}}\label{eq:GtoGGCrosSec}\\
 & \times\left[\left(1+2z\bar{z}\frac{\left(\boldsymbol{p}_{1}\cdot\boldsymbol{p}_{2}\right)}{\boldsymbol{q}^{2}}\right)\Phi_{gg\rightarrow gg}^{\left(1\right)}\left(\boldsymbol{k}\right)-z\bar{z}\frac{\left(\boldsymbol{p}_{1}\cdot\boldsymbol{p}_{2}\right)}{\boldsymbol{q}^{2}}\Phi_{gg\rightarrow gg}^{\left(2\right)}\left(\boldsymbol{k}\right)\right], \nonumber 
\end{align}
where a factor $1/2$ is added due to the symmetry. This result coincides exactly with Eq.~\eqref{g2gg-ITMD} by using Eqs. \eqref{g2gg_1} and \eqref{g2gg_2}. 

Let us now consider the dilute limit of the cross section in the $g\to gg$ channel. Once the proper color representations of this channel are plugged into the dilute limit of the generic cross section given in Eq. \eqref{eq:DilCSuPDF}, we get
\begin{align}
\frac{d\sigma_{g\rightarrow gg}^{gA\sim0}}{dy_{1}dy_{2}d^{2}\boldsymbol{p}_{1}d^{2}\boldsymbol{p}_{2}} & =\frac{\alpha_{s}}{16\left(N_{c}^{2}-1\right)\left(p_{0}^{+}\right)^{2}}\left(2\pi\right)p_{0}^{+}\delta\left(p_{1}^{+}+p_{2}^{+}-p_{0}^{+}\right)\frac{\mathcal{G}_{ac}\left(\boldsymbol{k}\right)}{\boldsymbol{k}^{2}}\left(\phi^{i}\phi^{j\ast}\right)\label{eq:gggdilini}\\
 & \times\mathrm{Tr}\left\{ \left[f^{a_{0}a_{1}b}f^{ba_{2}a}\left(\frac{\boldsymbol{p}_{1}^{i}}{\boldsymbol{p}_{1}^{2}}-\frac{\boldsymbol{q}^{i}}{\boldsymbol{q}^{2}}\right)+f^{a_{0}a_{2}b}f^{ba_{1}a}\left(\frac{\boldsymbol{p}_{2}^{i}}{\boldsymbol{p}_{2}^{2}}+\frac{\boldsymbol{q}^{i}}{\boldsymbol{q}^{2}}\right)\right]\right.\nonumber \\
 & \left.\times\left[f^{a_{0}a_{1}d}f^{da_{2}c}\left(\frac{\boldsymbol{p}_{1}^{j}}{\boldsymbol{p}_{1}^{2}}-\frac{\boldsymbol{q}^{j}}{\boldsymbol{q}^{2}}\right)+f^{a_{0}a_{2}d}f^{da_{1}c}\left(\frac{\boldsymbol{p}_{2}^{j}}{\boldsymbol{p}_{2}^{2}}+\frac{\boldsymbol{q}^{j}}{\boldsymbol{q}^{2}}\right)\right]\right\} \nonumber 
\end{align}
which after some color algebra leads to
\begin{align}
\frac{d\sigma_{g\rightarrow gg}^{gA\sim0}}{dy_{1}dy_{2}d^{2}\boldsymbol{p}_{1}d^{2}\boldsymbol{p}_{2}} & =2\alpha_{s}^{2}\frac{N_{c}^{2}}{N_{c}^{2}-1}p_{0}^{+}\delta\left(p_{1}^{+}+p_{2}^{+}-p_{0}^{+}\right)\frac{\left(1-z\bar{z}\right)^{2}}{\boldsymbol{p}_{1}^{2}\boldsymbol{p}_{2}^{2}}\mathcal{G}\left(\boldsymbol{k}\right)\left(1+\frac{z\bar{z}\left(\boldsymbol{p}_{1}\cdot\boldsymbol{p}_{2}\right)}{\boldsymbol{q}^{2}}\right), \label{eq:gggdilfin}
\end{align}
with the symmetry factor of $1/2$.
As a last comment for this channel, we would like to emphasize that a comparison between Eqs. (\ref{eq:GtoGGCrosSec}), (\ref{eq:gggdilfin}) and the iTMD results lead to the same matching condition between the iTMD scheme and the dilute limit:
\begin{equation}
\sigma_{g\rightarrow gg}^{gA\sim0}=\left.\sigma_{g\rightarrow q\bar{q}}^{WW}\right|_{\Phi_{g\rightarrow gg}^{\left(1\right)}=\Phi_{g\rightarrow gg}^{\left(2\right)}=\mathcal{G}}.\label{eq:gggDilvsTMD}
\end{equation}

%%%%%%%%%%%%%%%%%%%%%%%%%%%%%%%%%%%%%%%%%%%%%%%%%%%%%%%%%%%%%%%%%%%%%%%%%%%%%

\subsection{$\gamma\to q\bar q$ channel \label{sec:photoqqres}}

We have used this channel as an example to study the corrections to the back-to-back correlation limit in subsection~\ref{subsec:GenExp}. In this subsection, we generalize that study by using the generic expressions for the kinemtic twist resummed cross section and the dilute limit of the generic CGC cross section. The amplitude is given by Eq. \eqref{eq:GammatoQQbarini} from which we can read off the tensor part: 
\begin{align}
\phi_{\mu}^{\left(\gamma\rightarrow q\bar{q}\right)} & =i\frac{e_{q}}{2\pi}\varepsilon_{p\perp}^{\sigma}\bar{u}_{p_{q}}\left[2zg_{\perp\mu\sigma}-\left(\gamma_{\perp\mu}\gamma_{\perp\sigma}\right)\right]\gamma^{+}v_{p_{\bar{q}}}\label{eq:GammatoQQbartens}
\end{align}
The square of the tensor part for an unpolarized observable can be calculated easily and the result reads
\begin{align}
\phi^{i\left(\gamma\rightarrow q\bar{q}\right)}\phi^{i^{\prime}\ast\left(\gamma\rightarrow q\bar{q}\right)} & =\delta^{ii^{\prime}}\left(\frac{e_{q}}{2\pi}\right)^{2}\left(p_{0}^{+}\right)^{2}16z\bar{z}\left(z^{2}+\bar{z}^{2}\right)\label{eq:GammatoQQbartens2}
\end{align}
\begin{figure}[hbt]
\begin{centering}
\includegraphics[scale=0.5]{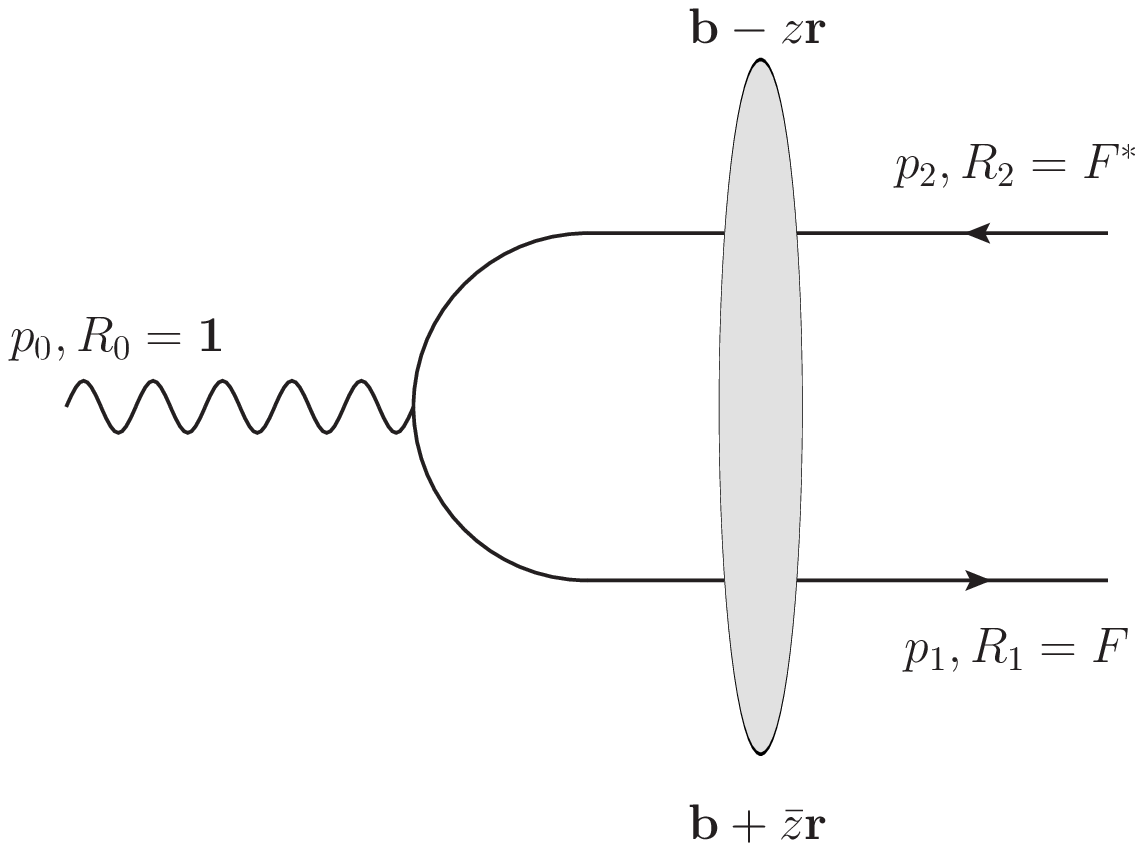}
\par\end{centering}
\caption{$\gamma\to q\bar q$ amplitude in an external shockwave background with the appropriate color representations.}
\label{fig:GammatoQQbar}  
\end{figure}
The color structure for this channel is demonstrated in Fig.~\ref{fig:GammatoQQbar} and it is given by setting  $U_{\boldsymbol{b}}^{R_{1}}=U_{\boldsymbol{b}},$ $U_{\boldsymbol{b}}^{R_{2}}=U_{\boldsymbol{b}}^{\dagger}$ and $T^{R_{0}}=1$. This color structure leads to the following TMD operators: 
\begin{align}
\mathcal{O}_{1}^{\left(\gamma\rightarrow q\bar{q}\right)} & =-\left(\partial^{j}U_{\boldsymbol{b}}\right)U_{\boldsymbol{b}}^{\dagger}\left(\partial^{j^{\prime}}U_{\boldsymbol{b}^{\prime}}\right)U_{\boldsymbol{b}^{\prime}}^{\dagger}\nonumber \\
\mathcal{O}_{2}^{\left(\gamma\rightarrow q\bar{q}\right)} & =\left(\partial^{j}U_{\boldsymbol{b}}\right)U_{\boldsymbol{b}}^{\dagger}\left(\partial^{j^{\prime}}U_{\boldsymbol{b}^{\prime}}\right)U_{\boldsymbol{b}^{\prime}}^{\dagger}\label{eq:GammatoQQbarope}\\
\mathcal{O}_{3}^{\left(\gamma\rightarrow q\bar{q}\right)} & =\left(\partial^{j}U_{\boldsymbol{b}}\right)U_{\boldsymbol{b}}^{\dagger}\left(\partial^{j^{\prime}}U_{\boldsymbol{b}^{\prime}}\right)U_{\boldsymbol{b}^{\prime}}^{\dagger}\nonumber \\
\mathcal{O}_{4}^{\left(\gamma\rightarrow q\bar{q}\right)} & =-\left(\partial^{j}U_{\boldsymbol{b}}\right)U_{\boldsymbol{b}}^{\dagger}\left(\partial^{j^{\prime}}U_{\boldsymbol{b}^{\prime}}\right)U_{\boldsymbol{b}^{\prime}}^{\dagger}.\nonumber 
\end{align}
The trace over the color indices can be performed in a straightforward manner and one can easily conclude that this channel involves only one TMD $\mathcal{F}_{gg}^{\left(3\right)}$ which is also referred  to as the Weizs\"{a}cker-Williams TMD defined in Eq. \eqref{F3_gg}. Using this result and the square of the tensor part given in Eq. \eqref{eq:GammatoQQbartens2}, we can write the kinematic twist resummed cross section for this channel as
\begin{align}
\frac{d\sigma_{\gamma\rightarrow q\bar{q}}^{WW}}{dy_{1}dy_{2}d^{2}\boldsymbol{p}_{1}d^{2}\boldsymbol{p}_{2}} & =\alpha_{\mathrm{em}}\alpha_{s}p_{0}^{+}\delta\left(p_{1}^{+}+p_{2}^{+}-p_{0}^{+}\right)\frac{z\bar{z}\left(z^{2}+\bar{z}^{2}\right)}{\boldsymbol{p}_{1}^{2}\boldsymbol{p}_{2}^{2}}\mathcal{F}_{gg}^{\left(3\right)}\left(\boldsymbol{k}\right), \label{eq:GammatoQQbarCrosSec}
\end{align}
which coincides exactly with Eq. \eqref{gamma2qbarq-ITMD} by using Eq. \eqref{gamma2qbarq_1}.

Using the proper color representations for this channel and the dilute limit of the generic CGC cross section given in Eq. \eqref{eq:DilCSuPDF}, we can simply write the dilute limit of the cross section for the $\gamma\to q\bar q$ channel as 
\begin{align}
\frac{d\sigma_{\gamma\rightarrow q\bar{q}}^{gA\sim0}}{dy_{1}dy_{2}d^{2}\boldsymbol{p}_{1}d^{2}\boldsymbol{p}_{2}} & =\frac{\alpha_{s}}{16\left(p_{0}^{+}\right)^{2}}\left(2\pi\right)p_{0}^{+}\delta\left(p_{1}^{+}+p_{2}^{+}-p_{0}^{+}\right)\frac{\mathcal{G}\left(\boldsymbol{k}\right)}{\boldsymbol{k}^{2}}\left(\phi^{i}\phi^{j\ast}\right)\label{eq:GammaToQQdilini}\\
 & \times\mathrm{Tr}\left[t^{a}\left(\frac{\boldsymbol{p}_{1}^{i}}{\boldsymbol{p}_{1}^{2}}-\frac{\boldsymbol{q}^{i}}{\boldsymbol{q}^{2}}\right)+t^{a}\left(\frac{\boldsymbol{p}_{2}^{i}}{\boldsymbol{p}_{2}^{2}}+\frac{\boldsymbol{q}^{i}}{\boldsymbol{q}^{2}}\right)\right]\left[t^{c}\left(\frac{\boldsymbol{p}_{1}^{j}}{\boldsymbol{p}_{1}^{2}}-\frac{\boldsymbol{q}^{j}}{\boldsymbol{q}^{2}}\right)+t^{c}\left(\frac{\boldsymbol{p}_{2}^{j}}{\boldsymbol{p}_{2}^{2}}+\frac{\boldsymbol{q}^{j}}{\boldsymbol{q}^{2}}\right)\right],\nonumber 
\end{align}
which, after a simple color algebra and using the result for the square of the tensor part given in Eq. \eqref{eq:GammatoQQbartens2}, leads to
\begin{align}
\frac{d\sigma_{\gamma\rightarrow q\bar{q}}^{gA\sim0}}{dy_{1}dy_{2}d^{2}\boldsymbol{p}_{1}d^{2}\boldsymbol{p}_{2}} & =\alpha_{\mathrm{em}}\alpha_{s}p_{0}^{+}\delta\left(p_{1}^{+}+p_{2}^{+}-p_{0}^{+}\right)\frac{z\bar{z}\left(z^{2}+\bar{z}^{2}\right)}{\boldsymbol{p}_{1}^{2}\boldsymbol{p}_{2}^{2}}\mathcal{G}\left(\boldsymbol{k}\right).\label{eq:GammaToQQdilfin}
\end{align}
Finally, we would like to mention that a comparison between   Eq.~(\ref{eq:GtoGGCrosSec}) and (\ref{eq:gggdilfin}) suggests a similar matching between the iTMD scheme and the dilute limit of the CGC calculation:
\begin{equation}
\sigma_{\gamma\rightarrow q\bar{q}}^{gA\sim0}=\left.\sigma_{\gamma\rightarrow q\bar{q}}^{WW}\right|_{\mathcal{F}_{gg}^{\left(3\right)}=\mathcal{G}}.\label{eq:gammaqqDilvsTMD}
\end{equation}
%

%%%%%%%%%%%%%%%%%%%%%%%%%%%%%%%%%%%%%%%%%%%%%%%%%%%%%%%%%%%%%%%%%%%%%%%%%%%%%%%%%%%%%%%%

\subsection{$q\rightarrow q\gamma$ channel \label{sec:DYres}}

The last channel we consider is the $q\to q\gamma$ one. The CGC amplitude for this channel is given by Eq. \eqref{eq:QtoQGammaini} from which we can read off the tensor part as 
\begin{align}
\phi_{\mu}^{\left(q\rightarrow q\gamma\right)} & =\frac{-ie_{q}}{2\pi}\varepsilon_{p_{\gamma}\perp}^{\sigma\ast}\bar{u}_{p_{q}}\left[2zg_{\perp\mu\sigma}+\bar{z}\left(\gamma_{\perp\mu}\gamma_{\perp\sigma}\right)\right]\gamma^{+}u_{p}\label{eq:QtoQGammatens}
\end{align}
Its square for an unpolarized observable can be written as 
\begin{align}
\phi^{i\left(q\rightarrow q\gamma\right)}\phi^{i^{\prime}\ast\left(q\rightarrow q\gamma\right)} & =\delta^{ii^{\prime}}\left(\frac{e_{q}}{2\pi}\right)^{2}\left(p_{0}^{+}\right)^{2}16z\left(1+z^{2}\right).\label{eq:QtoQGammatens2}
\end{align}
\begin{figure}[hbt]
\begin{centering}
\includegraphics[scale=0.5]{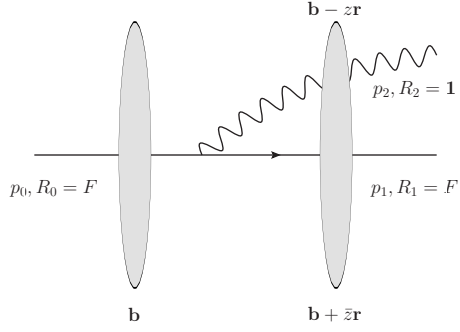}
\par\end{centering}
\caption{$q\to q\gamma$ amplitude in an external shockwave background with the appropriate color representations.}
\label{fig:QtoQGamma}  
\end{figure}
As it can be seen from Fig.~\ref{fig:QtoQGamma}, the color structure of this channel is quite simple. One gets the proper color structure by setting $U_{\boldsymbol{b}}^{R_{1}}=U_{\boldsymbol{b}},$ $U_{\boldsymbol{b}}^{R_{2}}=\boldsymbol{1}$ and $T^{R_{0}}=1$. With this simple color structure, only one TMD operator appears in this channel which reads
\begin{equation}
\mathcal{O}^{\left(q\rightarrow q\gamma\right)}=\left(\partial^{j}U_{\boldsymbol{b}}\right)\left(\partial^{j^{\prime}}U_{\boldsymbol{b}^{\prime}}^{\dagger}\right).\label{eq:QtoQGammaope}
\end{equation}
Performing the trace over color indices leads to $\mathcal{F}_{qg}^{\left(1\right)}$ TMD which has been introduced in Eq. \eqref{F1_qg}. Plugging these results into Eq. \eqref{eq:genCrossSec}, we get the cross section for $q\to q\gamma$ channel:
\begin{align}
\frac{d\sigma_{q\rightarrow q\gamma}^{WW}}{dy_{1}dy_{2}d^{2}\boldsymbol{p}_{1}d^{2}\boldsymbol{p}_{2}} & =\frac{\alpha_{\mathrm{em}}\alpha_{s}}{N_{c}}p_{0}^{+}\delta\left(p_{1}^{+}+p_{2}^{+}-p_{0}^{+}\right)\frac{z\bar{z}^{2}\left(1+z^{2}\right)}{\boldsymbol{p}_{2}^{2}\boldsymbol{q}^{2}}\mathcal{F}_{qg}^{\left(1\right)}\left(\boldsymbol{k}\right),\label{eq:QtoQGammaCrosSec}
\end{align}
which coincides exactly with Eq. \eqref{q2qgamma-ITMD} by using Eq. \eqref{q2qgamma_1}. 

Before we continue with the dilute limit for this channel we would like to mention that Eq.(\ref{eq:QtoQGammaCrosSec}) is exact and it resums not only the kinematic twists but all twists for this process, i.e. no higher-body twist correction is expected for the $q\to q\gamma$ channel. This is due to the fact that one of the Wilson line operators is trivial for this process and there is no other TMD operator involved.  

Inserting the simple color structure of this process into the dilute limit of the generic CGC cross section given in Eq. \eqref{eq:DilCSuPDF}, we get the dilute limit of the cross section for the $q\to q\gamma$ channel:
\begin{align}
\frac{d\sigma_{q\rightarrow q\gamma}^{gA\sim0}}{dy_{1}dy_{2}d^{2}\boldsymbol{p}_{1}d^{2}\boldsymbol{p}_{2}} & =\frac{\alpha_{s}}{16N_{c}\left(p_{0}^{+}\right)^{2}}\left(2\pi\right)p_{0}^{+}\delta\left(p_{1}^{+}+p_{2}^{+}-p_{0}^{+}\right)\frac{\mathcal{G}_{ac}\left(\boldsymbol{k}\right)}{\boldsymbol{k}^{2}}\left(\phi^{i}\phi^{j\ast}\right)\label{eq:qqgammadilini}\\
 & \times\mathrm{Tr}\left[t^{a}\left(\frac{\boldsymbol{p}_{2}^{i}}{\boldsymbol{p}_{2}^{2}}+\frac{\boldsymbol{q}^{i}}{\boldsymbol{q}^{2}}\right)\right]\left[t^{c}\left(\frac{\boldsymbol{p}_{2}^{j}}{\boldsymbol{p}_{2}^{2}}+\frac{\boldsymbol{q}^{j}}{\boldsymbol{q}^{2}}\right)\right],\nonumber 
\end{align}
which leads to
\begin{align}
\frac{d\sigma_{q\rightarrow q\gamma}^{gA\sim0}}{dy_{1}dy_{2}d^{2}\boldsymbol{p}_{1}d^{2}\boldsymbol{p}_{2}} & =\frac{\alpha_{\mathrm{em}}\alpha_{s}}{N_{c}}p_{0}^{+}\delta\left(p_{1}^{+}+p_{2}^{+}-p_{0}^{+}\right)\frac{z\bar{z}^{2}\left(1+z^{2}\right)}{\boldsymbol{p}_{2}^{2}\boldsymbol{q}^{2}}\mathcal{G}\left(\boldsymbol{k}\right).\label{eq:qqgammadilfin}
\end{align}
As in the case of the other channels, comparing  Eq.~(\ref{eq:QtoQGammaCrosSec}) and (\ref{eq:qqgammadilfin}), we also get a straightforward matching between the improved TMD scheme and the dilute scheme:
\begin{equation}
\sigma_{q\rightarrow q\gamma}^{gA\sim0}=\left.\sigma_{q\rightarrow q\gamma}^{WW}\right|_{\mathcal{F}_{qg}^{\left(1\right)}=\mathcal{G}}.\label{eq:gammaqqDilvsTMD-1}
\end{equation}

\section{Discussions \label{sec:Discussions}}

Earlier studies have shown that for certain observables the small-$x$ limit of the TMD framework and the so-called "correlation limit" of the CGC framework overlap. In particular, two particle production (such as dijet or photon+jet) in forward pp and pA collisions, gluon TMDs can be recovered from the CGC calculations in the correlation limit. This specific limit corresponds to the case when the total transverse momentum of the produced particles $\boldsymbol{k}$ is much smaller than the hard scale $Q$. On the other hand, it is also well known that in the dilute limit of the CGC framework, that is in the limit when the total transverse momentum of the  produced particles are of the same order as the hard scale, one recovers the BFKL results. Recently, the small-$x$ improved TMD (iTMD) formalism has been developed to interpolate between these two limits.  

In this paper we studied two cases. First, by studying the correlation limit of the CGC amplitude for a generic $1\to 2$ process, we identified the kinematic twist contributions from higher order terms in the Taylor expansion of the transverse size of the pair of particles produced in that process, resummed those twist corrections in the Wandzura-Wilczek approximation, i.e. neglecting all genuine twist corrections. The kinematic-twist-resummed cross section for a generic $1\to 2$ process, Eq. \eqref{eq:genCrossSec}, is then used to compare the results obtained in the iTMD framework for different channels in forward pp and pA collisions. The perfect matching between these frameworks proves that the iTMD formalism is fully obtained from the CGC formalism by taking the Wandzura-Wilczek approximation. 

Second, we considered the dilute limit of the CGC amplitude for a generic $1\to2$ process. The BFKL amplitudes obtained by taking the dilute limit of the CGC amplitudes are shown to match iTMD results by simply setting the different TMD distributions to the unintegrated parton distribution function that defines the target.   

One of the most striking results of this study is the perfect matching between the hard parts of the kinematic twist resummed cross section and the dilute limit of the CGC one. 
This can be explained in the following way. The kinematic twist resummation procedure that has been developed in this paper isolates and resums the parts of the higher-body contributions which can be rewritten as gauge invariance fixing counterterms to the 1-body hard part. The remaining terms are the genuine twist contributions which vanish in the dilute limit since they account for multiple scatterings. 
In that sense, we resum the terms which do not vanish in the dilute limit. Thus, the difference between a rigorous twist resummation and the dilute expansion does not lie in the hard parts. Instead, it is linked to the way the hard parts couple to the distributions.

While iTMD distinguishes distributions depending on their gauge link structures, therefore extending its validity range in terms of $\left|\boldsymbol{k}\right|/Q_s$ when compared to BFKL, both formalisms rely on  the Wandzura-Wilczek approximation in the CGC.

With the previous observations, two origins of saturation can be expected. First of all, the difference between BFKL and iTMD is related to the distinction between gauge link structures, which account for multiple scattering from low $x$ gluons. As discussed earlier, all distributions are equal at large $|\boldsymbol{k}|/Q_s$ and distinct at low $|\boldsymbol{k}|/Q_s$, were saturation is expected. In that sense, the saturation scale $Q_s$ is the parameter which controls the importance of multiple scatterings via gauge links.
On the other hand, BFKL and iTMD both rely on the Wandzura-Wilczek approximation when compared to the CGC. It will be very instructive to compare predictions from iTMD and full CGC once genuine twists are extracted from the CGC as well~\citep{genuine}. This would probe $Q_s$ as the parameter which controls the importance of multiple scattering via genuine twists.

As a natural extension of this study, we plan to perform a similar analysis for more complex observables where not only the unpolarized TMD distributions but their linearly polarized partners appear. The two immediate observables that we are planning to study are the heavy quark production \cite{Marquet:2017xwy} and three-particle production such as dijet+photon production \cite{Altinoluk:2018byz}.

Last but not least, we would like to mention that recently there have been several studies devoted to understand the subeikonal corrections in the CGC framework \cite{Altinoluk:2014oxa, Altinoluk:2015gia, Altinoluk:2015xuy, Balitsky:2015qba, Balitsky:2016dgz, Balitsky:2017flc, Chirilli:2018kkw, Kovchegov:2015pbl, Kovchegov:2016zex, Kovchegov:2018znm}. Comparing those to the future moderate-$x$ corrections to the iTMD scheme would be also a natural extension of our study. 

\section*{Acknowledgments}
We thank P. Taels for stimulating discussions. TA gratefully acknowledges  the  support  from Bourses  du  Gouvernement  Fran\c{c}ais (BGF)-S\'{e}jour de recherche, and expresses his gratitude to CPHT, Ecole  Polytechnique, and  to the Institute of Nuclear Physics, Polish Academy of Sciences, for hospitality when part of this work was done. The work of TA is supported by Grant No. 2017/26/M/ST2/01074 of the National Science Centre, Poland. RB is grateful to NCBJ for hospitality when this project was started. The work of RB is supported by the  National Science Centre, Poland, grant No.2015/17/B/ST2/01838, by the U.S. Department of Energy, Office of Nuclear Physics, under Contracts No. DE-SC0012704 and by an LDRD grant from Brookhaven Science Associates. The work of PK is partially supported by Polish National Science Centre grant no. DEC-2017/27/B/ST2/01985. This work received additional partial support by Polish National Science Centre grant no. DEC-2017/27/B/ST2/01985. 

\appendix

\section{Effective Feynman rules in the external CGC shockwave field\label{sec:Feymanrules}}

In this appendix, we list the effective Feynman rules that are used to calculate the CGC amplitudes in Appendix~\ref{sec:CGCamplitudes}.  \\

\hspace{-0.8cm}{\it Outgoing quark line:}
\begin{eqnarray}
\bar{u}\left(p_{q},\,z_{0}\right) & = & \frac{1}{2}\left(\frac{p_{q}^{+}}{2\pi}\right)^{\frac{d}{2}}\int d^{d}\boldsymbol{x}_{1}e^{ip_{q}^{+}\left(z_{0}^{-}-\frac{\left(\boldsymbol{x}_{1}-\boldsymbol{z}_{0}\right)^{2}}{2z_{0}^{+}}+i0\right)-i\left(\boldsymbol{p}_{q}\cdot\boldsymbol{x}_{1}\right)+i\frac{z_{0}^{+}}{2p_{q}^{+}}\left(m^{2}+i0\right)}\left[U_{\boldsymbol{x}_{1}}\theta\left(-z_{0}^{+}\right)+\theta\left(z_{0}^{+}\right)\right]\nonumber \\
 &  & \times\left(\frac{i}{z_{0}^{+}}\right)^{\frac{d}{2}}\bar{u}_{p_{q}}\gamma^{+}\left(\gamma^{-}-\frac{\hat{x}_{1\perp}-\hat{z}_{0\perp}}{z_{0}^{+}}+\frac{m}{p_{q}^{+}}\right)\label{eq:QuarkLine}
\end{eqnarray}

\hspace{-0.8cm}{\it Outgoing antiquark line:}
\begin{eqnarray}
v\left(p_{\bar{q}},\,z_{0}\right) & = & \frac{1}{2}\left(\frac{p_{\bar{q}}^{+}}{2\pi}\right)^{\frac{d}{2}}\int d^{d}\boldsymbol{x}_{2}e^{ip_{\bar{q}}^{+}\left(z_{0}^{-}-\frac{\left(\boldsymbol{x}_{2}-\boldsymbol{z}_{0}\right)^{2}}{2z_{0}^{+}}+i0\right)-i\left(\boldsymbol{p}_{\bar{q}}\cdot\boldsymbol{x}_{2}\right)+i\frac{z_{0}^{+}}{2p_{\bar{q}}^{+}}\left(m^{2}+i0\right)}\left[U_{\boldsymbol{x}_{2}}^{\dagger}\theta\left(-z_{0}^{+}\right)+\theta\left(z_{0}^{+}\right)\right]\nonumber \\
 &  & \times\left(\frac{i}{z_{0}^{+}}\right)^{\frac{d}{2}}\left(\gamma^{-}-\frac{\hat{x}_{2\perp}-\hat{z}_{0\perp}}{z_{0}^{+}}-\frac{m}{p_{\bar{q}}^{+}}\right)\gamma^{+}v_{p_{\bar{q}}}\label{eq:AntiquarkLine}
\end{eqnarray}

\hspace{-0.8cm}{\it Incoming gluon line:}
\begin{eqnarray}
\varepsilon_{\mu_{0}}^{b_{0}a_{0}}\left(p_{0},\,z_{0}\right) & = & \left(\frac{p_{0}^{+}}{2\pi}\right)^{\frac{d}{2}}\int d^{d}\boldsymbol{x}_{0}e^{-ip_{0}^{+}\left(z_{0}^{-}-\frac{\left(\boldsymbol{x}_{0}-\boldsymbol{z}_{0}\right)^{2}}{2z_{0}^{+}}-i0\right)+i\left(\boldsymbol{p}_{0}\cdot\boldsymbol{x}_{0}\right)}\left[U_{\boldsymbol{x}_{0}}^{b_{0}a_{0}}\theta\left(z_{0}^{+}\right)+\delta^{a_{0}b_{0}}\theta\left(-z_{0}^{+}\right)\right]\nonumber \\
 &  & \times\left(\frac{-i}{z_{0}^{+}}\right)^{\frac{d}{2}}\left(g_{\perp\mu_{0}\sigma_{0}}+\frac{x_{0\perp\sigma_{0}}-z_{0\perp\sigma_{0}}}{z_{0}^{+}}n_{2\mu_{0}}\right)\varepsilon_{p_{0}\perp}^{\sigma_{0}}\label{eq:IncomingGluonLine}
\end{eqnarray}

\hspace{-0.8cm}{\it Outgoing gluon line:}
\begin{eqnarray}
\varepsilon_{\mu}^{ba\ast}\left(p_{g},\,z_{0}\right) & = & \left(\frac{p_{g}^{+}}{2\pi}\right)^{\frac{d}{2}}\int d^{d}\boldsymbol{x}_{2}e^{ip_{g}^{+}\left(z_{0}^{-}-\frac{\left(\boldsymbol{x}_{2}-\boldsymbol{z}_{0}\right)^{2}-i0}{2z_{0}^{+}}\right)-i\left(\boldsymbol{p}_{g}\cdot\boldsymbol{x}_{2}\right)}\left[U_{\boldsymbol{x}_{2}}^{ab}\theta\left(-z_{0}^{+}\right)+\delta^{ab}\theta\left(z_{0}^{+}\right)\right]\nonumber \\
 &  & \times\left(\frac{i}{z_{0}^{+}}\right)^{\frac{d}{2}}\left(g_{\perp\mu\sigma}+\frac{x_{2\perp\sigma}-z_{0\perp\sigma}}{z_{0}^{+}}n_{2\mu}\right)\varepsilon_{p_{g}\perp}^{\sigma\ast}\label{eq:OutgoingGluonLine}
\end{eqnarray}

\section{CGC amplitudes for all channels\label{sec:CGCamplitudes}}

In this appendix we list the CGC amplitudes calculated by using the effective Feynman rules listed in Appendix~\ref{sec:Feymanrules}. \\

\hspace{-0.8cm}{\it $q\rightarrow qg$ channel for forward dijet production in $pp$
and $pA$ collisions:\label{subsec:qqgamp}}
\begin{align}
\mathcal{A}_{q\rightarrow qg} & =\frac{ig_{s}}{2\pi}\varepsilon_{p_{g}\perp}^{\sigma\ast}\left(2\pi\right)\delta\left(p_{q}^{+}+p_{g}^{+}-p^{+}\right)\int d^{2}\boldsymbol{b}d^{2}\boldsymbol{r}e^{-i\left(\boldsymbol{q}\cdot\boldsymbol{r}\right)-i\left(\boldsymbol{k}\cdot\boldsymbol{b}\right)}\nonumber \\
 & \times\frac{r_{\perp}^{\mu}}{\boldsymbol{r}^{2}}\left[\left(U_{\boldsymbol{b}+\bar{z}\boldsymbol{r}}t^{b}U_{\boldsymbol{b}-z\boldsymbol{r}}^{ab}\right)-\left(t^{b}\delta^{ab}U_{\boldsymbol{b}}\right)\right]\label{eq:QtoQGini}\\
 & \times\bar{u}_{p_{q}}\left[2zg_{\perp\mu\sigma}+\bar{z}\left(\gamma_{\perp\mu}\gamma_{\perp\sigma}\right)\right]\gamma^{+}u_{p}\nonumber 
\end{align}

\hspace{-0.8cm}{\it $g\rightarrow q\bar{q}$ channel for forward dijet production in
$pp$ and $pA$ collisions: \label{subsec:gqqamp}}
\begin{align}
\mathcal{A}_{g\rightarrow q\bar{q}} & =-i\frac{g_{s}}{2\pi}\varepsilon_{p\perp}^{\sigma}\left(2\pi\right)\delta\left(p_{q}^{+}+p_{\bar{q}}^{+}-p_{g}^{+}\right)\int d^{2}\boldsymbol{b}d^{2}\boldsymbol{r}e^{-i\left(\boldsymbol{q}\cdot\boldsymbol{r}\right)-i\left(\boldsymbol{k}\cdot\boldsymbol{b}\right)}\nonumber \\
 & \times\frac{r_{\perp}^{\mu}}{\boldsymbol{r}^{2}}\left[\left(U_{\boldsymbol{b}+\bar{z}\boldsymbol{r}}t^{b}U_{\boldsymbol{b}-z\boldsymbol{r}}^{\dagger}\delta^{ab}\right)-\left(t^{b}U_{\boldsymbol{b}}^{ba}\right)\right]\label{eq:GtoQQbarini}\\
 & \times\bar{u}_{p_{q}}\left[2zg_{\perp\mu\sigma}-\left(\gamma_{\perp\mu}\gamma_{\perp\sigma}\right)\right]\gamma^{+}v_{p_{\bar{q}}}\nonumber 
\end{align}

\hspace{-0.8cm}{\it $g\rightarrow gg$ channel for forward dijet production in $pp$
and $pA$ collisions:\label{subsec:gggamp}}
\begin{align}
\mathcal{A}_{g\rightarrow gg} & =\frac{2g_{s}p^{+}}{\pi}\varepsilon_{p\perp}^{\sigma_{0}}\varepsilon_{p_{g}\perp}^{\sigma_{1}\ast}\varepsilon_{q_{g}\perp}^{\sigma_{2}\ast}\left(2\pi\right)\delta\left(p_{g}^{+}+q_{g}^{+}-p^{+}\right)\int d^{2}\boldsymbol{b}d^{2}\boldsymbol{r}e^{-i\left(\boldsymbol{q}\cdot\boldsymbol{r}\right)-i\left(\boldsymbol{k}\cdot\boldsymbol{b}\right)}\nonumber \\
 & \times\frac{r_{\perp}^{\mu}}{\boldsymbol{r}^{2}}\left[f^{b_{0}b_{1}b_{2}}\delta^{b_{0}a_{0}}U_{\boldsymbol{b}+\bar{z}\boldsymbol{r}}^{b_{1}a_{1}}U_{\boldsymbol{b}-z\boldsymbol{r}}^{b_{2}a_{2}}-f^{b_{0}b_{1}b_{2}}\delta^{b_{1}a_{1}}\delta^{b_{2}a_{2}}U_{\boldsymbol{b}}^{b_{0}a_{0}}\right]\label{eq:GtoGGini}\\
 & \times\left[zg_{\perp\sigma_{0}\sigma_{1}}g_{\perp\mu\sigma_{2}}-z\bar{z}g_{\perp\sigma_{1}\sigma_{2}}g_{\perp\mu\sigma_{0}}+\bar{z}g_{\perp\sigma_{0}\sigma_{2}}g_{\perp\mu\sigma_{1}}\right]\nonumber 
\end{align}

\hspace{-0.8cm}{\it Production of a forward photon-jet pair in $pp$ and $pA$ collisions:\label{subsec:DYamp}}
\begin{align}
\mathcal{A}_{q\rightarrow q\gamma} & =\frac{ig_{s}}{2\pi}\varepsilon_{p_{\gamma}\perp}^{\sigma\ast}\left(2\pi\right)\delta\left(p_{q}^{+}+p_{g}^{+}-p^{+}\right)\int d^{2}\boldsymbol{b}d^{2}\boldsymbol{r}e^{-i\left(\boldsymbol{q}\cdot\boldsymbol{r}\right)-i\left(\boldsymbol{k}\cdot\boldsymbol{b}\right)}\nonumber \\
 & \times\frac{r_{\perp}^{\mu}}{\boldsymbol{r}^{2}}\left(U_{\boldsymbol{b}+\bar{z}\boldsymbol{r}}-U_{\boldsymbol{b}}\right)\bar{u}_{p_{q}}\left[2zg_{\perp\mu\sigma}+\bar{z}\left(\gamma_{\perp\mu}\gamma_{\perp\sigma}\right)\right]\gamma^{+}u_{p}\label{eq:QtoQGammaini}
\end{align}

\section{The integral\label{sec:Integral}}

In this appendix, we present the details of the calculation of the following integral 
\begin{align}
I^{ij}\left(\boldsymbol{p}\right) & \equiv\int d^{d}\boldsymbol{r}\frac{\boldsymbol{r}^{i}\boldsymbol{r}^{j}}{\boldsymbol{r}^{2}}\frac{e^{-i\left(\boldsymbol{p}\cdot\boldsymbol{r}\right)}-1}{\left(\boldsymbol{p}\cdot\boldsymbol{r}\right)}e^{-i\left(\boldsymbol{q}\cdot\boldsymbol{r}\right)}.\label{eq:Integral}
\end{align}
This integral is a symmetric tensor, hence we can decompose it in
a 3-dimensional basis. Let us choose 
\begin{equation}
\left(\delta^{ij},\frac{\boldsymbol{p}^{i}\boldsymbol{q}^{j}+\boldsymbol{q}^{j}\boldsymbol{p}^{i}}{\boldsymbol{p}\cdot\boldsymbol{q}},\frac{\boldsymbol{p}^{i}\boldsymbol{p}^{j}}{\boldsymbol{p}^{2}}\right),\label{eq:IntegralBasis}
\end{equation}
and write 
\begin{equation}
I^{ij}\left(\boldsymbol{p}\right)=I_{0}\delta^{ij}+I_{1}\frac{\boldsymbol{p}^{i}\boldsymbol{q}^{j}+\boldsymbol{q}^{i}\boldsymbol{p}^{j}}{\boldsymbol{p}\cdot\boldsymbol{q}}+I_{2}\frac{\boldsymbol{p}^{i}\boldsymbol{p}^{j}}{\boldsymbol{p}^{2}}.\label{eq:IntegralinBasis}
\end{equation}
This relations inverts to 
\begin{align}
I_{0} & =I^{ii}\left(\boldsymbol{p}\right)-\frac{\boldsymbol{p}^{i}\boldsymbol{p}^{j}}{\boldsymbol{p}^{2}}I^{ij}\left(\boldsymbol{p}\right)\nonumber \\
\nonumber \\
I_{1} & =-\frac{\left(\boldsymbol{p}\cdot\boldsymbol{q}\right)^{2}}{\boldsymbol{p}^{2}\boldsymbol{q}^{2}-\left(\boldsymbol{p}\cdot\boldsymbol{q}\right)^{2}}\frac{\boldsymbol{p}^{i}\boldsymbol{p}^{j}}{\boldsymbol{p}^{2}}I^{ij}\left(\boldsymbol{p}\right)+\frac{\left(\boldsymbol{p}\cdot\boldsymbol{q}\right)}{\boldsymbol{p}^{2}\boldsymbol{q}^{2}-\left(\boldsymbol{p}\cdot\boldsymbol{q}\right)^{2}}\boldsymbol{p}^{i}\boldsymbol{q}^{j}I^{ij}\left(\boldsymbol{p}\right)\label{eq:InvertingBasis}\\
\nonumber \\
I_{2} & =-I^{ii}\left(\boldsymbol{p}\right)+\frac{2\boldsymbol{p}^{2}\boldsymbol{q}^{2}}{\boldsymbol{\boldsymbol{p}}^{2}\boldsymbol{q}^{2}-\left(\boldsymbol{p}\cdot\boldsymbol{q}\right)^{2}}\frac{\boldsymbol{p}^{i}\boldsymbol{p}^{j}}{\boldsymbol{p}^{2}}I^{ij}\left(\boldsymbol{p}\right)-\frac{2\left(\boldsymbol{p}\cdot\boldsymbol{q}\right)}{\boldsymbol{p}^{2}\boldsymbol{q}^{2}-\left(\boldsymbol{p}\cdot\boldsymbol{q}\right)^{2}}\boldsymbol{p}^{i}\boldsymbol{q}^{j}I^{ij}\left(\boldsymbol{p}\right).\nonumber 
\end{align}
Thus in order to compute $I^{ij}$, it is sufficient to compute $J_{0}=\delta^{ij}I^{ij}$
and $J_{1}^{j}\equiv\boldsymbol{\boldsymbol{p}}^{i}I^{ij}.$ One can
actually show that $J_{0}=0$. This becomes apparent  by going to spherical coordinates,
integrating $\left|\boldsymbol{r}\right|$ out (taking into account
the phase regulators $i0$ in the exponent from the effective rules
in Appendix~\ref{sec:Feymanrules}) and checking that the remaining
angular integral is null.

$J_{1}^{j}$ is obtained easily with the usual Schwinger representation
tricks and reads: 
\begin{equation}
J_{1}^{j}=-2i\pi\left(\frac{\boldsymbol{q}^{j}+\boldsymbol{\boldsymbol{p}}^{j}}{\left(\boldsymbol{q}+\boldsymbol{p}\right)^{2}}-\frac{\boldsymbol{q}^{j}}{\boldsymbol{q}^{2}}\right).\label{eq:J1}
\end{equation}
Finally plugging Eq.~(\ref{eq:J1}) in Eq.~(\ref{eq:InvertingBasis})
then in Eq.~(\ref{eq:IntegralinBasis}), one obtains 
\begin{equation}
I^{ij}\left(\boldsymbol{k}\right)=-2\frac{i\pi}{\boldsymbol{p}^{2}}\left[\left(\frac{\left(\boldsymbol{p}\cdot\boldsymbol{q}\right)}{\boldsymbol{q}^{2}}-\frac{\left(\boldsymbol{p}\cdot\boldsymbol{q}\right)+\boldsymbol{\boldsymbol{p}}^{2}}{\left(\boldsymbol{q}+\boldsymbol{p}\right)^{2}}\right)\delta^{ij}+\left(\frac{1}{\left(\boldsymbol{q}+\boldsymbol{p}\right)^{2}}-\frac{1}{\boldsymbol{q}^{2}}\right)\left(\boldsymbol{p}^{i}\boldsymbol{q}^{j}+\boldsymbol{q}^{i}\boldsymbol{p}^{j}\right)+2\frac{\boldsymbol{p}^{i}\boldsymbol{p}^{j}}{\left(\boldsymbol{q}+\boldsymbol{p}\right)^{2}}\right],\label{eq:IntegralResult}
\end{equation}
which leads to the expression given in Eq.~(\ref{eq:Integralfin}).

%-----------------------------------------------------------------------------

%-----------------------------------------------------------------------------


\begin{thebibliography}{99}


\bibitem{Gribov:1972ri} 
  V.~N.~Gribov and L.~N.~Lipatov,
  \emph{{Deep inelastic e p scattering in perturbation theory}}, 
  {\emph {Sov.\ J.\ Nucl.\ Phys.}  {\bf 15}, 438 (1972)}
  [{\emph {Yad.\ Fiz.}  {\bf 15}, 781 (1972)}].
  
  
  \bibitem{Altarelli:1977zs} 
  G.~Altarelli and G.~Parisi,
  \emph{{Asymptotic Freedom in Parton Language}}, 
  \href{https://doi.org/10.1016/0550-3213(77)90384-4}
  {\emph{Nucl.\ Phys.\ B} {\bf 126}, 298 (1977)}.
  
  
  \bibitem{Dokshitzer:1977sg}
  Y.~L.~Dokshitzer,
  \emph{{Calculation of the Structure Functions for Deep Inelastic Scattering and e+ e- Annihilation by Perturbation Theory in Quantum Chromodynamics}},
  {\emph {Sov.\ Phys.\ JETP} {\bf 46} (1977) 641}
   [{\emph {Zh.\ Eksp.\ Teor.\ Fiz.} {\bf 73} (1977) 1216}].
  


  
\bibitem{Kuraev:1977fs} 
  E.~A.~Kuraev, L.~N.~Lipatov and V.~S.~Fadin,
  \emph{{The Pomeranchuk Singularity in Nonabelian Gauge Theories}}, 
  {\emph {Sov.\ Phys.\ JETP} {\bf 45}, 199 (1977)}
  [{\emph{Zh.\ Eksp.\ Teor.\ Fiz.}  {\bf 72}, 377 (1977)}].
    
  \bibitem{Balitsky:1978ic} 
  I.~I.~Balitsky and L.~N.~Lipatov,
  \emph{{The Pomeranchuk Singularity in Quantum Chromodynamics}}, 
  {\emph{ Sov.\ J.\ Nucl.\ Phys.} {\bf 28}, 822 (1978)}
  [{\emph {Yad.\ Fiz.} {\bf 28}, 1597 (1978)}].
 

  
\bibitem{Mueller:1989st} 
  A.~H.~Mueller,
  \emph{{Small x Behavior and Parton Saturation: A QCD Model}}, 
\href{https://doi.org/10.1016/0550-3213(90)90173-B} 
  {\emph{Nucl.\ Phys.\ B} {\bf 335}, 115 (1990)}.
  
  \bibitem{Mueller:1993rr} 
  A.~H.~Mueller,
  \emph{{Soft gluons in the infinite momentum wave function and the BFKL pomeron}},
 \href{https://doi.org/10.1016/0550-3213(94)90116-3}
 {\emph{ Nucl.\ Phys.\ B }{\bf 415}, 373 (1994)}.
 
 \bibitem{Mueller:1994gb} 
  A.~H.~Mueller,
  \emph{{Unitarity and the BFKL pomeron}},
  \href{https://doi.org/10.1016/0550-3213(94)00480-3}
  {\emph{Nucl.\ Phys.\ B} {\bf 437}, 107 (1995)}, 
  [\href{https://arxiv.org/abs/hep-ph/9408245 }{{\tt hep-ph/9408245}}].
  
  
   

 
 \bibitem{Balitsky:1995ub} 
  I.~Balitsky,
  \emph{{Operator expansion for high-energy scattering}}, 
  \href{https://doi.org/10.1016/0550-3213(95)00638-9}
  {\emph{Nucl.\ Phys.\ B} {\bf 463}, 99 (1996)}, 
  [\href{https://arxiv.org/abs/hep-ph/9509348}{{\tt hep-ph/9509348}}].
  
  \bibitem{Balitsky:1998kc} 
  I.~Balitsky,
  \emph{{Factorization for high-energy scattering}}, 
  \href{https://doi.org/10.1103/PhysRevLett.81.2024}
  {\emph{Phys.\ Rev.\ Lett.}  {\bf 81}, 2024 (1998)}, 
  [\href{https://arxiv.org/abs/hep-ph/9807434}{{\tt hep-ph/9807434}}].
 
 \bibitem{Balitsky:1998ya} 
  I.~Balitsky,
  \emph{{Factorization and high-energy effective action}}, 
  \href{https://doi.org/10.1103/PhysRevD.60.014020}
  {\emph{Phys.\ Rev.\ D} {\bf 60}, 014020 (1999)}, 
  [\href{https://arxiv.org/abs/hep-ph/9812311}{{\tt hep-ph/9812311}}].
  
 

 
 
 
\bibitem{McLerran:1993ni} 
  L.~D.~McLerran and R.~Venugopalan,
 \emph{{Computing quark and gluon distribution functions for very large nuclei}}, 
 \href{https://doi.org/10.1103/PhysRevD.49.2233}
  {\emph{Phys.\ Rev.\ D} {\bf 49}, 2233 (1994)}, 
  [\href{https://arxiv.org/abs/hep-ph/9309289}{{\tt hep-ph/9309289}}].
  
  \bibitem{McLerran:1993ka} 
  L.~D.~McLerran and R.~Venugopalan,
  \emph{{Gluon distribution functions for very large nuclei at small transverse momentum}},
  \href{https://doi.org/10.1103/PhysRevD.49.3352}
  {\emph{Phys.\ Rev.\ D} {\bf 49}, 3352 (1994)}, 
  [\href{https://arxiv.org/abs/hep-ph/9311205}{{\tt hep-ph/9311205}}].

\bibitem{McLerran:1994vd} 
  L.~D.~McLerran and R.~Venugopalan,
  \emph{{Green's functions in the color field of a large nucleus}},
  \href{https://doi.org/10.1103/PhysRevD.50.2225}
  {\emph{Phys.\ Rev.\ D} {\bf 50}, 2225 (1994)}, 
  [\href{https://arxiv.org/abs/hep-ph/9402335}{{\tt hep-ph/9402335}}].
 
 






 
 \bibitem{JalilianMarian:1997jx} 
  J.~Jalilian-Marian, A.~Kovner, A.~Leonidov and H.~Weigert,
  \emph{{The BFKL equation from the Wilson renormalization group}}, 
  \href{https://doi.org/10.1016/S0550-3213(97)00440-9}
  {\emph{Nucl.\ Phys.\ B} {\bf 504}, 415 (1997)}, 
  [\href{https://arxiv.org/abs/hep-ph/9701284}{{\tt hep-ph/9701284}}].
 
  
 \bibitem{JalilianMarian:1997gr} 
  J.~Jalilian-Marian, A.~Kovner, A.~Leonidov and H.~Weigert,
  \emph{{The Wilson renormalization group for low x physics: Towards the high density regime}}, 
  \href{https://doi.org/10.1103/PhysRevD.59.014014}
  {\emph{Phys.\ Rev.\ D} {\bf 59}, 014014 (1998)}, 
  [\href{https://arxiv.org/abs/hep-ph/9706377}{{\tt hep-ph/9706377 }}].
  
  \bibitem{JalilianMarian:1997dw} 
  J.~Jalilian-Marian, A.~Kovner and H.~Weigert,
  \emph{{The Wilson renormalization group for low x physics: Gluon evolution at finite parton density}}, 
  \href{https://doi.org/10.1103/PhysRevD.59.014015}
  {\emph{Phys.\ Rev.\ D} {\bf 59}, 014015 (1998)},
  [\href{https://arxiv.org/abs/hep-ph/9709432}{{\tt hep-ph/9709432}}].
  
  \bibitem{Kovner:1999bj} 
  A.~Kovner and J.~G.~Milhano,
  \emph{{Vector potential versus color charge density in low x evolution}}, 
  \href{https://doi.org/10.1103/PhysRevD.61.014012}
  {\emph{Phys.\ Rev.\ D} {\bf 61}, 014012 (2000)}, 
  [\href{https://arxiv.org/abs/hep-ph/9904420}{{\tt hep-ph/9904420}}].
  
  \bibitem{Kovner:2000pt} 
  A.~Kovner, J.~G.~Milhano and H.~Weigert,
  \emph{{Relating different approaches to nonlinear QCD evolution at finite gluon density}}, 
  \href{https://doi.org/10.1103/PhysRevD.62.114005}
  {\emph{Phys.\ Rev.\ D }{\bf 62}, 114005 (2000)}, 
  [\href{https://arxiv.org/abs/hep-ph/0004014}{{\tt hep-ph/0004014}}].
  
  \bibitem{Weigert:2000gi} 
  H.~Weigert,
  \emph{{Unitarity at small Bjorken x}}, 
  \href{https://doi.org/10.1016/S0375-9474(01)01668-2}
  {\emph{Nucl.\ Phys.\ A }{\bf 703}, 823 (2002)}, 
  [\href{https://arxiv.org/abs/hep-ph/0004044}{{\tt hep-ph/0004044}}].
 
 

 
 \bibitem{CGC}
 E.~Iancu, A.~Leonidov and L.~D.~McLerran,
  \emph{{Nonlinear gluon evolution in the color glass condensate. 1}}, 
  \href{https://doi.org/10.1016/S0375-9474(01)00642-X}
  {\emph{Nucl.\ Phys.\ A} {\bf 692}, 583 (2001)}, 
  [\href{https://arxiv.org/abs/hep-ph/0011241}{{\tt hep-ph/0011241}}].

\bibitem{Ferreiro:2001qy} 
  E.~Ferreiro, E.~Iancu, A.~Leonidov and L.~McLerran,
  \emph{{Nonlinear gluon evolution in the color glass condensate. 2}}, 
  \href{https://doi.org/10.1016/S0375-9474(01)01329-X}
  {\emph{Nucl.\ Phys.\ A} {\bf 703}, 489 (2002)}, 
  [\href{https://arxiv.org/abs/hep-ph/0109115}{{\tt hep-ph/0109115}}].
 
 
 \bibitem{Kovchegov}
 Y.~V.~Kovchegov,
  \emph{{Small x F(2) structure function of a nucleus including multiple pomeron exchanges}}, 
  \href{https://doi.org/10.1103/PhysRevD.60.034008}
  {\emph{Phys.\ Rev.\ D} {\bf 60}, 034008 (1999)}, 
  [\href{https://arxiv.org/abs/hep-ph/9901281}{{\tt hep-ph/9901281}}].
   
 \bibitem{Mueller_BFKL}
 A.~H.~Mueller and B.~Patel,
 \emph{{Single and double BFKL pomeron exchange and a dipole picture of high-energy hard processes}}, 
 \href{https://doi.org/10.1016/0550-3213(94)90284-4}
  {\emph{Nucl.\ Phys.\ B} {\bf 425}, 471 (1994)}, 
  [\href{https://arxiv.org/abs/hep-ph/9403256}{{\tt hep-ph/9403256}}].
 
 \bibitem{Chen:1995pa} 
  Z.~Chen and A.~H.~Mueller,
  \emph{{The Dipole picture of high-energy scattering, the BFKL equation and many gluon compound states}}, 
  \href{https://doi.org/10.1016/0550-3213(95)00350-2}
  {\emph{Nucl.\ Phys.\ B} {\bf 451}, 579 (1995)}.
  
 
 \bibitem{Caron-Huot:2013fea} 
  S.~Caron-Huot,
  \emph{{When does the gluon reggeize?}}, 
  \href{https://doi.org/10.1007/JHEP05(2015)093}
  {\emph{JHEP} {\bf 1505}, 093 (2015)}, 
  [\href{https://arxiv.org/abs/1309.6521}{{\tt hep-th/1309.6521}}].
  
  \bibitem{Balitsky:2008zza} 
I.~Balitsky and G.~A.~Chirilli,
 \emph{{Next-to-leading order evolution of color dipoles}}, 
 \href{https://doi.org/10.1103/PhysRevD.77.014019}
  {\emph{Phys.\ Rev.\ D} {\bf 77}, 014019 (2008)}, 
  [\href{https://arxiv.org/abs/0710.4330}{{\tt hep-ph/ 0710.4330}}].
    
\bibitem{Fadin:2007ee} 
  V.~S.~Fadin, R.~Fiore and A.~Papa,
  \emph{{The Dipole form of the quark part of the BFKL kernel}}, 
  \href{https://doi.org/10.1016/j.physletb.2007.02.022}
  {\emph{Phys.\ Lett.\ B} {\bf 647}, 179 (2007)}, 
  [\href{https://arxiv.org/abs/hep-ph/0701075}{{\tt hep-ph/0701075}}].
 
 \bibitem{Fadin:2007de} 
  V.~S.~Fadin, R.~Fiore, A.~V.~Grabovsky and A.~Papa,
\emph{{The Dipole form of the gluon part of the BFKL kernel}}, 
\href{https://doi.org/10.1016/j.nuclphysb.2007.06.002}
  {\emph{Nucl.\ Phys.\ B }{\bf 784}, 49 (2007)}, 
  [\href{https://arxiv.org/abs/0705.1885}{{\tt hep-ph/0705.1885}}].
    
  
  
   
\bibitem{Fadin:2011jg}
V.~S.~Fadin, R.~Fiore, A.~V.~Grabovsky and A.~Papa,
 \emph{{Connection between complete and Moebius forms of gauge invariant operators}}, 
 \href{https://doi.org/10.1016/j.nuclphysb.2011.11.008}
  {\emph{Nucl.\ Phys.\ B} {\bf 856}, 111 (2012)}, 
  [\href{https://arxiv.org/abs/1109.6634}{{\tt hep-th/1109.6634 }}].
 

\bibitem{Collins:2011zzd}
J.~Collins, \emph{{Foundations of perturbative QCD}}, vol.~32.
\newblock Cambridge Univ. Press, 2011.

  %\cite{Collins:1983pk}
\bibitem{Collins:1983pk} 
  J.~C.~Collins, D.~E.~Soper and G.~F.~Sterman,
  \emph{{Relation of Parton Distribution Functions in {Drell-Yan} Process to Deeply Inelastic Scattering}}, 
  \href{https://doi.org/10.1016/0370-2693(83)90606-8}
  {\emph{Phys.\ Lett.}\  {\bf 126B}, 275 (1983)}, 


  
  %\cite{Brodsky:2002cx}
\bibitem{Brodsky:2002cx} 
S.~J.~Brodsky, D.~S.~Hwang and I.~Schmidt,
 \emph{{Final state interactions and single spin asymmetries in semiinclusive deep inelastic scattering}}, 
 \href{https://doi.org/10.1016/S0370-2693(02)01320-5}
  {\emph{Phys.\ Lett.\ B} {\bf 530}, 99 (2002)}, 
  [\href{https://arxiv.org/abs/hep-ph/0201296}{{\tt hep-ph/0201296]}}].
  
\bibitem{Collins:2002kn} 
  J.~C.~Collins,
  \emph{{Leading twist single transverse-spin asymmetries: Drell-Yan and deep inelastic scattering}}, 
  \href{https://doi.org/10.1016/S0370-2693(02)01819-1}
  {\emph{Phys.\ Lett.\ B} {\bf 536}, 43 (2002)}, 
  [\href{https://arxiv.org/abs/hep-ph/0204004}{{\tt hep-ph/0204004}}].
    

 
\bibitem{Belitsky:2002sm} 
  A.~V.~Belitsky, X.~Ji and F.~Yuan,
  \emph{{Final state interactions and gauge invariant parton distributions}}, 
  \href{https://doi.org/10.1016/S0550-3213(03)00121-4}
  {\emph{Nucl.\ Phys.\ B} {\bf 656}, 165 (2003)}, 
  [\href{https://arxiv.org/abs/hep-ph/0208038}{{\tt hep-ph/0208038}}].
 

\bibitem{Bomhof:2004aw} 
  C.~J.~Bomhof, P.~J.~Mulders and F.~Pijlman,
  \emph{{Gauge link structure in quark-quark correlators in hard processes}},
  \href{https://doi.org/10.1016/j.physletb.2004.06.100}
  {\emph{Phys.\ Lett.\ B} {\bf 596}, 277 (2004)}, 
  [\href{https://arxiv.org/abs/hep-ph/0406099}{{\tt hep-ph/0406099 }}].


\bibitem{Boer:1999si} 
  D.~Boer and P.~J.~Mulders,
  \emph{{Color gauge invariance in the Drell-Yan process}}, 
  \href{https://doi.org/10.1016/S0550-3213(99)00719-1}
  {\emph{Nucl.\ Phys.\ B }{\bf 569}, 505 (2000)}, 
  [\href{https://arxiv.org/abs/hep-ph/9906223}{{\tt hep-ph/9906223 }}].
    
 
\bibitem{TMD}
  C.~J.~Bomhof, P.~J.~Mulders and F.~Pijlman,
 \emph{{The Construction of gauge-links in arbitrary hard processes}}, 
  \href{https://doi.org/10.1140/epjc/s2006-02554-2}
  {\emph{ Eur.\ Phys.\ J.\ C} {\bf 47}, 147 (2006)}, 
  [\href{https://arxiv.org/abs/hep-ph/0601171}{{\tt hep-ph/0601171}}].
  
 \bibitem{Dominguez:2010xd} 
 F.~Dominguez, B.~W.~Xiao and F.~Yuan,
  \emph{{$k_t$-factorization for Hard Processes in Nuclei}}, 
  \href{https://doi.org/10.1103/PhysRevLett.106.022301}
  {\emph{Phys.\ Rev.\ Lett.}  {\bf 106}, 022301 (2011)}, 
  [\href{https://arxiv.org/abs/1009.2141}{{\tt hep-ph/1009.2141}}].
 
  \bibitem{Dominguez:2011wm} 
  F.~Dominguez, C.~Marquet, B.~W.~Xiao and F.~Yuan,
  \emph{{Universality of Unintegrated Gluon Distributions at small x}}, 
  \href{https://doi.org/10.1103/PhysRevD.83.105005}
  {\emph{Phys.\ Rev.\ D }{\bf 83}, 105005 (2011)}, 
  [\href{https://arxiv.org/abs/1101.0715}{{\tt hep-ph/1101.0715}}].
   
  
  
  \bibitem{Marquet:2016cgx} 
  C.~Marquet, E.~Petreska and C.~Roiesnel,
  \emph{{Transverse-momentum-dependent gluon distributions from JIMWLK evolution}}, 
  \href{https://doi.org/10.1007/JHEP10(2016)065}
  {\emph{JHEP} {\bf 1610}, 065 (2016)}, 
  [\href{https://arxiv.org/abs/1608.02577}{{\tt hep-ph/1608.02577 }}].
   
  \bibitem{Metz:2011wb}
  A.~Metz and J.~Zhou,
  \emph{Distribution of linearly polarized gluons inside a large nucleus},
  \href{https://doi.org/10.1103/PhysRevD.84.051503}
  {\emph{Phys.\ Rev.\ D }{\bf 84}, 051503 (2011)},
  [\href{https://arxiv.org/abs/1105.1991}{{\tt hep-ph/1105.1991}}].
  
  \bibitem{Akcakaya:2012si}
  E.~Akcakaya, A.~Sch\"{a}fer and J.~Zhou,
  \emph{Azimuthal asymmetries for quark pair production in pA collisions},
  \href{https://doi.org/10.1103/PhysRevD.87.054010}
  {\emph{Phys.\ Rev.\ D }{\bf 87}, 054010 (2013)},
  [\href{https://arxiv.org/abs/1208.4965}{{\tt hep-ph/1208.4965}}].

  \bibitem{Dumitru:2016jku}
  A.~Dumitru and V.~Skokov,
  \emph{$\cos(4\varphi)$ azimuthal anisotropy in small-$x$ DIS dijet production beyond the leading power TMD limit},
  \href{https://doi.org/10.1103/PhysRevD.94.014030}
  {\emph{Phys.\ Rev.\ D }{\bf 94}, 014030 (2016)},
  [\href{https://arxiv.org/abs/1605.02739}{{\tt hep-ph/1605.02739}}]. 
  
   
  \bibitem{Boer:2017xpy}
  D.~Boer, P.~J.~Mulders, J.~Zhou and Y.~Zhou,
  \emph{Suppression of maximal linear gluon polarization in angular asymmetries},
  \href{https://doi.org/10.1007/JHEP10(2017)196}
  {\emph{JHEP} {\bf 1710}, 196 (2017)}, 
  [\href{https://arxiv.org/abs/arXiv:1702.08195}{{\tt hep-ph/1702.08195}}].
   
   
   \bibitem{Marquet:2017xwy} 
  C.~Marquet, C.~Roiesnel and P.~Taels,
  \emph{{Linearly polarized small-$x$ gluons in forward heavy-quark pair production}}, 
  \href{https://doi.org/10.1103/PhysRevD.97.014004}
  {\emph{ Phys.\ Rev.\ D} {\bf 97}, no. 1, 014004 (2018)}, 
  [\href{https://arxiv.org/abs/1710.05698}{{\tt hep-ph/1710.05698}}].
 
  \bibitem{Altinoluk:2018byz} 
  T.~Altinoluk, R.~Boussarie, C.~Marquet and P.~Taels,
  \emph{{TMD factorization for dijets $+$ photon production from the dilute-dense CGC framework}}, 
  [\href{https://arxiv.org/abs/1810.11273}{{\tt hep-ph/1810.11273}}]. 
  
\bibitem{Petreska:2018cbf} 
  E.~Petreska,
 \emph{{TMD gluon distributions at small x in the CGC theory}}, 
 \href{https://doi.org/10.1142/S0218301318300035}
  {\emph {Int.\ J.\ Mod.\ Phys.\ E} {\bf 27}, no. 05, 1830003 (2018)}, 
  [\href{https://arxiv.org/abs/1804.04981}{{\tt hep-ph/1804.04981}}].
   

\bibitem{GolecBiernat:1998js} 
 K.~J.~Golec-Biernat and M.~Wusthoff,
  \emph{{Saturation effects in deep inelastic scattering at low Q**2 and its implications on diffraction}}, 
  \href{https://doi.org/10.1103/PhysRevD.59.014017}
  {\emph{Phys.\ Rev.\ D} {\bf 59}, 014017 (1998)}, 
  [\href{https://arxiv.org/abs/hep-ph/9807513}{{\tt hep-ph/9807513}}].

 \bibitem{Kotko:2015ura} 
  P.~Kotko, K.~Kutak, C.~Marquet, E.~Petreska, S.~Sapeta and A.~van Hameren,
  \emph{{Improved TMD factorization for forward dijet production in dilute-dense hadronic collisions}}, 
  \href{https://doi.org/10.1007/JHEP09(2015)106}
  {\emph{ JHEP} {\bf 1509}, 106 (2015)}, 
  [\href{https://arxiv.org/abs/1503.03421}{{\tt hep-ph/1503.03421}}].
  
  \bibitem{vanHameren:2016ftb} 
  A.~van Hameren, P.~Kotko, K.~Kutak, C.~Marquet, E.~Petreska and S.~Sapeta,
 \emph{{Forward di-jet production in p+Pb collisions in the small-x improved TMD factorization framework}}, 
 \href{https://doi.org/10.1007/JHEP12(2016)034}
 {\emph {JHEP} {\bf 1612}, 034 (2016)}, 
 [\href{https://arxiv.org/abs/1607.03121}{{\tt hep-ph/1607.03121}}].
 
 \bibitem{Bartels:2009tu}
 J.~Bartels, K.~Golec-Biernat and L.~Motyka,
 \href{https://doi.org/10.1103/PhysRevD.81.054017}
 {\emph{Phys.\ Rev.\ D} {\bf 81}, 054017 (2010)}, 
 [\href{http://arxiv.org/abs/0911.1935}{{\tt hep-ph/0911.1935}}]
  
  \bibitem{Balitsky:1987bk}
 I.~I.~Balitsky and V.~M.~Braun,
  \emph{{Evolution Equations for QCD String Operators}}, 
  \href{https://doi.org/10.1016/0550-3213(89)90168-5}
  {\emph{ Nucl.\ Phys.\ B} {\bf 311}, 541 (1989)}.
  
\bibitem{Wandzura:1997qf}
S.~Wandzura and F.~Wilczek,
 \emph{{Sum Rules for Spin Dependent Electroproduction: Test of Relativistic Constituent Quarks}}, 
  \href{https://doi.org/10.1016/0370-2693(77)90700-6}
 {\emph {Phys.\ Lett.\ } {\bf 72B}, 195 (1977)}.
  
  \bibitem{Altinoluk:2013rua} 
  T.~Altinoluk, C.~Contreras, A.~Kovner, E.~Levin, M.~Lublinsky and A.~Shulkin,
  \emph{{QCD Reggeon Calculus From KLWMIJ/JIMWLK Evolution: Vertices, Reggeization and All}}, 
  \href{https://doi.org/10.1007/JHEP09(2013)115}
  {\emph {JHEP} {\bf 1309}, 115 (2013)}, 
  [\href{https://arxiv.org/abs/1306.2794}{{\tt hep-ph/1306.2794}}].
  
  
  
%%%%%%%%%%%%%%%%%%%%%%%%%%%%%%%%%%%%

\bibitem{Dumitru:2005gt}
A.~Dumitru, A.~Hayashigaki and J.~Jalilian-Marian, \emph{{The color glass
  condensate and hadron production in the forward region}},
  \href{http://dx.doi.org/10.1016/j.nuclphysa.2005.11.014}{\emph{Nucl. Phys. A}
  {\bf 765} (feb, 2006) 464--482},
  [\href{http://arxiv.org/abs/hep-ph/0506308}{{\tt hep-ph/0506308}}].

   
   \bibitem{Altinoluk:2011qy} 
   T.~Altinoluk and A.~Kovner,
  \emph{{Particle Production at High Energy and Large Transverse Momentum - 'The Hybrid Formalism' Revisited}}, 
   \href{https://doi.org/10.1103/PhysRevD.83.105004}
   {\emph {Phys.\ Rev.\ D} {\bf 83} (2011) 105004}, 
 [\href{https://arxiv.org/abs/1102.5327}{{\tt hep-ph/102.5327}}].
    
\bibitem{Chirilli:2011km} 
 G.~A.~Chirilli, B.~W.~Xiao and F.~Yuan,
  \emph{{One-loop Factorization for Inclusive Hadron Production in $pA$ Collisions in the Saturation Formalism}}, 
  \href{https://doi.org/10.1103/PhysRevLett.108.122301} 
  {\emph{Phys.\ Rev.\ Lett.}  {\bf 108}, 122301 (2012)},
  [\href{https://arxiv.org/abs/1112.1061}{{\tt hep-ph/112.1061}}]. 
  
 \bibitem{Chirilli:2012jd} 
  G.~A.~Chirilli, B.~W.~Xiao and F.~Yuan,
 \emph{{ Inclusive Hadron Productions in pA Collisions}}, 
  \href{https://doi.org/10.1103/PhysRevD.86.054005}
  {\emph{Phys.\ Rev.\ D}{\bf 86}, 054005 (2012)}, 
  [\href{https://arxiv.org/abs/1203.6139}{{\tt hep-ph/1203.6139}}].
 
\bibitem{Stasto:2013cha} 
  A.~M.~Stasto, B.~W.~Xiao and D.~Zaslavsky,
  \emph{{Towards the Test of Saturation Physics Beyond Leading Logarithm}}, 
  \href{https://doi.org/10.1103/PhysRevLett.112.012302} 
  {\emph{Phys.\ Rev.\ Lett.}  {\bf 112}, no. 1, 012302 (2014)}, 
  [\href{https://arxiv.org/abs/1307.4057}{{\tt hep-ph/1307.4057}}].
 
 \bibitem{Stasto:2014sea} 
  A.~M.~Stasto, B.~W.~Xiao, F.~Yuan and D.~Zaslavsky,
  \emph{{ Matching collinear and small $x$ factorization calculations for inclusive hadron production in $pA$ collisions}}, 
  \href{https://doi.org/10.1103/PhysRevD.90.014047}
  {\emph{Phys.\ Rev.\ D} {\bf 90}, no. 1, 014047 (2014)}, 
  [\href{https://arxiv.org/abs/1405.6311}{{\tt hep-ph/1405.6311}}].
  
\bibitem{Altinoluk:2014eka} 
   T.~Altinoluk, N.~Armesto, G.~Beuf, A.~Kovner and M.~Lublinsky,
  \emph{{Single-inclusive particle production in proton-nucleus collisions at next-to-leading order in the hybrid formalism}}, 
  \href{https://doi.org/10.1103/PhysRevD.91.094016}
  {\emph{Phys.\ Rev.\ D} {\bf 91}, no. 9, 094016 (2015)}, 
  [\href{https://arxiv.org/abs/1411.2869}{{\tt hep-ph/1411.2869}}].
  
 \bibitem{Watanabe:2015tja} 
  K.~Watanabe, B.~W.~Xiao, F.~Yuan and D.~Zaslavsky,
  \emph{{Implementing the exact kinematical constraint in the saturation formalism}}, 
  \href{https://doi.org/10.1103/PhysRevD.92.034026}
  {\emph{Phys.\ Rev.\ D} {\bf 92}, no. 3, 034026 (2015)}, 
  [\href{https://arxiv.org/abs/1505.05183}{{\tt hep-ph/1505.05183 }}].
 
 \bibitem{Ducloue:2016shw} 
 B.~DuclouŽ, T.~Lappi and Y.~Zhu,
  \emph{{Single inclusive forward hadron production at next-to-leading order}}, 
 \href{https://doi.org/10.1103/PhysRevD.93.114016}
  {\emph{Phys.\ Rev.\ D} {\bf 93}, no. 11, 114016 (2016)}, 
  [\href{https://arxiv.org/abs/1604.00225}{{\tt hep-ph/1604.00225}}].
 
\bibitem{Iancu:2016vyg} 
 E.~Iancu, A.~H.~Mueller and D.~N.~Triantafyllopoulos,
 \emph{{CGC factorization for forward particle production in proton-nucleus collisions at next-to-leading order}}, 
  \href{https://doi.org/10.1007/JHEP12(2016)041}
  {\emph{JHEP} {\bf 1612}, 041 (2016)}, 
  [\href{https://arxiv.org/abs/1608.05293}{{\tt hep-ph/1608.05293}}].
  
\bibitem{Ducloue:2017mpb} 
  B.~DuclouŽ, T.~Lappi and Y.~Zhu,
  \emph{{Implementation of NLO high energy factorization in single inclusive forward hadron production}}, 
 \href{https://doi.org/10.1103/PhysRevD.95.114007}
  {\emph{Phys.\ Rev.\ D} {\bf 95}, no. 11, 114007 (2017)},
  [\href{https://arxiv.org/abs/1703.04962}{{\tt  hep-ph/1703.04962}}]. 

 \bibitem{Altinoluk:2015vax} 
T.~Altinoluk, N.~Armesto, G.~Beuf, A.~Kovner and M.~Lublinsky,
 \emph{{Heavy quarks in proton-nucleus collisions - the hybrid formalism}}, 
  \href{https://doi.org/10.1103/PhysRevD.93.054049}
  {\emph{Phys.\ Rev.\ D} {\bf 93}, no. 5, 054049 (2016)}, 
  [\href{https://arxiv.org/abs/1511.09415}{{\tt hep-ph/1511.09415}}].

 \bibitem{Altinoluk:2018uax} 
 T.~Altinoluk, N.~Armesto, A.~Kovner, M.~Lublinsky and E.~Petreska,
  \emph{{Soft photon and two hard jets forward production in proton-nucleus collisions}}, 
 \href{https://doi.org/10.1007/JHEP04(2018)063}
  {\emph{JHEP} {\bf 1804}, 063 (2018)}, 
  [\href{https://arxiv.org/abs/1802.01398}{{\tt hep-ph/1802.01398}}].
 
 \bibitem{Iancu:2018hwa} 
E.~Iancu and Y.~Mulian,
  \emph{{Forward trijet production in proton-nucleus collisions}}, 
  [\href{https://arxiv.org/abs/1809.05526}{{\tt hep-ph/1809.05526}}]. 
  


%%%%%%%%%%%%%%%%%%%%%%%%%%%%%%%%%%%%


\bibitem{Antonov:2004hh}
E.~Antonov, I.~Cherednikov, E.~Kuraev and L.~Lipatov, \emph{{Feynman rules for
  effective Regge action}},
  \href{http://dx.doi.org/10.1016/j.nuclphysb.2005.05.013}{\emph{Nucl. Phys. B}
  {\bf 721} (aug, 2005) 111--135}, [\href{http://arxiv.org/abs/0411185}{{\tt
  0411185}}].

\bibitem{VanHameren2012}
A.~{Van Hameren}, P.~Kotko and K.~Kutak, \emph{{Multi-gluon helicity amplitudes
  with one off-shell leg within high energy factorization}}, {\emph{J. High
  Energy Phys.} {\bf 2012} (2012) },
  [\href{http://arxiv.org/abs/1207.3332}{{\tt 1207.3332}}].

\bibitem{VanHameren2013a}
A.~{Van Hameren}, P.~Kotko and K.~Kutak, \emph{{Helicity amplitudes for
  high-energy scattering}}, {\emph{J. High Energy Phys.} {\bf 2013} (2013) }.

\bibitem{vanHameren:2014iua}
A.~{Van Hameren}, \emph{{BCFW recursion for off-shell gluons}},
  \href{http://dx.doi.org/10.1007/JHEP07(2014)138}{\emph{J. High Energy Phys.}
  {\bf 2014} (jul, 2014) 138}, [\href{http://arxiv.org/abs/1404.7818}{{\tt
  1404.7818}}].

\bibitem{vanHameren:2015bba}
A.~van Hameren and M.~Serino, \emph{{BCFW recursion for TMD parton
  scattering}}, \href{http://dx.doi.org/10.1007/JHEP07(2015)010}{\emph{J. High
  Energy Phys.} {\bf 2015} (jul, 2015) 10}.

\bibitem{VanHameren2017}
A.~van Hameren, \emph{{Calculating off-shell one-loop amplitudes for
  $k_T$-dependent factorization: a proof of concept}},
  \href{http://arxiv.org/abs/1710.07609}{{\tt 1710.07609}}.

\bibitem{Kotko2014a}
P.~Kotko, \emph{{Wilson lines and gauge invariant off-shell amplitudes}},
  \href{http://dx.doi.org/10.1007/JHEP07(2014)128}{\emph{J. High Energy Phys.}
  {\bf 2014} (jul, 2014) 128}, [\href{http://arxiv.org/abs/1403.4824}{{\tt
  1403.4824}}].

\bibitem{Mangano:1990by}
M.~L. Mangano and S.~J. Parke, \emph{{Multi-parton amplitudes in gauge
  theories}}, \href{http://dx.doi.org/10.1016/0370-1573(91)90091-Y}{\emph{Phys.
  Rep.} {\bf 200} (feb, 1991) 301--367},
  [\href{http://arxiv.org/abs/hep-th/0509223}{{\tt hep-th/0509223}}].

\bibitem{Bury2018}
M.~Bury, P.~Kotko and K.~Kutak, \emph{{TMD gluon distributions for multiparton
  processes}},  \href{http://arxiv.org/abs/1809.08968}{{\tt 1809.08968}}.

\bibitem{Deak:2009xt}
M.~Deak, F.~Hautmann, H.~Jung and K.~Kutak, \emph{{Forward jet production at
  the Large Hadron Collider}},
  \href{http://dx.doi.org/10.1088/1126-6708/2009/09/121}{\emph{J. High Energy
  Phys.} {\bf 2009} (sep, 2009) 121--121},
  [\href{http://arxiv.org/abs/0908.0538}{{\tt 0908.0538}}].





\bibitem{Rogers:2010dm}
T.~C. Rogers and P.~J. Mulders, \emph{{No generalized transverse momentum
  dependent factorization in the hadroproduction of high transverse momentum
  hadrons}}, \href{http://dx.doi.org/10.1103/PhysRevD.81.094006}{\emph{Phys.
  Rev. D} {\bf 81} (may, 2010) 094006},
  [\href{http://arxiv.org/abs/1001.2977}{{\tt 1001.2977}}].

\bibitem{Balitsky2015a}
I.~Balitsky and A.~Tarasov, \emph{{Rapidity evolution of gluon TMD from low to
  moderate x}}, \href{http://dx.doi.org/10.1007/JHEP10(2015)017}{\emph{J. High
  Energy Phys.} {\bf 2015} (oct, 2015) 17}.

\bibitem{Balitsky2016}
I.~Balitsky and A.~Tarasov, \emph{{Gluon TMD in particle production from low to
  moderate x}}, \href{http://dx.doi.org/10.1007/JHEP06(2016)164}{\emph{J. High
  Energy Phys.} {\bf 2016} (mar, 2016) 164},
  [\href{http://arxiv.org/abs/1603.06548}{{\tt 1603.06548}}].


\bibitem{Kotko2017b}
P.~Kotko, K.~Kutak, S.~Sapeta, A.~M. Stasto and M.~Strikman, \emph{{Estimating
  nonlinear effects in forward dijet production in ultra-peripheral heavy ion
  collisions at the LHC}},
  \href{http://dx.doi.org/10.1140/epjc/s10052-017-4906-6}{\emph{Eur. Phys. J.
  C} {\bf 77} (may, 2017) 353}, [\href{http://arxiv.org/abs/1702.03063}{{\tt
  1702.03063}}].
  
 
 
 
 \bibitem{Altinoluk:2014oxa} 
  T.~Altinoluk, N.~Armesto, G.~Beuf, M.~Mart'nez and C.~A.~Salgado,
  \emph{{Next-to-eikonal corrections in the CGC: gluon production and spin asymmetries in pA collisions}}, 
  \href{https://doi.org/10.1007/JHEP07(2014)068}
  {\emph{JHEP }{\bf 1407}, 068 (2014)}, 
  [\href{https://arxiv.org/abs/1404.2219}{{\tt hep-ph/404.2219}}].
  
  \bibitem{Altinoluk:2015gia} 
  T.~Altinoluk, N.~Armesto, G.~Beuf and A.~Moscoso,
  \emph{{Next-to-next-to-eikonal corrections in the CGC}}, 
  \href{https://doi.org/10.1007/JHEP01(2016)114}
  {\emph{JHEP} {\bf 1601}, 114 (2016)}, 
  [\href{https://arxiv.org/abs/1505.01400}{{\tt hep-ph/1505.01400}}]. 
 
 \bibitem{Altinoluk:2015xuy} 
  T.~Altinoluk and A.~Dumitru,
  \emph{{Particle production in high-energy collisions beyond the shockwave limit}}, 
  \href{https://doi.org/10.1103/PhysRevD.94.074032}
  {\emph{Phys.\ Rev.\ D} {\bf 94}, no. 7, 074032 (2016)}, 
  [\href{https://arxiv.org/abs/1512.00279}{{\tt hep-ph/1512.00279}}].
  
  
  \bibitem{Balitsky:2015qba} 
  I.~Balitsky and A.~Tarasov,
  \emph{{Rapidity evolution of gluon TMD from low to moderate x}},
  \href{https://doi.org/10.1007/JHEP10(2015)017}
  {\emph{JHEP} {\bf 1510}, 017 (2015)}, 
  [\href{https://arxiv.org/abs/1505.02151}{{\tt hep-ph/1505.02151}}].
  
  \bibitem{Balitsky:2016dgz} 
  I.~Balitsky and A.~Tarasov,
  \emph{{Gluon TMD in particle production from low to moderate x}}, 
  \href{https://doi.org/10.1007/JHEP06(2016)164}
  {\emph{JHEP} {\bf 1606}, 164 (2016)}, 
  [\href{https://arxiv.org/abs/1603.06548}{{\tt hep-ph/1603.06548}}].
  
  \bibitem{Balitsky:2017flc} 
  I.~Balitsky and A.~Tarasov,
  \emph{{Higher-twist corrections to gluon TMD factorization}},
  \href{https://doi.org/10.1007/JHEP07(2017)095}
  {\emph{JHEP} {\bf 1707}, 095 (2017)}, 
  [\href{https://arxiv.org/abs/1706.01415}{{\tt  hep-ph/1706.01415}}].
  
  \bibitem{Chirilli:2018kkw} 
  G.~A.~Chirilli,
  \emph{{Sub-eikonal corrections to scattering amplitudes at high energy}}, 
  [\href{https://arxiv.org/abs/1807.11435}{{\tt hep-ph/1807.11435}}].
 
 \bibitem{Kovchegov:2015pbl}  
  Y.~V.~Kovchegov, D.~Pitonyak and M.~D.~Sievert,
  \emph{{Helicity Evolution at Small-x}}
  \href{https://doi.org/10.1007/JHEP01(2016)072}
  {\emph {JHEP} {\bf 1601}, 072 (2016)}, 
  \href{https://doi.org/10.1007/JHEP10(2016)148}
  {\emph{Erratum:}} [\emph{ {JHEP} {\bf 1610}, 148 (2016)}], 
  [\href{https://arxiv.org/abs/1511.06737}{{\tt hep-ph/1511.06737}}].
  
  \bibitem{Kovchegov:2016zex} 
  Y.~V.~Kovchegov, D.~Pitonyak and M.~D.~Sievert,
  \emph{{Helicity Evolution at Small $x$: Flavor Singlet and Non-Singlet Observables}}, 
  \href{https://doi.org/10.1103/PhysRevD.95.014033}
  {\emph{Phys.\ Rev.\ D} {\bf 95}, no. 1, 014033 (2017)}, 
  [\href{https://arxiv.org/abs/1610.06197}{{\tt hep-ph/1610.06197}}].
  
  \bibitem{Kovchegov:2018znm} 
  Y.~V.~Kovchegov and M.~D.~Sievert
  \emph{Small-$x$ Helicity Evolution: an Operator Treatment}, 
  [\href{https://arxiv.org/abs/1808.09010}{{\tt hep-ph/1808.09010}}]. 
   
  \bibitem{genuine}
  T.~Altinoluk, R.~Boussarie, \textit{in preparation}.
  
\end{thebibliography}
\end{document}